\documentclass[twocolumn,floats,floatfix,showpacs,prd,superscriptaddress,nofootinbib]{revtex4-2}

\usepackage{graphicx,epsfig}
\usepackage{amssymb,amsmath,amsthm,amsfonts}
\usepackage{bm}
\usepackage[inline]{enumitem}
\usepackage{tensor}
\usepackage[linktocpage]{hyperref}
\usepackage[usenames,dvipsnames]{xcolor}
\usepackage{url}
\usepackage[inline]{enumitem}
\usepackage{xspace}
\usepackage{comment}
\usepackage[caption=false, justification=RaggedRight,singlelinecheck=false]{subfig}
\usepackage{placeins}
\usepackage{accents}

\newcommand{\dbtilde}[1]{\accentset{\approx}{#1}}
\newcommand*{\bolddot}[1]{\accentset{\mbox{\large\bfseries .}}{#1}}
\newcommand*{\boldddot}[1]{\accentset{\mbox{\large\bfseries ..}}{#1}}

\def\newacronym#1#2#3{\gdef#1{\gdef#1{#2\xspace}#3 (#2)\xspace}}
\newacronym{\amr}{AMR}{adaptive mesh refinement}
\def\bh#1{black hole#1 (BH#1)\gdef\bh{BH}}
\newacronym{\bbh}{BBH}{black-hole binary}
\newacronym{\gr}{GR}{General Relativity}
\newacronym{\bssn}{BSSN}{Baumgarte-Shapiro-Shibata-Nakamura}
\newacronym{\sGB}{sGB}{scalar Gauss--Bonnet}
\newacronym{\dCS}{dCS}{dynamical Chern--Simons}
\newacronym{\EdGB}{EdGB}{Einstein-dilaton-Gauss--Bonnet}

\newacronym{\BL}{BL}{Boyer-Lindquist}
\newacronym{\EFT}{EFT}{effective field theory}
\newacronym{\EFTs}{EFTs}{effective field theories}
\newacronym{\EHT}{EHT}{Event Horizon Telescope}
\newacronym{\GW}{GW}{gravitational-wave}
\newacronym{\LVK}{LVK}{LIGO-Virgo-KAGRA}
\newacronym{\RL}{RL}{refinement level}

\def\Canuda{\textsc{Canuda}\,}
\def\CanudaAD{\textsc{Canuda}--\textsc{AxiDil}\,}
\def\CanudadCS{\textsc{Canuda}--\textsc{dCS}\,}
\def\CanudaEdGB{\textsc{Canuda}--\textsc{EdGB}\,}
\def\ETK{\textsc{Einstein Toolkit}\,}

\def\Carpet{\textsc{Carpet}\xspace}
\def\Cactus{\textsc{Cactus}\xspace}

\def\kuibit{\textsc{kuibit}\xspace}

\def\dif{\textrm{d}}
\def\p{\partial}

\def\rex{r_{\rm{ex}}}

\def\as{\alpha_{\rm s}}
\def\aCS{\alpha_{\rm CS}}
\def\aGB{\alpha_{\rm GB}}
\def\ha{\hat{\alpha}_{\rm s}}
\def\haGB{\hat{\alpha}_{\rm{GB}}}
\def\haCS{\hat{\alpha}_{\rm{CS}}}
\def\BoxGR{\overline{\Box}}
\def\nablaGR{\overline{\nabla}}
\def\RGBGR{\overline{\mathcal{R}}_{\rm{GB}}}
\def\RCSGR{\overline{\mathcal{R}}_{\rm{CS}}}
\def\RGR{\overline{R}}
\def\RGB{\mathcal{R}_{\rm{GB}}}
\def\CGB{\mathcal{C}^{\rm{GB}}}
\def\RCS{\mathcal{R}_{\rm{CS}} }
\def\CCS{\mathcal{C}^{\rm{CS}}}

\def\dualR{\tilde{R}}
\def\ddualR{\dbtilde{R}}
\def\Kphi{K_{\rm \Phi}}
\def\Ktheta{K_{\rm \Theta}}
\def\gph{g_{\Phi}}
\def\dgph{g'_{\Phi}}
\def\fph{f_{\Phi}}
\def\dfph{f'_{\Phi}}
\def\ddfph{f''_{\Phi}}
\def\fth{f_{\Theta}}
\def\dfth{\bolddot{f}_{\Theta}}
\def\ddfth{\boldddot{f}_{\Theta}}

\def\Vph{V_{\Phi}}
\def\dVph{V'_{\Phi}}
\def\Vth{V_{\Theta}}
\def\dVth{\bolddot{V}_{\Theta}}

\def\Lie{\mathcal{L}}

\def\D{\mathcal{D}}
\def\E{\mathcal{E}}
\def\F{\mathcal{F}}
\def\G{\mathcal{G}}
\def\H{\mathcal{H}}

\def\M{\mathcal{M}}

\def\R{\mathcal{R}}
\def\S{\mathcal{S}}

\def\tgam{\tilde{\gamma}}
\def\tGam{\tilde{\Gamma}}
\def\tA{\tilde{A}}
\def\tB{\tilde{B}}
\def\tD{\tilde{D}}
\def\tE{\tilde{E}}

\def\tscF{\tilde{\F}}
\def\tscG{\tilde{\G}}
\def\tR{\tilde{R}}

\def\BL{Boyer-Lindquist}
\def\rBL{r_{\rm{BL}}}
\def\rBLp{r_{\rm{BL,+}}}

\begin{document}

\title{Growing black-hole hair in nonminimally coupled biscalar gravity}

\author{Chloe Richards}\email{chloer3@illinois.edu}
\affiliation{The Grainger College of Engineering,
Department of Physics \& Illinois Center for Advanced Studies of the Universe, University of Illinois Urbana-Champaign, Urbana, Illinois 61801, USA}

\author{Alexandru Dima}\email{alexandru.dima@uniroma1.it}
\affiliation{The Grainger College of Engineering,
Department of Physics \& Illinois Center for Advanced Studies of the Universe, University of Illinois Urbana-Champaign, Urbana, Illinois 61801, USA}
\affiliation{Dipartimento di Fisica, Sapienza Università 
	di Roma, Piazzale Aldo Moro 5, 00185, Roma, Italy}
\affiliation{INFN, Sezione di Roma, Piazzale Aldo Moro 2, 00185, Roma, Italy}

\author{Deborah Ferguson}\email{dferg@illinois.edu}
\affiliation{The Grainger College of Engineering,
Department of Physics \& Illinois Center for Advanced Studies of the Universe, University of Illinois Urbana-Champaign, Urbana, Illinois 61801, USA}

\author{Helvi Witek}\email{hwitek@illinois.edu}
\affiliation{The Grainger College of Engineering,
Department of Physics \& Illinois Center for Advanced Studies of the Universe, University of Illinois Urbana-Champaign, Urbana, Illinois 61801, USA}

\begin{abstract}
Black holes offer a unique laboratory for fundamental physics and are crucial for probing theories beyond Einstein's theory of General Relativity.
In this paper, we consider 4D effective field theories with scalar fields.
We focus on axi-dilaton gravity, a quadratic gravity theory with two kinetically coupled scalar fields, an axion and a dilaton.
To evolve these fields around black holes,
we introduce \CanudaAD, the first open-source, parameterized numerical relativity code for quadratic and bi-scalar gravity. 
Using this code, we perform single black hole simulations to show the dynamical formation of axion and dilaton hairs.
Through these simulations, we measure the impact of black-hole spin and curvature coupling strength on the axion and dilaton, and show that a kinetic coupling between the fields increases the observed deviations from General Relativity.
Furthermore, we simulate the axion and dilaton fields 
around a binary black hole coalescence
demonstrating
the growth of axion hair during the inspiral and  
the production of
radiative modes for both fields.

\end{abstract}

\maketitle
\tableofcontents

\clearpage
\newpage

\section{Introduction}\label{sec:intro}

The measurement of \GW signals 
over the last decade
has not only confirmed the existence of \bbh coalescences~\cite{LIGOScientific:2016aoc, LIGOScientific:2018mvr, LIGOScientific:2020ibl, LIGOScientific:2021usb, KAGRA:2021vkt, 
Poggiani:2024aat},
but has also spurred a new interest in using \bh{s} as laboratories to search for new signatures of fundamental physics~\cite{Yunes:2013dva, Berti:2015itd, LISA:2022kgy, Yunes:2024lzm}.
Current, ground-based \GW detectors are only the beginning; the potential to 
search for signatures of new 
phenomena
using \bh{s}
becomes increasingly accessible as new experiments
come online~\cite{Sathyaprakash:2019yqt,Kalogera:2021bya}.

In addition, collaboration across the gravity and high energy
physics communities continues to push our understanding of fundamental physics to new frontiers.
One area of interest shared by these communities is using compact objects to probe for corrections to 
\gr by considering 
additional degrees of freedom coupled to curvature.    
Recent results by the
\LVK~\cite{LIGOScientific:2014pky, VIRGO:2014yos, KAGRA:2018plz} 
and NANOGrav collaborations~\cite{NANOGrav:2023gor},
and upcoming \GW observatories such as
the recently approved LISA mission~\cite{LISA:2017pwj},
aim to constrain deviations from \gr~\cite{Berti:2015itd,Yunes:2013dva,LIGOScientific:2021sio,Ghosh:2022xhn,Yunes:2024lzm, Barausse:2020rsu, LISA:2022kgy}. 
The prospect of detecting
high curvature deviations from \gr{} increases as we detect smaller \bh{s};
experiments such as LIGO \cite{LIGOScientific:2022hai} and Cosmic Explorer \cite{Reitze:2019iox}
would be able to detect sub-solar mass \bh{s} and probe extreme regions of curvature.

While many beyond \gr theories have been proposed, we 
focus on bi-scalar--tensor models that are contained in the four-dimensional (4D)
string-inspired \EFT of gravity \cite{Kanti:1995cp, Cano:2021rey}. 
Here, we focus on axi-dilaton gravity that consists of 
a metric and two scalar fields
that are kinetically coupled to each other and 
nonminimally coupled to spacetime curvature.
We adopt a phenomenological approach to these 
models
(despite their high-energy origin)
and, thus, we do not assume the coupling length to be the Planck scale.

The 4D \EFT of gravity considered in this paper
contains \dCS gravity \cite{Jackiw:2003pm,Alexander:2009tp} and \sGB gravity \cite{Kanti:1995vq} as special cases.
They are popular theories in the context of high curvature corrections to \gr{.}
In both \dCS and \sGB gravity 
an additional scalar degree of freedom 
couples to a quadratic curvature invariant 
to form \bh{} hair~\cite{Stein:2014xba, Sotiriou:2015pka,Herdeiro:2015waa,Blazquez-Salcedo:2016yka,Benkel:2016rlz,
Silva:2017uqg,Doneva:2017bvd,Dima:2020yac,Doneva:2021tvn,HegadeKR:2022xij,Doneva:2022ewd,Doneva:2022yqu}.
In \dCS gravity, a pseudo-scalar field
(or axion)
couples to the Pontryagin density which sources axion hair.
In \sGB gravity, a scalar field known as the dilaton 
couples to the Gauss-Bonnet invariant which sources dilaton hair.

The effect of coupling to higher curvature terms
on \GW{s} produced in \dCS gravity has been studied with 
post-Newtonian methods \cite{Yagi:2011xp,Loutrel:2018rxs,Loutrel:2018ydv,Loutrel:2022tbk, Li:2022grj, Jenks:2023pmk, Li:2023lqz}, numerical simulations using an order-by-order expansion scheme~\cite{Okounkova:2017yby,Okounkova:2019dfo,Okounkova:2019zjf}, and \bh{}  perturbation theory \cite{Yunes:2007ss,Cardoso:2009pk,Molina:2010fb,Kimura:2018nxk,Owen:2021eez,Wagle:2021tam,Li:2022pcy,Wagle:2023fwl} to probe the inspiral, merger, and ringdown phases of a \bbh coalescence, respectively.
Furthermore, \bbh merger simulations with similar parameters as those from \LVK data have been performed to constrain \dCS gravity~\cite{Okounkova:2022grv}.

Simulations of single \bh{s} have been performed to investigate hairy, rotating \bh{s} in \dCS gravity~\cite{Konno:2014qua,Stein:2014xba,McNees:2015srl,Okounkova:2018abo, Okounkova:2018pql,Doneva:2021dcc}.
Previous studies have also considered a massive axion potential around single \bh{s}
to explore 
the quasi-bound states formed by an ultra-light axion field coupled to the \dCS curvature invariant \cite{Molina:2010fb,Macedo:2018txb,Alexander:2022avt,Richards:2023xsr}.

Similarly, the effect on \GW propagation in \sGB gravity has been considered.
Post-Newtonian methods have been used to study the inspiral phase of a \bbh coalescence \cite{Julie:2019sab,Shiralilou:2020gah,Shiralilou:2021mfl,Julie:2022qux,Bernard:2023eul,Julie:2024fwy}.
Numerical relativity simulations in the order-by-order expansion have been performed to estimate the effect during merger \cite{Witek:2018dmd,Witek:2020uzz,Silva:2020acr,Elley:2022ept}.
Furthermore, recent results of a well-posed formulation of \sGB gravity \cite{Kovacs:2020pns,Kovacs:2020ywu} have been implemented 
to perform \bbh simulations
and compute the \GW{} signal in the full theory~\cite{East:2020hgw, East:2021bqk,AresteSalo:2022hua,Corman:2022xqg,AresteSalo:2023mmd, Corman:2024cdr}.
The  ringdown phase of the \bbh coalescence has been studied using \bh{} perturbation theory \cite{Blazquez-Salcedo:2018jnn,Blazquez-Salcedo:2020rhf,Blazquez-Salcedo:2022omw,Luna:2024spo}.

In the decoupling approximation, single \bh{} simulations have been performed to demonstrate the formation of dilaton hair in nonrotating and rotating \bh{} backgrounds~\cite{Benkel:2016rlz,Benkel:2016kcq, Witek:2018dmd,Doneva:2021dqn,HegadeKR:2022xij}.
Beyond this approximation, single \bh{} simulations have been performed 
to include backreaction from dilaton hair growth~\cite{Ripley:2019aqj,Doneva:2023oww}.
Other work has determined conditions on predictivity,
and loss thereof,
during gravitational collapse
in \sGB gravity~\cite{R:2022hlf}. 

As summarized above, significant work has
gone into
studying beyond \gr effects in models where a single field is coupled to curvature, i.e., either \dCS or \sGB gravity.
It may even seem natural to study them separately as the axion is parity-odd while the dilaton is parity-even, thus probing different phenomena of coupling to curvature. 
In principle, however, both
extensions of \gr{} appear as leading-order corrections in the 4D \EFT of gravity~\cite{Kanti:1995cp, Cano:2021rey},
and therefore should
be considered together.

Recent works considering both the axion and dilaton, but no coupling between them, have explored the  fields in the context of
\bh{} hair formation \cite{R:2022tqa} 
and
testing slowly-rotating solutions via astrophysical observations
\cite{Cano:2023qqm}.
Other works have combined both fields via a kinetic coupling 
to study \GW{} propagation in the presence of both parity-violating and parity-conserving corrections
\cite{Daniel:2024lev} 
and
a wider set of cosmological applications~\cite{Burgess:2021qti,Brax:2022vlf,Brax:2023qyp,Smith:2024ibv}.

\begin{figure}
    \centering
    \includegraphics[width = \columnwidth]{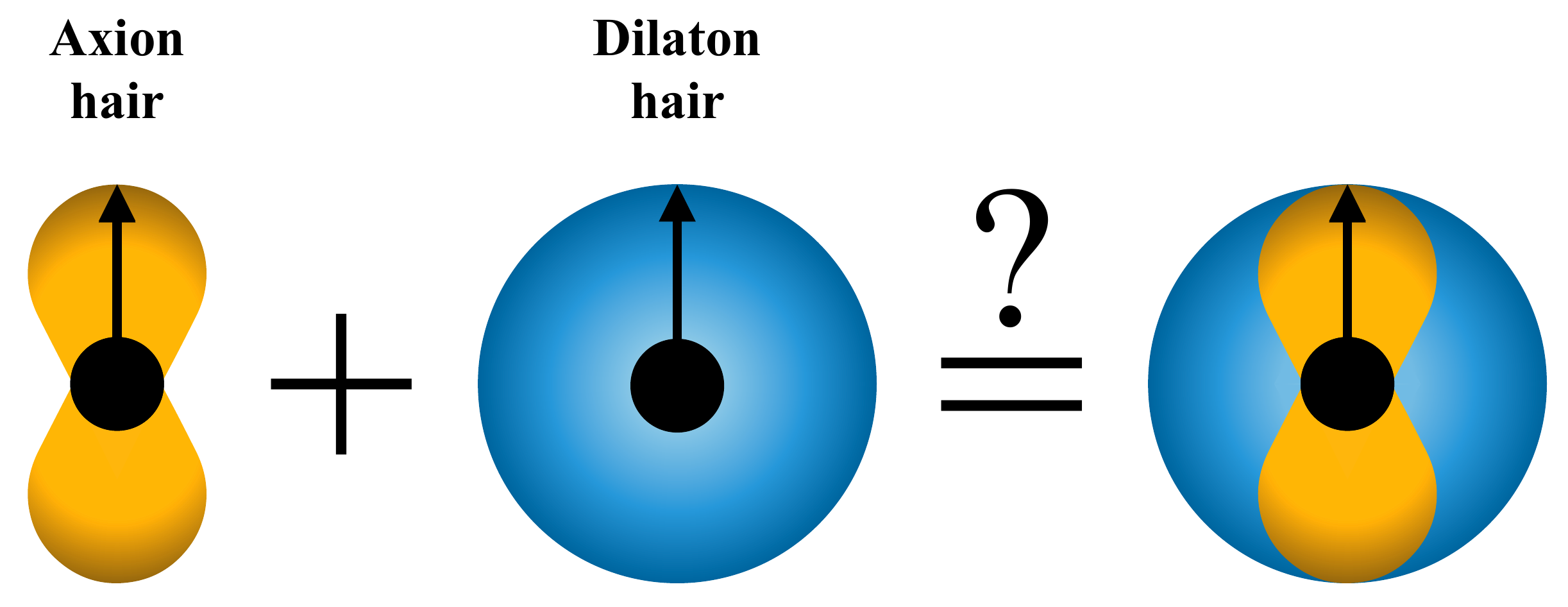}
    \caption{Sketch of coupled axion (yellow dipole) and dilaton (blue monopole) hairs around a spinning black hole, 
    with the arrow indicating its axis of rotation.
    }
    \label{fig:AxiDil_sketch}
\end{figure}

In this paper, we perform numerical relativity simulations 
to study 
axi-dilaton gravity which
combines \dCS and \sGB models through a kinetic coupling between the fields. 
We simulate
both single and binary \bh{} backgrounds 
and investigate the evolution of the axion and dilaton hairs. 
The sketch in Fig.~\ref{fig:AxiDil_sketch} illustrates the main idea:
on the left hand side of the 
pseudo-equation, we show the axion
and dilaton 
hairs after they evolve to a stationary profile around a 
single, spinning \bh{} where the axis of rotation is indicated by the arrow. 
On the right hand side of the pseudo-equation, we illustrate
how the axion and dilaton hairs 
might evolve together
when they are kinetically coupled.
We aim to solve this pseudo-equation by the end of this paper.

To explore the evolution of the right-hand side of the pseudo-equation in Fig.~\ref{fig:AxiDil_sketch},
we have developed the \CanudaAD code~\cite{CanudaAxiDil_repo} as a new module of the open-source \Canuda infrastructure, a numerical relativity code for fundamental physics \cite{witek_helvi_2023_7791842}. 
We implement the evolution equations of the axion and dilaton fields in axi-dilaton gravity in the decoupling approximation, 
i.e., the 
fields do not backreact onto the metric. 
\CanudaAD{}
provides the first, open-source 
parameterized numerical relativity code
capable to simulate
a large class of quadratic gravity and (bi-)scalar--tensor theories, including \dCS and \sGB gravity.

We present simulations of the axion and dilaton fields evolving in the background of single and binary \bh{s}. 
Our simulations show the dynamical formation of axion and dilaton hairs in both backgrounds.
We perform a series of single \bh{} runs
to address the following questions: 
(i) how does the \bh{} spin impact the evolution of the 
axion and dilaton hairs?
(ii) how does the coupling strength to the curvature invariants affect the 
axion and dilaton hairs? 
(iii) what is the effect of the kinetic coupling between the axion and dilaton fields on 
their respective
hairs? 

The \bbh simulation
provides a proof-of-principle computation
of the axion and dilaton fields' evolution 
in axi-dilaton gravity with a \bbh{} background.
Through this simulation, we investigate the formation of axion and dilaton hairs around \bbh{s} and compute the radiative multipoles produced by the inspiraling scalar charges.

The paper is organized as follows:
In Sec.~\ref{sec:QuadraticGravityTheory}, we define the action and field equations 
of the 4D \EFT{.}
In Sec.~\ref{sec:GravityModels} we
describe the (bi-)scalar--tensor theories and quadratic gravity theories that are captured by our generalized model.
In Sec.~\ref{sec:TimeEvolutionProblem}, we present the time evolution formulation in axi-dilaton gravity.
In Sec.~\ref{sec:NRFramework}, we
introduce our numerical relativity framework \CanudaAD{,} and summarize the set-up of the simulations presented in this paper.
In Sec.~\ref{sec:ResultsSingleBH} we present results of single \bh{} simulations,
and we show the proof-of-principle \bbh{} simulation in Sec.~\ref{sec:BBHSimulations}.
Finally we conclude in Sec.~\ref{sec:Summary}. 

Animations of the simulations presented in this paper are available on the Canuda  
YouTube channel 
\href{https://www.youtube.com/@canudanumericalrelativity1634}{www.youtube.com/@canudanumericalrelativity1634}.
The \CanudaAD code is publicly available at
\href{https://bitbucket.org/canuda/canuda_axidilaton}{https://bitbucket.org/canuda/canuda\_axidilaton}.
A sample simulation 
is available on the Einstein Toolkit website at \href{http://einsteintoolkit.org/gallery/axidilaton}{http://einsteintoolkit.org/gallery/axidilaton}.

In this paper we adopt the mostly-plus signature convention $(-,+,+,+)$ and use geometric units $G = c = 1$.

\section{Quadratic gravity in a nutshell}\label{sec:QuadraticGravityTheory}
\subsection{Action and field equations}\label{ssec:ActionAndEoMs}

We consider a generalization of 
bi-scalar--tensor
theories of gravity~\cite{Kanti:1995cp,Cano:2021rey} in which a pseudo-scalar, $\Theta$, and a scalar, $\Phi$, are nonminimally coupled to gravity.
The action is
\begin{align}
\label{eq:ActionAxiDil}
S & =  \kappa \int d^4x \sqrt{-g} \left( 
    ^{(4)}R 
    + \frac{\aCS}{4} \fth \, \RCS 
    + \frac{\aGB}{4}\fph\,\RGB  
\right. \nonumber\\ & \qquad\quad
    -  \gph^{2} \left( \frac{1}{2}(\nabla\Theta)^2  + \Vth \right)
    - \frac{1}{2}(\nabla\Phi)^2 - \Vph 
\left. \right)
\,,
\end{align}
where 
$^{(4)}R$ is the four-dimensional Ricci scalar constructed from the metric $g_{\mu\nu}$,
$\kappa=(16\pi G)^{-1}$,
$\fth\equiv\fth(\Theta)$, $\fph\equiv\fph(\Phi)$, and $\gph\equiv\gph(\Phi)$ are functions of the fields,
$\Vth\equiv\Vth(\Theta)$ and $\Vph\equiv\Vph(\Phi)$ are the fields' potentials,
and $\aCS$ and $\aGB$ are dimensionful coupling constants with units of length squared, $[\aCS]=[\aGB]=[L]^2$,
such that the (pseudo-)scalar fields are dimensionless, $[\Theta]=[L]^0$ and $[\Phi]=[L]^{0}$. 

These coupling constants are labeled with the ``Chern-Simons" (CS) and ``Gauss-Bonnet" (GB) acronyms respectively,
since the individual terms that they multiply 
correspond to \dCS or \sGB gravity; 
see Sec.~\ref{sec:GravityModels} for details.
The pseudo-scalar (or axion) field, $\Theta$, couples to the Pontryagin topological invariant
\begin{align}
\label{eq:Pontryagin}
\RCS= &\dualR^{abcd}R_{bacd} = -\frac{1}{2}\tensor{\epsilon}{^c^d_e_f}R^{abef}R_{abcd}
\,,
\end{align}
via the function $\fth$,
where 
$\dualR^{abcd}:=1/2 \epsilon^{cd}{}_{ef} R^{abef}$
is the dual of the Riemann tensor.
The scalar (or dilaton) field, $\Phi$, couples to the Gauss--Bonnet invariant
\begin{align}
\label{eq:GBInvariant}
\RGB = & -\ddualR^{abcd}R_{bacd}
\nonumber\\
= &  R^{2} - 4 R_{ab} R^{ab} + R_{abcd} R^{abcd}
\,,
\end{align}
via the function $\fph$,
where 
$\ddualR^{abcd}= 1/4 \epsilon^{ab}{}_{hg}\epsilon^{cd}{}_{ef}R^{efhg}$
is the double-dual of the Riemann tensor.

The pseudo-scalar and scalar fields 
are coupled to each other via the function
$\gph=\gph(\Phi)$.
For example,
choosing $\gph=\text{constant}$ decouples the fields,
while $\gph=1/\fph$ selects axi-dilaton gravity.
Depending on the choice of the (dimensionless) coupling functions, 
$\fth$ and $\fph$,
and on the self-interaction potentials, $\Vth=\Vth(\Theta)$ and $\Vph=\Vph(\Phi)$, different models of gravity can be selected.
In the following derivation we keep the coupling functions generic
and only specify them for the numerical simulations.
For brevity, we introduce the notation
\begin{align}
\bolddot\, =\dif/\dif\Theta
\,,\quad{\textrm{and}}\quad
' = \dif/\dif\Phi
\,,    
\end{align}
to indicate differentiation with respect to the pseudo-scalar $\Theta$ and scalar field $\Phi$.

Varying the action, Eq.~\eqref{eq:ActionAxiDil}, with respect to the pseudo-scalar, scalar, and metric
gives the field equations,
\begin{subequations}\label{eq:eoms}
\begin{align}
\label{eq:axionKG}
\Box\Theta & - \dVth + \frac{\aCS}{4}\frac{\dfth}{\gph^2}\RCS
 + 2 \frac{\dgph}{\gph} \nabla_{\mu}\Phi \nabla^{\mu}\Theta  = 0
\,,\\
\label{eq:dilatonKG}
\Box\Phi & - \dVph + \frac{\aGB}{4}\dfph\RGB - \dgph\gph\left[ (\nabla\Theta)^2 + 2 \Vth\right]
  = 0
\,,\\
\label{eq:Einstein}
R_{ab} & -\frac{1}{2}g_{ab}R - \frac{1}{2}T^{\rm eff}_{ab} = 0
\,,
\end{align}
\end{subequations}
where we define the effective energy--momentum tensor
\begin{align}
\label{eq:Tmneff}
T^{\rm eff}_{ab} & =  -2\, \left( \aCS\CCS_{ab} + \aGB\CGB_{ab} \right) 
\\ & \quad
    + \nabla_a\Phi\nabla_b\Phi - \frac{1}{2} g_{ab} \left((\nabla\Phi)^2 + 2 \Vph\right) 
\nonumber\\ & \quad
    + \gph^2\left( \nabla_a\Theta\nabla_b\Theta - \frac{1}{2} g_{ab}\left((\nabla\Theta)^2 + 2 \Vth\right)\right)
\,.\nonumber
\end{align}
Note that Eq.~\eqref{eq:axionKG} has been normalized by dividing by $\gph$, which is strictly allowed only if $\gph \neq 0$.
The $\mathcal{C}$-tensors that contain the coupling between the (pseudo-)scalar fields 
and the curvature tensors, are~\footnote{Note that the definition of $\CGB_{ab}$ differs from the one given in Ref.~\cite{Witek:2018dmd} by a factor $1/8$. Here, we absorb the factor in $\CGB_{ab}$.}
\begin{subequations}\label{eq:ctensor}
\begin{align}
\label{eq:CCS}
\CCS_{ab} := &  \E_c\epsilon^{cd}{}_{e(a}\nabla^e R_{b)d} +  \F^{cd}\,(^*R_{c(ab)d})
\,,\\
\label{eq:CGB}
\CGB_{ab}  = & R_{acbd} \G^{cd} + 2 R_{c(a} \G^{c}{}_{b)} - g_{ab} R_{cd} \G^{cd}
\nonumber\\ &
 -  G_{ab} {\rm{tr}}(\G) - \frac{1}{2} R\, \G_{ab}
\,,
\end{align}
\end{subequations}
where ${\rm{tr}}(\G)=g^{ab}\G_{ab}$,
and the auxiliary tensors are
\begin{subequations}
\label{eq:ctensorAux}
\begin{align}
\label{eq:Etensor}
\E_{a} := & \dfth \nabla_a\Theta 
\,,\\
\label{eq:Ftensor}
\F_{ab} := & \dfth \nabla_a\nabla_b\Theta
+ \ddfth \nabla_a\Theta\nabla_b\Theta 
\,,\\
\label{eq:Gtensor}
\G_{ab} := & \dfph \nabla_{a}\nabla_{b} \Phi +\ddfph \nabla_{a}\Phi \nabla_{b}\Phi
\,.
\end{align}
\label{eq:tensorAux}
\end{subequations}
In the next section we perform the $3+1$ decomposition of the field equations, where we
express the curvature invariants in terms of the gravito-electric and gravito-magnetic decomposition of the Weyl tensor
\begin{equation}
\label{eq:DefWeyl}
W_{abcd}:=R_{abcd}-(g_{a[c}R_{d]b}-g_{b[c}R_{d]a})+\frac{1}{3}g_{a[c}g_{d]b}R\,,
\end{equation}
and its dual, $^*W^{abcd}:=1/2\tensor{\epsilon}{^c^d_e_f}W^{abef}$.
Therefore we here rewrite the topological invariants as
\begin{subequations}\label{eq:TopInWeyl}
\begin{align}
\label{eq:RCS}
 \RCS & = \,^*W_{abcd}W^{abcd}
\,,\\
\label{eq:RGB}
\RGB & = W_{abcd}W^{abcd} - R_{ab}R^{ab} + \frac{2}{3} R^2
\,.
\end{align}
\end{subequations}
Note that Eqs.~\eqref{eq:TopInWeyl} can be proven to be always true regardless of the background~\cite{Grumiller:2007rv}.
Applying the Bianchi identities to Eq.~\eqref{eq:DefWeyl}, one can derive  
\begin{equation}
\nabla_a\tensor{W}{^a_b_c_d}=\nabla_{[c}R_{d]b}+\frac{1}{6}g_{b[c}\nabla_{d]}R\,,
\end{equation}
to recast Eq.~\eqref{eq:ctensor} in terms of the 
Weyl tensor and its dual,
\begin{subequations} 
\begin{align}
\CCS_{ab}\equiv &  2\E^c \nabla^d\,^*W_{d(ab)c} + \F^{cd}\,^*W_{d(ab)c}
\,,\\
\CGB_{ab}\equiv &  W_{acbd} \G^{cd} - \frac{1}{2}g_{ab} R_{cd} \G^{cd} + R_{c(a} \G^{c}{}_{b)} 
\nonumber\\ &
 - \frac{1}{2} {\rm{tr}}(\G) R_{ab} - \frac{1}{3} R\, \left(\G_{ab} - g_{ab} {\rm{tr}}(\G) \right)
\,.
\end{align}
\end{subequations}
One can check that $\CCS = g^{ab}\CCS_{ab} = 0$. 

\subsection{Decoupling limit}\label{ssec:decoupling}

The bi-scalar--tensor theories of gravity we focus on
likely lack a well-posed
formulation of the initial value problem when considered as a ``complete'' theory. 
This is mainly due to the axion sector of the model,
which has been conjectured to be ill-posed outside the
\EFT treatment~\cite{Delsate:2014hba}.
Thus, we work in the decoupling approximation of the theory 
in which we neglect the back-reaction of the axion and dilaton fields onto the spacetime metric.
Then, the background spacetime is determined by Einstein's equations in vacuum and can be a stationary or a time-dependent \gr solution.
This approximation enables us to investigate the (pseudo-) scalar fields' dynamics
in a range of interesting scenarios:
the formation of (pseudo-) scalar 
\bh{} hairs
and their scalar radiation generated in \bbh{s},
or the onset of \bh{} scalarization
and dynamical (de-)scalarization in \bbh{s}.
In the decoupling approximation, the field equations~\eqref{eq:eoms} reduce to
\begin{subequations}\label{eq:eoms_decoup}
\begin{align}
\label{eq:axionKG_decoup}
\BoxGR\Theta & - \dVth + \frac{\aCS}{4} \frac{\dfth}{\gph^2} \RCSGR
 + 2 \frac{\dgph}{\gph} \nablaGR_{\mu}\Phi \nablaGR^{\mu}\Theta  = 0
\,,\\
\label{eq:dilatonKG_decoup}
\BoxGR\Phi & - \dVph + \frac{\aGB}{4} \dfph \RGBGR - \dgph \gph \left[ (\nablaGR\Theta)^2 + 2 \Vth\right]
  = 0
\,,\\
\label{eq:EinsteinGR}
\RGR_{ab} & -\frac{1}{2}\bar{g}_{ab}\RGR  = 0
\,,
\end{align}
\end{subequations}
where the bar denotes operators and tensors constructed from the \gr background metric, $\bar{g}_{\mu\nu}$, and its derivatives. 
In the decoupling limit, in a vacuum \gr background for which $\RGR_{ab}=0$, one can express 
the Pontryagin density and Gauss--Bonnet invariant, Eqs.~\eqref{eq:TopInWeyl}, as 
\begin{align}
\label{eq:TopInWeylGR}
\overline{\R}_{\rm CS} =\, ^{\ast} \overline{W}_{abcd}\overline{W}^{abcd}
\,,\quad
\overline{\mathcal{R}}_{\rm GB} & = \overline{W}_{abcd}\overline{W}^{abcd}
\,.
\end{align}
For the remainder of the paper, 
we drop the bar
to avoid cumbersome notation.
The curvature tensors, derivatives, and metric refer to the background, i.e., vacuum \gr quantities.

To further simplify the discussion,
we refer to both the pseudo-scalar, $\Theta$, and scalar, $\Phi$, as ``scalars''.

\section{Bi-scalar--tensor models of gravity}\label{sec:GravityModels}

The field equations in the decoupling approximation, Eqs.~\eqref{eq:eoms_decoup},
represent a generalized scalar--tensor model of gravity in which (up to two) scalars are minimally or nonminimally coupled to gravity.
The generalized equations reduce to well-known models of modified gravity
through simple parameter choices.
In the following, we briefly describe the different models, determine their field equations,
and identify the appropriate choice of parameters and coupling functions for the model selection;
see also Table~\ref{tab:parameter choices} in App.~\ref{appsec:code_description}
on how to select them in the \CanudaAD code.
\subsection{Single minimally coupled scalar field}\label{ssec:MinimalScalar}

The simplest case  included in Eqs.~\eqref{eq:eoms_decoup}
is that of a single scalar field $\Phi$, with a potential $\Vph$, minimally coupled to gravity.
This corresponds to setting
\begin{align}\label{eq:EKG_couplings}
\Theta =0 \,,\quad
\fph = 0 \,,\quad
\fth = \Vth=0
\,.
\end{align}
Then, the action in Eq.~\eqref{eq:ActionAxiDil}
reduces to
\begin{align}
\label{eq:EKG_action}
S = & \kappa \int d^4x \sqrt{-g} \left( ^{(4)}R - \frac{1}{2}(\nabla\Phi)^2 - \Vph \right)
\,,
\end{align}
and the scalar's equation of motion
reduces to the Klein-Gordon equation,
\begin{align}
\label{eq:EKG_dilatonKG}
\Box \Phi & - \dVph 
  = 0
\,.
\end{align}

\subsection{Dynamical Chern-Simons gravity}\label{ssec:dCSgravity}

Staying with single scalar field models, we now consider the special case of \dCS gravity.
Here, the axion, $\Theta$, is nonminimally coupled to gravity via the Pontryagin density $\RCS$, while the dilaton field vanishes.
This corresponds to setting
\begin{align}\label{eq:dCS_couplings}
\Phi = 0 \,,
\quad
\fph = \Vph = 0 \,,
\quad
\gph = 1 \,.
\end{align}
With this choice, the action in Eq.~\eqref{eq:ActionAxiDil} reduces to
\begin{align}
\label{eq:dCS_action}
S = & \kappa \int d^4x \sqrt{-g} \left( ^{(4)}R - \frac{1}{2}(\nabla\Theta)^2  - \Vth + \frac{\aCS}{4} \fth \,\RCS \right)
\,.
\end{align}
The axion's field equation
becomes
\begin{align}
\label{eq:dCS_axioneom}
\Box \Theta & - \dVth + \frac{\aCS}{4}\dfth \RCS  = 0
\,.
\end{align}
The choices of coupling function $\fth$ available in the numerical implementation are
\begin{subequations}
\label{eq:dCS_CouplingFunctions}
\begin{align}
\label{eq:SS_dCS}
{\textrm{Shift-symmetric:}} &\quad
\fth = \Theta\,,\\ 
\label{eq:Cubic_dCS}
{\textrm{Cubic:}} & \quad
\fth = \Theta^3\,.
\end{align}
\end{subequations}
The first choice yields a shift-symmetric coupling function and is the standard for \dCS gravity~\cite{Alexander:2009tp}.
In principle, one could also consider higher-order polynomials as in the second choice, Eq.~\eqref{eq:Cubic_dCS},
that might yield new phenomena 
like spontaneous scalarization of \bh{s}~\cite{Doneva:2021dcc}. 

Typically, the potential $\Vth$ is set to zero. However, symmetry breaking in the axion section
may introduce a small mass-term \cite{Macedo:2018txb,Alexander:2022avt,Richards:2023xsr}.
The latter choice is also available in our numerical implementation.

\subsection{Scalar-Gauss-Bonnet gravity}
\label{sssec:sGBgravity}

Our model of quadratic gravity, furthermore, includes \sGB gravity as a special case.
Here, a single scalar field is nonminimally coupled to gravity via the Gauss-Bonnet invariant, $\RGB$.
This corresponds to the choice
\begin{align}
\label{eq:sGB_couplings}
\Theta = 0\,,\quad
\fth = \Vth = 0
\,.
\end{align}
Then, the action in Eq.~\eqref{eq:ActionAxiDil} reduces to 
\begin{align}
\label{eq:sGB_action}
S = & \kappa \int d^4x \sqrt{-g} \left( ^{(4)}R - \frac{1}{2}(\nabla\Phi)^2 - \Vph + \frac{\aGB}{4}\fph\,\RGB \right)
\,.
\end{align}
The resulting scalar field equation of motion is
\begin{align}
\label{eq:sGB_scalareom}
\Box \Phi & - \dVph + \frac{\aGB}{4} \dfph \RGB
= 0
\,.
\end{align}
This includes the original \EdGB{} gravity model~\cite{Mignemi:1992nt, Kanti:1995vq, Alexeev:1996vs, Torii:1996yi, Kanti:1997br}, 
the widely used shift-symmetric \sGB model \cite{Sotiriou:2013qea, Sotiriou:2014pfa},
as well as models with quadratic coupling functions that allow for scalarized \bh{s} \cite{Doneva:2017bvd, Silva:2017uqg}.
The respective choices are
\begin{subequations}
\label{eq:sGB_CouplingFunctions}
\begin{align}
\label{eq:sGB_DilatonCoupling}
{\textrm{Dilaton:}} &\quad
\fph = e^{\lambda_{\text{GB}} \Phi}\,,\\
\label{eq:sGB_LinearCoupling}
{\textrm{Shift-symmetric:}} &\quad
\fph = \Phi\,,\\ 
\label{eq:sGB_QuadraticCoupling}
{\textrm{Quadratic:}} & \quad
\fph = \Phi^2\,,
\end{align}
\end{subequations}
where $\lambda_{\text{GB}}$ is a dimensionless constant.
Note that the sign conventions in Ref.~\cite{Kanti:1995vq,Pani:2009wy} correspond to $\lambda_{\text{GB}}=+1$ in our model.
Typically, the potential is set to $\Vph=0$,
although including a mass term may yield new interesting phenomena~\cite{vanGemeren:2024bzf}.

\subsection{Minimally coupled bi-scalar field model}

Next, let us consider a model in which both the axion $\Theta$ and dilaton $\Phi$ fields are present and coupled to each other,
but they are minimally coupled to gravity.
The latter is achieved by setting
\begin{align}\label{eq:Emulti_couplings}
\fth =  0 
\,,\quad
\fph = 0 
\,.
\end{align}
Then, the action then becomes
\begin{align}
\label{eq:Emulti_action}
S = & \kappa \int d^4x \sqrt{-g} \left( 
    ^{(4)}R 
    - \frac{1}{2}(\nabla\Phi)^2 - \Vph
\right.\nonumber\\ & \left.\qquad\qquad\qquad
    - \gph^{2} \left[ \frac{1}{2}(\nabla\Theta)^2  + \Vth \right]
 \right)
\,,
\end{align}
where the function $\gph$ determines the coupling between the two fields.
The resulting field equations are
\begin{subequations}\label{eq:Emulti_eoms}
\begin{align}
\label{eq:Emulti_axioneom}
\Box\Theta & - \dVth 
 + 2 \frac{\dgph}{\gph} \nabla_{\mu}\Phi \nabla^{\mu}\Theta  = 0
\,,\\
\label{eq:Emulti_dilatoneom}
\Box\Phi & - \dVph - \dgph\gph\left[ (\nabla\Theta)^2 + 2 \Vth\right]
  = 0
\,.
\end{align}
\end{subequations}
The action~\eqref{eq:Emulti_action} is a special case of the more general multi-scalar-tensor theories~\cite{Damour:1992we,Horbatsch:2015bua}.
In these models, multiple scalars are coupled via a target-space metric, which reduces to $\textrm{diag}(1\,, \gph^2)$ in our case (e.g., compare with Eq.~(2.1) in \cite{Horbatsch:2015bua}).
Furthermore, the bi-scalar field action can describe dark energy models investigated in cosmological contexts~\cite{Burgess:2021qti,Brax:2022vlf,Brax:2023qyp,Smith:2024ibv}.

\subsection{Axi-dilaton gravity}\label{ssec:4dstring}

The generalized quadratic gravity action, Eq.~\eqref{eq:ActionAxiDil},
also includes axi-dilaton gravity \cite{Kanti:1995cp,Cano:2021rey}
and corresponds to
\begin{align}
\label{eq:axidilmodel}
& \fph = e^{\lambda_{\text{GB}}\Phi}
\,,\quad
\gph = e^{-\lambda_{\text{AD}}\Phi}
\,,\quad
\fth = \Theta\,,
\nonumber\\
&\Vph=0
\,,\quad
\Vth=0
\,,
\end{align}
where $\lambda_{\text{GB}}$ and $\lambda_{\text{AD}}$
are dimensionless constants.
We typically set $\lambda_{\text{GB}}=\lambda_{\text{AD}}=\lambda=+1$ 
in our simulations~\cite{Kanti:1997br,Yunes:2009hc}. 
In axi-dilaton gravity, the coupling constants are related as $\aCS=\aGB =\as$.\footnote{We note that the model in Eq.~\eqref{eq:axidilmodel} reduces to the one derived in~\cite{Cano:2021rey} when $2\aGB=-2\aCS=\alpha'$ and $\lambda=-1$ as the authors use the signature (+,--,--,--).}

Inserting the model parameters into Eq.~\eqref{eq:ActionAxiDil},
and keeping the parameter $\lambda$,
the axi-dilaton action becomes
\begin{align}
\label{eq:AxiDil_action}
S = & \kappa \int d^4x \sqrt{-g} \left( ^{(4)}R
    - \frac{1}{2}(\nabla\Phi)^2 
    - \frac{1}{2} e^{-2\lambda\Phi} (\nabla\Theta)^2  
\right. \nonumber\\ &  \left. 
    + \frac{\as}{4}\left( e^{\lambda\Phi}\,\RGB
        + \Theta \, \RCS\right)
\right)
\,.
\end{align}
Note, that $\as$ is a dimensionful coupling constant with units of length squared, $[\as]=[L]^{2}$.
For the numerical implementation it is useful to introduce dimensionless parameter,
$\ha = (\as /\textrm{M}^2)$.
Then, the axi-dilaton field equations evaluated on a \gr background are given by
\begin{subequations}
\label{eq:AxiDil_eoms_with_ahat}
\begin{align}
\label{eq:AxiDil_gen_axioneom}
\Box\Theta & + \frac{\ha \textrm{M}^2}{4}e^{2\lambda\Phi}\RCS
 - 2 \lambda \nabla_{\mu}\Phi \nabla^{\mu}\Theta  = 0
\,,\\
\label{eq:AxiDil_gen_dilatoneom}
\Box\Phi & + \frac{\ha \textrm{M}^2\lambda}{4}e^{\lambda\Phi}\RGB + \lambda e^{-2\lambda\Phi} (\nabla\Theta)^2
  = 0
\,.
\end{align}
\end{subequations}
The present paper focuses on the evolution of single and binary \bh{s} in axi-dilaton gravity, i.e., we adopt Eqs.~\eqref{eq:AxiDil_eoms_with_ahat}
for the remainder of this paper unless stated otherwise. 

\section{Time evolution formulation}\label{sec:TimeEvolutionProblem}

In this section, we present the time evolution formulation for our numerical simulations.
We 
describe the
background spacetime in Sec.~\ref{ssec:3p1SplitAndBackground},
provide the scalar field initial data in Sec.~\ref{ssec:ScalarInitialData},
and present the $3+1$ decomposition of the field equations in Sec.~\ref{ssec:ScalarEvolutionEquations}.

\subsection{Background spacetime}\label{ssec:3p1SplitAndBackground}

To conduct the numerical simulations of the scalar fields,
we rewrite Eqs.~\eqref{eq:eoms_decoup} as a time evolution problem.
We obtain the
time evolution 
formulation by foliating the four-dimensional spacetime $\left(\M,g_{ab}\right)$ into a set of three-dimensional, spatial hypersurfaces
$\left(\Sigma_{t},\gamma_{ij}\right)$.
Each hypersurface $\Sigma_t$ is labeled by the time parameter $t$ and 
its geometry is determined by the induced 
(or spatial)
metric
$\gamma_{ab}=g_{ab} + n_{a}n_{b}$.
Here, $n^{a}$ is the timelike unit vector normal to the hypersurface,
$\gamma^{a}{}_{b} n^{b} = 0$,
with normalization $n^{a}n_{a} = -1$.
Furthermore, the spatial metric defines a projection operator
\begin{align}
\label{eq:ProjOp}
\gamma^{a}{}_{b} = & \delta^{a}{}_{b} + n^{a} n_{b}
\,.
\end{align}
The line element takes the form
\begin{align}
\label{eq:LineElement3p1}
\dif s^{2} = & g_{ab} \dif x^{a} \dif x^{b}
 \\
        = & - \left(\alpha^{2} - \beta^{k} \beta_{k} \right) \dif t^{2}
            + 2 \gamma_{ij} \beta^{i} \dif t \dif x^{j}
            + \gamma_{ij} \dif x^{i} \dif x^{j}
\,. \nonumber
\end{align}
where $\alpha$ and $\beta^{i}$ are, respectively, the lapse function and shift vector.
We denote the covariant derivative and Ricci tensor associated with the 3-metric $\gamma_{ij}$ as $D_{i}$ and $R_{ij}$.
The extrinsic curvature $K_{ab}$ describes how a hypersurface
is embedded in the spacetime manifold,
and is defined as
\begin{align}
\label{eq:DefKij}
K_{ab} = & - \gamma^{c}{}_{a} \gamma^{d}{}_{b} \nabla_{c} n_{d}
        = - \frac{1}{2} \Lie_{n} \gamma_{ab}
\,,
\end{align}
where
$\Lie_{n}$ is the Lie derivative along $n^{a}$. 

We simulate the formation and time evolution of axion and dilaton fields
in the background of
1) a single rotating \bh{}, 
and
2) a dynamical spacetime consisting of a \bbh{} coalescence.
Please refer to Sec.~II~B of Ref.~\cite{Richards:2023xsr} and Sec.~III~B of Ref.~\cite{Witek:2018dmd}, respectively, for details.

\subsection{Scalar field initial data}\label{ssec:ScalarInitialData}

In this paper, we focus on two types of initial data for the axion and dilaton fields.
They are either initialized as zero 
to follow the formation of axion and dilaton hairs,
or they are initialized by the 
small-spin, small-coupling approximation~\cite{Cano:2021rey}
to determine their new end states for large \bh{} spins
and large coupling strengths. 
We also demonstrate the robustness of the end state and its independence from the initial data.

The \CanudaAD{} code provides additional choices, such as Gaussian profiles or quasi-bound states,
and we present a complete list 
in App.~\ref{appssec:SFID}.
Furthermore,
the initial data profiles can be mixed and matched for the axion and dilaton fields, i.e., the fields do not have to have the same type of initial data.

\noindent{\bf{Type I:} Zero fields}\label{eq:IDTypeI}
The first choice of initial data initializes both the axion and dilaton to be zero,
\begin{subequations}
\label{eq:IDZero}    
\begin{align}
\label{eq:IDZeroAxion}
\Theta|_{t=0}=0
\,,\qquad & 
\Ktheta|_{t=0}=0
\,, \\
\label{eq:IDZeroDilaton} 
\Phi|_{t=0}=0
\,,\qquad & 
\Kphi|_{t=0}=0
\,.
\end{align}
\end{subequations}

\noindent{\bf{Type II:} Axion and dilaton hairs in the small-spin, small-coupling approximation}
The second choice of initial data for the axion and dilaton implements the 
analytical solutions around single \bh{s}
in axi-dilaton gravity
in the small-spin, small-coupling approximation~\cite{Cano:2021rey}.
The fields' conjugated momenta vanish initially, as follows from their definition~\eqref{eq:DefKPhiKTheta}
and the symmetries of the background spacetime.

Then, the initial axion profile
in spherical-polar coordinated $\{r,\theta,\phi\}$
is given by
\begin{subequations}
\label{eq:IDAxionHair}
\begin{align}
\label{eq:IDAxionHairTheta}
\Theta|_{t=0} & \simeq  
    2 \ha \chi \cos\theta 
    \left[ \vartheta_{1} 
      - \chi^2 \left( \vartheta_{2}  
        + \vartheta_{3} \cos^2\theta \right) 
    \right] 
\,,\\
\label{eq:IDAxionHairKTheta}
\Ktheta|_{t=0} & = 0
\,,
\end{align}
\end{subequations}
where we neglected terms of order
$\mathcal{O}(\ha \chi^{5})$ and $\mathcal{O}(\ha^2 \chi)$,
and introduced
\begin{subequations}
\label{eq:IDAxionHair_vartheta}
\begin{align}
\vartheta_{1} & = \left(\frac{5 {M}^2}{16\rBL^2} + \frac{5 {M}^3}{8\rBL^3} + \frac{9 {M}^4}{8\rBL^4} \right) 
\,,\\
\vartheta_{2} & = \left(\frac{ {M}^2}{32\rBL^2} + \frac{ {M}^3}{16\rBL^3} + \frac{3 {M}^4}{40\rBL^4} + \frac{ {M}^5}{20\rBL^5}\right)
\,,\\
\vartheta_{3} & = \left(\frac{3 {M}^4}{8\rBL^4} + \frac{3 {M}^5}{2\rBL^5} + \frac{25 {M}^6}{6\rBL^6}\right)
\,.
\end{align}
\end{subequations}
Here, $\rBL$ denotes the Boyer-Lindquist radial coordinate.
It is related to the isotropic radial coordinate,
\begin{align}
\label{eq:IsoRad}
\rBL & = r\left(1+\frac{\rBLp}{4r}\right)^2
\,.
\end{align}
Here, $r_{\rm{BL,+}}/\textrm{M}= 1+\sqrt{1-\chi^2}$ denotes the outer horizon in \BL{} coordinates, and it corresponds to $r_{+}=\rBLp/4$ in isotropic radial coordinates.
The inner horizon, $r_{\rm{BL,-}}/\textrm{M}=1-\sqrt{1-\chi^2}$, is not covered by the isotropic coordinate chart.

The dilaton's initial profile is given by
\begin{subequations}
\label{eq:IDDilatonHair}
\begin{align}
\label{eq:IDDilatonHairPhi}
\Phi|_{t=0} & \simeq 2\lambda \ha 
    \left[ \varphi_{0}
        - \chi^2\left( \varphi_{1} + \varphi_{2}\cos^2\theta \right)
\right.\\ & \left.\quad\quad
        - \chi^4\left( \varphi_{3} - \varphi_{4} \cos^2\theta - \varphi_{5} \cos^4\theta\right)
    \right]
\,,\nonumber \\
\label{eq:IDDilatonHairKPhi}
\Kphi|_{t=0} & = 0
\,,
\end{align}
\end{subequations}
where we neglected terms of order
$\mathcal{O}(\ha\chi^{6})$ and $\mathcal{O}(\ha^2)$,
and introduced
\begin{subequations}
\label{eq:IDDilatonHair_varphi}
\begin{align}
\varphi_{0} & = \left( \frac{ {M}}{4\rBL} 
    + \frac{ {M}^2}{4\rBL^2} 
    + \frac{ {M}^3}{3\rBL^3} \right)
\,,\\
\varphi_{1} & = \left(\frac{ {M}}{16\rBL} 
    + \frac{ {M}^2}{16\rBL^2} 
    + \frac{ {M}^3}{20\rBL^3} 
    + \frac{ {M}^4}{40\rBL^4}\right)
\,,\\
\varphi_{2} & = \left( \frac{7 {M}^3}{20\rBL^3} 
    + \frac{21 {M}^4}{20\rBL^4} 
    + \frac{12 {M}^5}{5\rBL^5} \right)
\,,\\
\varphi_{3} & = \left(\frac{ {M}}{32\rBL}
    + \frac{ {M}^2}{32\rBL^2} 
    + \frac{3 {M}^3}{112\rBL^3} 
    + \frac{ {M}^4}{56\rBL^4} 
    + \frac{ {M}^5}{140\rBL^5}\right)
\,,\\
\varphi_{4} & = \left( \frac{ {M}^3}{56\rBL^3} 
    + \frac{3 {M}^4}{56\rBL^4}
    + \frac{3 {M}^5}{35\rBL^5} 
    + \frac{ {M}^6}{14\rBL^6} \right)
\,,\\
\varphi_{5} & = \left( \frac{11 {M}^5}{28\rBL^5}  
    + \frac{55 {M}^6}{28\rBL^6} 
    + \frac{45 {M}^7}{7\rBL^7} \right)
\,.
\end{align}
\end{subequations}

As a reminder,
we introduce the dimensionless constant 
$\lambda$ in Eq.~\eqref{eq:AxiDil_action}, and typically set $\lambda=+1$.
In passing, we also note that Eqs.~\eqref{eq:IDAxionHairTheta} and \eqref{eq:IDDilatonHairPhi} 
can be generalized by introducing dimensionless coupling parameters $\haCS = \aCS / \textrm{M}^2$ and $\haGB = \aGB / \textrm{M}^2$
such that $\haCS \neq \haGB$; please refer to App.~\ref{appsec:GeneralAxiDil3p1Eqs} for more details.

\subsection{Axi-dilaton scalar evolution equations}\label{ssec:ScalarEvolutionEquations}

In this project, we adopt the axi-dilaton model of gravity presented in Sec.~\ref{ssec:4dstring}.
We rewrite the scalar field equations, Eqs.~\eqref{eq:axionKG} and~\eqref{eq:dilatonKG}, in terms of the 3+1 variables
introduced in the previous subsection
and the conjugated momenta of the scalar fields
\begin{align}
\label{eq:DefKPhiKTheta}
\Ktheta & := - \Lie_{n} \Theta
\,,\quad 
\Kphi :=  - \Lie_{n} \Phi
\,.
\end{align}
This procedure yields evolution equations for the scalars and their conjugated momenta,
\begin{subequations}
\label{eq:AxiDil_gen_ADM}
\begin{align}
\label{eq:dttheta_gen_ADM}
\dif_{\rm t} \Theta        = & -   \alpha \Ktheta
\,,\\
\label{eq:dtphi_gen_ADM}
\dif_{\rm t} \Phi        = & -   \alpha \Kphi
\,,\\
\label{eq:dtKtheta_gen_ADM}
\dif_{t} \Ktheta =  & - \alpha D^{i} D_{i} \Theta - D^{i}\alpha D_{i} \Theta 
\nonumber\\
+ & \alpha \left( K \Ktheta - \frac{\ha \textrm{M}^2 \lambda}{4}e^{2\lambda\Phi}\RCS \right)
\nonumber\\
- & 2 \alpha\lambda\left( \frac{}{}\Ktheta \Kphi - D_i\Theta D^i\Phi \right)
\,,\\
\label{eq:dtKphi_gen_ADM}
\dif_{t} \Kphi =  & - \alpha D^{i} D_{i} \Phi  - D^{i}\alpha D_{i} \Phi
\nonumber\\
 + & \alpha \left( K  \Kphi - \frac{\ha \textrm{M}^2 \lambda}{4} e^{\lambda \Phi} \RGB \right)
\nonumber\\
+ & \alpha \lambda e^{-2\lambda\Phi}\left(\frac{}{} \Ktheta^2 - D_i\Theta D^i\Theta \right)
\,,
\end{align}
\end{subequations}
where $\dif_{\rm t} = \left(\p_{t} - \Lie_{\beta}\right)$, and $\Lie_{\beta}$ is the Lie derivative along the shift vector. 
For the evolution equations with general coupling functions $\fth, \fph, \gph$, see App.~\ref{appsec:GeneralAxiDil3p1Eqs}.

The topological invariants of the background spacetime, $\RGB$ and $\RCS$, can be written as,
\begin{subequations}
\label{eq:RCSRGBinEBWeyl}
\begin{align}
\label{eq:RCSinEBWeyl}
\RCS = & - 16  E^{ij}B_{ij}
\,,\\
\label{eq:RGBinEBWeyl}
\RGB = & 8\left( E^{ij} E_{ij} - B^{ij} B_{ij} \right)
\,,
\end{align}
\end{subequations}
where we have introduced the gravito-electric, $E_{ij}=\gamma^{a}{}_{i} \gamma^{b}{}_{j} n^c n^d W_{acbd}$, and the gravito-magnetic, $B_{ij}=\gamma^{a}{}_{i} \gamma^{b}{}_{j} n^c n^d \,^{\ast}W_{acbd}$, parts of the Weyl tensor.

In the decoupling approximation, and focusing on a vacuum \gr background, the gravito-electro and gravito-magnetic fields, $E_{ij}$ and $B_{ij}$, become
\begin{subequations}
\label{eq:EijBijInGR}
\begin{align}
\label{eq:EijInGR}
E_{ij} = & R^{\rm tf}_{ij} + \frac{1}{3} A_{ij} K - A_{i}{}^{k}A_{jk} + \frac{1}{3}\gamma_{ij} A_{kl}A^{kl}
\,,\\
\label{eq:BijInGR}
B_{ij} = & -\epsilon_{(i|}{}^{kl}D_{l}A_{|i)k} 
\,.
\end{align}
\end{subequations}
The components of the effective energy-momentum tensor associated with bi-scalar--tensor models of gravity are summarized in App.~\ref{appsec:GeneralAxiDil3p1Eqs}.

\section{Numerical relativity framework}\label{sec:NRFramework}

In this section, we provide a detailed description of our open-source numerical relativity framework \CanudaAD{} 
and describe 
the single and binary \bh{} simulations that we perform.

\subsection{Code description}\label{ssec:CodeDescription}
The core infrastructure for our numerical experiments is provided by the \ETK~\cite{Loffler:2011ay,Zilhao:2013hia,EinsteinToolkit:2022_11},
an open-source numerical relativity code for computational astrophysics.
The \ETK is based on the \Cactus computational toolkit~\cite{Cactuscode:web,Goodale:2002a}
and the \Carpet driver~\cite{Schnetter:2003rb,CarpetCode:web} to provide box-in-box
adaptive mesh refinement and MPI parallelization. 
We pair the \ETK with \Canuda~\cite{witek_helvi_2023_7791842,Okawa:2014nda,Witek:2018dmd,Silva:2020acr}, our open-source code for fundamental physics. 

As a new extension to \Canuda{'s} capabilities,
we
present the \CanudaAD{}~\cite{CanudaAxiDil_repo} module here.
It provides a framework for parametrized numerical relativity that incorporates a large class of
quadratic gravity models
and of bi-scalar fields that are minimally or nonminimally coupled to gravity.
\CanudaAD{} implements the field equations~\eqref{eq:axionKG_decoup} and~\eqref{eq:dilatonKG_decoup}
and enables the selection of the different models presented in Sec.~\ref{sec:QuadraticGravityTheory} via simple parameter choices; see details in Table~\ref{tab:parameter choices} of App.~\ref{appsec:code_description}. 
The evolution equations are implemented using
up to sixth-order finite-differences 
for the spatial derivatives.
Our standard choice for the time integration is a fourth-order Runge-Kutta scheme.
Up to seventh-order Kreiss-Oliger dissipation is employed to reduce the numerical noise produced at the refinement boundaries. 

The \CanudaAD module consists of three thorns.
The \textsc{AxiDil-Base} thorn sets up the necessary grid functions.
The \textsc{AxiDil-Init} thorn provides a variety of initial axion and dilaton profiles. 
They represent 
zero fields
or approximate analytical solutions
as introduced in Sec.~\ref{ssec:ScalarInitialData}.
In App.~\ref{appssec:SFID} we list additional initial data choices
that are available in the code but have not been used in this paper.
The time evolution of the scalar field sector is conducted by the \textsc{AxiDil-Evol} thorn, which implements the equations~\eqref{eq:axionKG_decoup} and~\eqref{eq:dilatonKG_decoup}
in the \bssn formulation given in App.~\ref{appsec:GeneralAxiDil3p1Eqs}.
This thorn provides all models described in Sec.~\ref{sec:GravityModels}. 
The parameter settings to select a specific model are summarized in App.~\ref{appssec:ModelSelection}.

For simulations of single \bh{s}, the background metric is fixed and consists of a stationary Kerr \bh{} in quasi-isotropic coordinates.
The \textsc{KerrQuasiIsotropic} thorn from the \Canuda{} library provides the appropriate metric and curvature initial data. 
For fixed backgrounds, we introduce a parameter called \texttt{set\_lapse\_to\_zero} in the \textsc{AxiDil-Init} thorn that enables setting the lapse to zero inside the horizon, $\alpha(r \leq r_+) = 0$.
We typically set this parameter to ``yes''
to minimize numerical growth arising from nonphysical gauge effects.

For the \bbh simulations, the \bh{} initial data is produced by \textsc{TwoPunctures}, a spectral solver for puncture data~\cite{Ansorg:2004ds}.
In this case, the metric is dynamical and evolved according to Einstein's equations in vacuum.
Specifically, we evolve Einstein's equations in the \bssn formulation as implemented in \Canuda's spacetime evolution solver \textsc{LeanBSSNMoL}~\cite{witek_helvi_2023_7791842},
adapted from the {\textsc{Lean}} code~\cite{Sperhake:2007}.
For both the scalars and the dynamical metric we apply radiative boundary conditions using the \textsc{NewRad} thorn~\cite{Alcubierre:2002kk}.

To compute the $(l,m)$ multipoles of $\Theta$ and $\Phi$, we use the {\textsc{Multipole}} thorn.
The grid functions representing the axion or dilaton field are interpolated onto
spheres of fixed extraction radii $\rex$ and projected onto $s=0$ spherical harmonics $Y_{lm}(\theta,\phi)$,
\begin{align}
\label{eq:ThetaYlmProjection}
\Theta_{lm}(t,\rex) = & \int\dif\Omega\, \Theta(t,\rex,\theta,\phi) Y^{\ast}_{lm} (\theta,\phi)
\,,
\end{align}
and likewise for the field $\Phi$.
We visualize our numerical data with 
\kuibit~\cite{kuibit}, a set of tools for post-processing data generated with the \ETK{.}

\subsection{Simulation setup}\label{ssec:Simulations}

\begin{table}[t!]
\begin{tabular}{|c|c|c|c|c|c|}
\hline
Run & Initial data & Model & $\ha$ & $\chi$ \\
\hline
\hline
    IDI\_AD\_chi01 & Type I & E & 0.1 & 0.1 \\
    IDI\_AD\_chi03 & Type I & E & 0.1 & 0.3 \\
    IDI\_AD\_chi05 & Type I & E & 0.1 & 0.5 \\
    IDI\_AD\_chi07 & Type I & E & 0.1 & 0.7 \\
    IDI\_AD\_chi09 & Type I & E & 0.1 & 0.9 \\
    \hline
    IDII\_AD\_chi01 & Type II & E & \{0.001,0.01,0.1\} & 0.1 \\
    IDII\_AD\_chi03 & Type II & E & \{0.001,0.01,0.1\} & 0.3 \\
    IDII\_AD\_chi05 & Type II & E & \{0.001,0.01,0.1\} & 0.5 \\
    IDII\_AD\_chi07 & Type II & E & \{0.001,0.01,0.1\} & 0.7 \\
    IDII\_AD\_chi09 & Type II & E & \{0.001,0.01,0.1\} & 0.9 \\
    \hline
    IDII\_dec\_chi01 & Type II & B\&C & 0.1 & 0.1 \\
    IDII\_dec\_chi03 & Type II & B\&C & 0.1 & 0.3 \\
    IDII\_dec\_chi05 & Type II & B\&C & 0.1 & 0.5 \\
    IDII\_dec\_chi07 & Type II & B\&C & 0.1 & 0.7 \\
    IDII\_dec\_chi09 & Type II & B\&C & 0.1 & 0.9 \\
\hline
\end{tabular}
\caption{Simulations of axion and dilaton fields about a single \bh{} with dimensionless spin $\chi$.
We initialize the fields
either as zero
(Type~I in Sec.~\ref{ssec:ScalarInitialData}),
or as the approximate analytical solution
(Type II in Sec.~\ref{ssec:ScalarInitialData}).
The first series of runs evolves axi-dilaton gravity, i.e., the axion and dilaton are kinetically coupled (Model E in Sec.~\ref{sec:GravityModels}) with coupling strength  $\ha$.
The second series simulates the axion and dilaton decoupled from each other
according to, respectively, \dCS gravity with a linear coupling function and \sGB gravity with a dilaton coupling function (Models B and C in Sec.~\ref{sec:GravityModels}).
} 
\label{tab:SBHSims}
\end{table}

In Table~\ref{tab:SBHSims}, we list all simulations performed in the background of a single \bh{} with dimensionless spin parameter $\chi$.
We give the axion and dilaton initial data (see Sec.~\ref{ssec:ScalarInitialData}), the model of gravity considered (see Sec.~\ref{sec:GravityModels}),
and the dimensionless coupling constant $\ha$.

For single \bh{} simulations with a coupling strength $\ha<0.1$, we employ a fully Cartesian box-in-box mesh refinement with the \bh{} placed at the origin.
In this case, the outermost level has a radius of $128$ M and a grid spacing of $1$ M.
We have a total of seven refinement levels with each inner refinement level halving the grid spacing.
The finest grid has a radius of $0.75$ M and a grid spacing of $1/64$ M.

The remaining single \bh{} simulations with a 
larger
coupling strength of $\ha=0.1$ are more demanding, so we use a box-in-box mesh refinement combined with a spherical multipatch grid provided by the \textsc{Llama} thorn~\cite{Pollney:2009yz, Reisswig:2012nc}.
We set up a total of eight Cartesian refinement levels with the coarsest level having an outer radius of $24$ M and a grid spacing of $1$ M.
The innermost (finest) refinement level has a radius of $0.75$ M and a grid spacing of $1/128$ M.
The inner boundary of the spherical grid is located at $24$ M and the outer boundary is at $200$ M with an angular resolution of $64$ points.

For the \bbh{} simulation, we set our initial punctures to be separated by $6$ M and use \textsc{TwoPunctures}~\cite{Ansorg:2004ds} to solve for the initial data.
The momenta are selected to produce a quasi-circular orbit.
The initial \bh{s} are nonspinning and have equal masses, $m_{1,2}=0.5$, where $m_{1,2}$ refer to the individual ADM masses of the punctues as defined in ~\cite{Ansorg:2004ds}. 
The grid is fully Cartesian with a total radius of $256$ M and a coarsest grid spacing of $\sim 1$ M. 
We use a total of eight refinement levels with the finest refinement level 
having a radius of $0.75$ M and a grid spacing of $\sim 1/128$ M.

\subsection{Code verification}\label{ssec:CodeVerification}

To verify our code, 
we benchmark simulations in \dCS gravity 
with the \CanudadCS module~\cite{Richards:2023xsr},
and benchmark simulations in \EdGB gravity
with the \CanudaEdGB module~\cite{Witek:2018dmd,Silva:2020acr}.
For the same physical parameters and simulation setup, we find that our simulations in \dCS gravity agree within $0.04\%$ with the \CanudadCS module and our simulations in \EdGB gravity agree within $0.3\%$ with the \CanudaEdGB module. 
Note that we adopt the default parameter setting \texttt{set\_lapse\_to\_zero = no} to match the \CanudaEdGB module. 

To confirm the validity of the simulations
and to estimate their numerical error,
we perform convergence tests for both the single and binary \bh{s}.
For the single \bh{} results, we show convergence for the run IDII\_AD\_chi09 with coupling strength $\ha = 0.1$ and present our results in App.~\ref{ssec:sbh_convergence}.
We find third order convergence,
and discretization errors at late times ($t > 150$M) of 
$\Delta \Theta_{10}/\Theta_{10,h}< 0.1\%$ and $\Delta \Phi_{00}/\Phi_{00,h} < 0.02\%$ for the leading-order contributions to the axion and dilaton fields, respectively.
For the \bbh{} results, our convergence tests are shown in App.~\ref{ssec:bbh_convergence}.
We find approximately third order convergence for the \bbh{} simulation as well.

\section{Formation of hairy black holes}\label{sec:ResultsSingleBH}
\begin{figure*}[ht!]
\centering
\includegraphics[width = 0.48\textwidth]{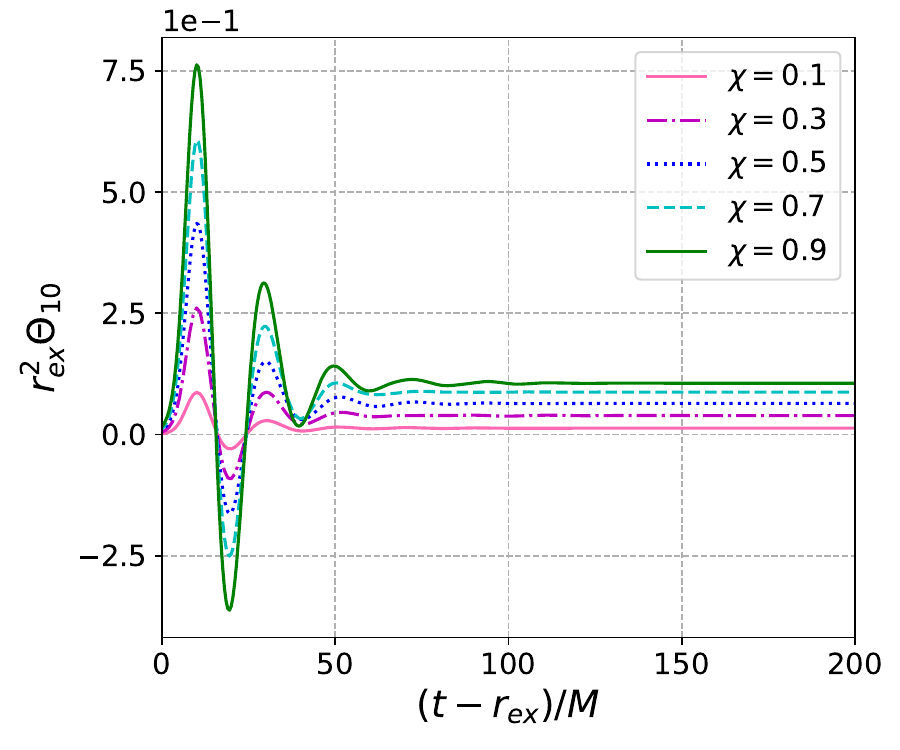}
\includegraphics[width = 0.48\textwidth]{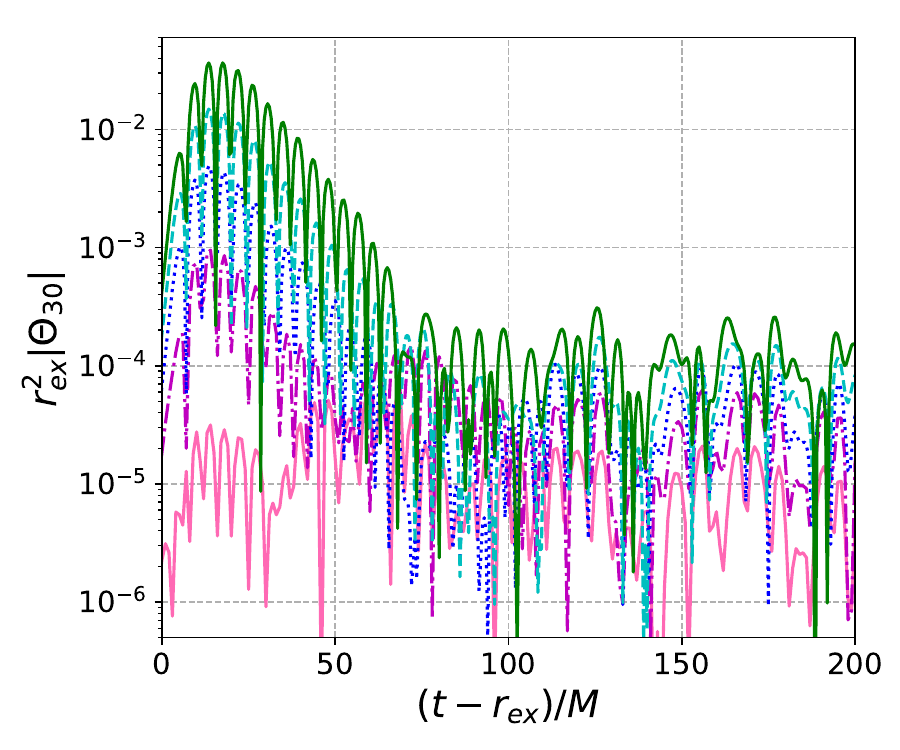}
\caption{Evolution of the axion's
$l=1,m=0$ (left panel) and $l=3,m=0$ (right panel)
multipoles extracted at $r_{ex} = 20M$.
The axion is initialized as zero, coupled to gravity with
coupling parameter $\ha = 0.1$,
and evolves in the background of a rotating \bh{} with dimensionless spin $\chi = 0.1$ (solid pink), $\chi = 0.3$ (dash-dotted magenta), $\chi=0.5$ (dotted blue), $\chi=0.7$ (dashed cyan), and $\chi=0.9$ (solid green). 
}
\label{fig:aSpt1_Theta10_ID0_comp_spins}    
\end{figure*}

\begin{figure*}[ht!]
    \centering
    \includegraphics[width = 0.48\textwidth]{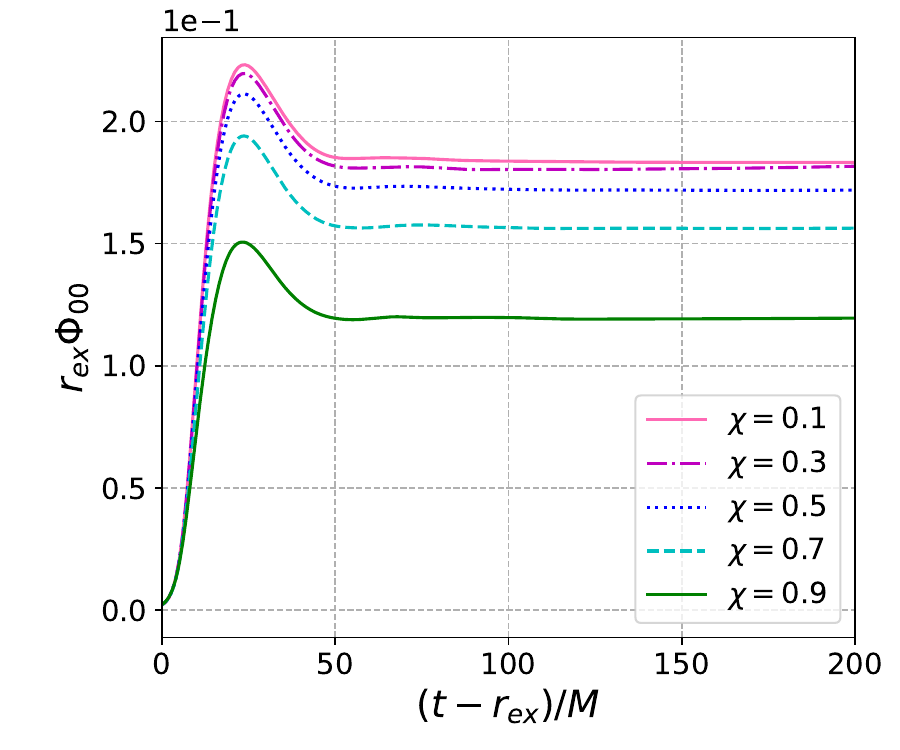}
    \includegraphics[width = 0.48\textwidth]{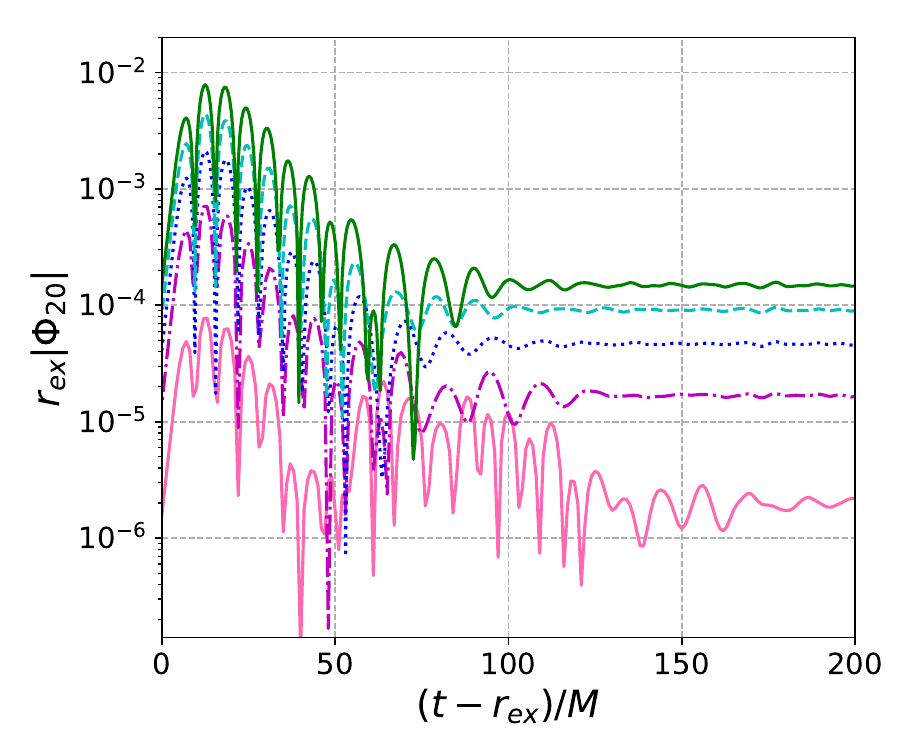}
    \caption{Same as Fig.~\ref{fig:aSpt1_Theta10_ID0_comp_spins} but for
    the dilaton's $l=m=0$ (left panel) and $l=2,m=0$ (right panel) multipoles.
}
    \label{fig:aSpt1_Phi00_ID0_comp_spins}
\end{figure*}

In this section, 
we present our numerical results of the axion and dilaton fields evolving in the background of a single \bh{} 
in axi-dilaton gravity.
We work in the decoupling approximation and capture the dynamical growth of the  axion and dilaton hairs 
as proposed in sketch~\ref{fig:AxiDil_sketch}.
The simulations presented here assist us 
to address the following questions:
(i) What is the effect of \bh{} spin on the axion and dilaton hairs
in axi-dilaton gravity? 
(ii) How does the strength of the couplings to curvature in axi-dilaton gravity affect the axion and dilaton hairs?
(iii) How does the kinetic coupling between the two fields affect their respective evolution as compared to their evolution in isolation?
Secs.~\ref{ssec:NumResEffectOfBHSpin}, ~\ref{ss:NumResCoupl}, and \ref{ssec:NumResCoupBTWFields} address each of these questions, respectively.

Many of the results presented in the following sections focus on the evolution of the $\Theta_{10}$ and $\Phi_{00}$ multipoles,
defined in Eq.~\eqref{eq:ThetaYlmProjection},
because they are the leading-order contribution to the axion and dilaton hairs.
This can be seen by re-writing the perturbative solutions of
Eqs.~\eqref{eq:IDAxionHair} and~\eqref{eq:IDDilatonHair} in terms of their leading harmonic content,
\begin{subequations}
\label{eq:IDHairMultipoleExpansion}
\begin{align}
\label{eq:IDAxionHairMultipole}
\Theta \simeq &
\sqrt{\frac{16\pi}{3}}\, \left[ \Theta_{10}(r) Y_{10} 
+ \Theta_{30}(r) Y_{30} + \ldots
\right]
\,,\\
\label{eq:IDDilatonHairMultipole}
\Phi \simeq & \sqrt{16\pi} \lambda \left[\Phi_{00}(r)\,Y_{00} + \Phi_{20}(r)\, Y_{20} + \dots \right]
\,,
\end{align}
\end{subequations}
where $Y_{lm}\equiv Y_{lm}(\theta,\phi)$ are spherical harmonics.
The fields' leading-order multipoles  are 
\begin{subequations}
\label{eq:IDHairMultipole}
\begin{align}
\label{eq:IDHairMultipoleTheta10}
\Theta_{10} & = \ha \chi\left[\vartheta_{1} - \chi^{2}\left( \vartheta_{2} +  \frac{3}{5} \vartheta_{3} \right) \right]
+ \mathcal{O}(\ha \chi^{4})
\,,\\
\label{eq:IDHairMultipoleTheta30}
\Theta_{30} & = - \frac{2}{5}\sqrt{\frac{3}{7}} \,\ha\,\chi^{3} \vartheta_{3}
+ \mathcal{O}(\ha \chi^{4})
\,,\\
\label{eq:IDHairMultipolePhi00}
\Phi_{00} & = \ha \left[ \varphi_{0}
- \chi^{2} \left( \varphi_{1} + \frac{1}{3} \varphi_{2}\right) \right]
+ \mathcal{O}(\ha \chi^{4})
\,,\\
\label{eq:IDHairMultipolePhi20}
\Phi_{20} & = - \frac{2}{3\sqrt{5}}\,\ha\, \chi^2 \varphi_{2}
+ \mathcal{O}(\ha \chi^{4})
\,,
\end{align}
\end{subequations}
where $\vartheta_{i}(r)$, $\varphi_{i}(r)$ are given by Eqs.~\eqref{eq:IDAxionHair_vartheta} and~\eqref{eq:IDDilatonHair_varphi}.

\subsection{Effect of black-hole spin on final hairs}
\label{ssec:NumResEffectOfBHSpin}
\begin{figure*}[!ht]
    \centering
    \includegraphics[width = 0.49\textwidth]{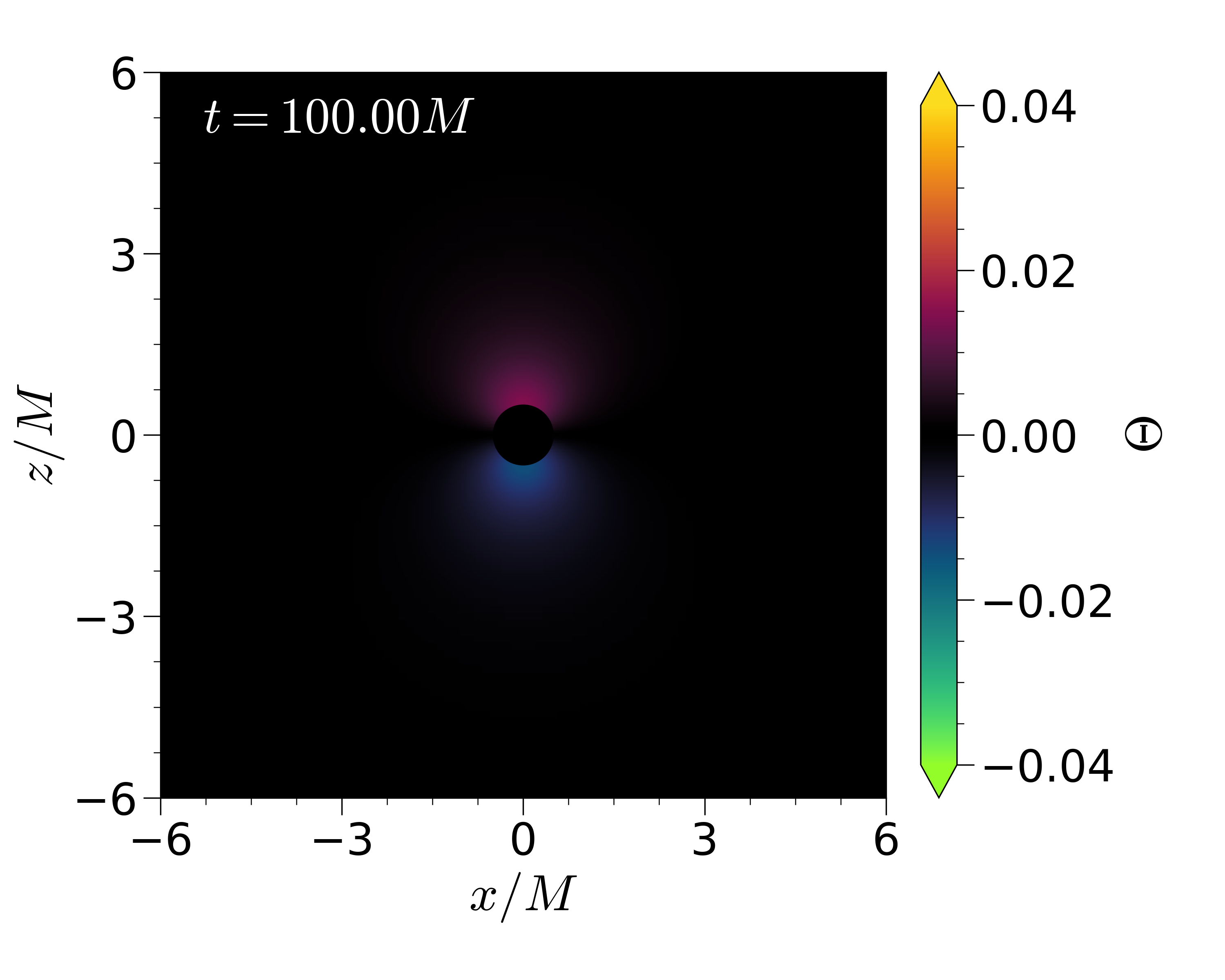}
    \includegraphics[width = 0.49\textwidth]{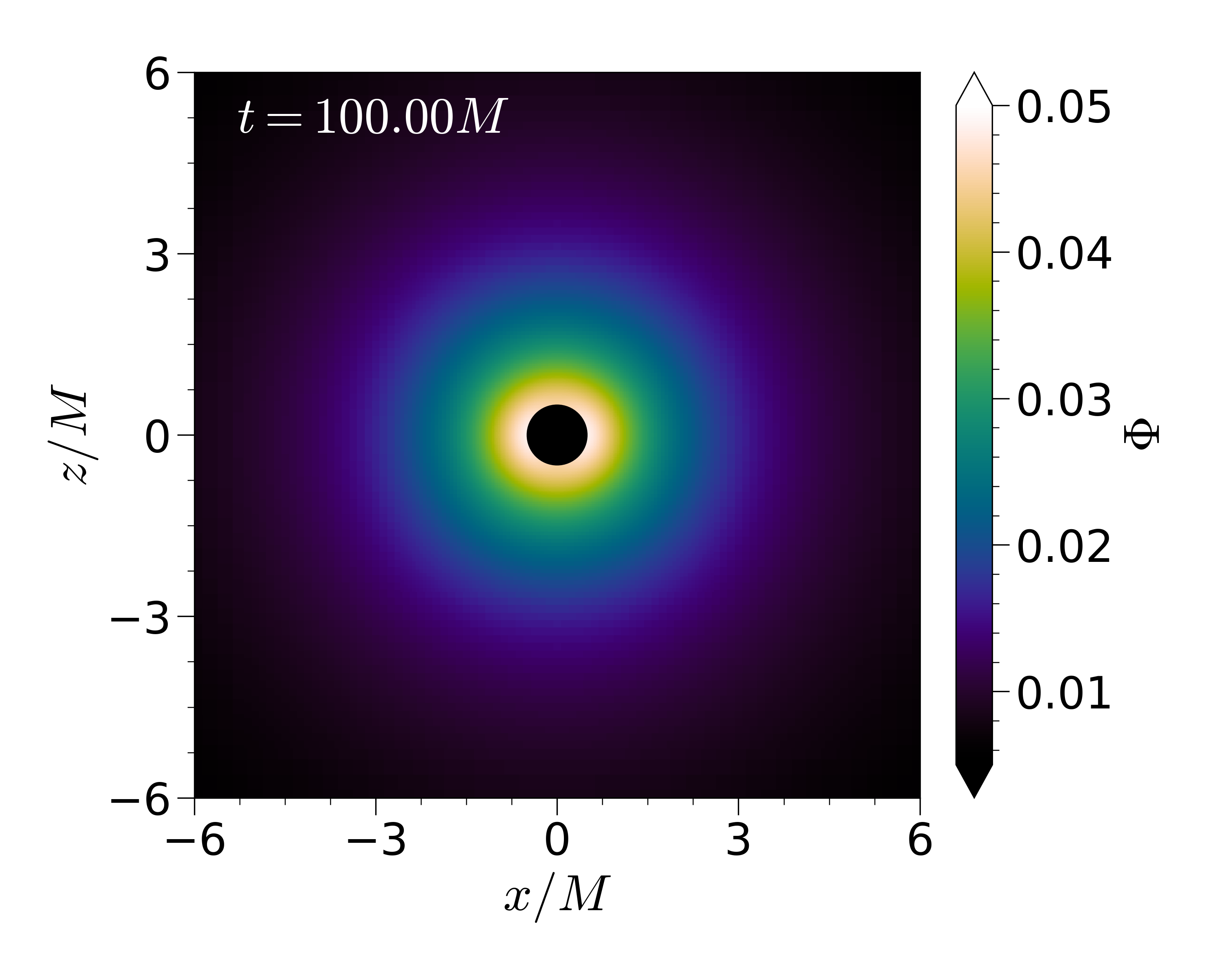}
    \caption{Snapshot of axion (left) and dilaton (right) fields with coupling strength $\ha = 0.1$ evolving around a single \bh{} with dimensionless spin $\chi = 0.3$ in the xz-plane, 
    i.e. the slice perpendicular to the equatorial plane, 
    at $t = 100M$. 
    The black circle indicates the \bh{'s} apparent horizon with a radius $r_{h} \sim 0.49 M$ (in isotropic coordinates).
    }
    \label{fig:2D_SBH_apt3_snaps}
\end{figure*}

\begin{figure*}[!ht]
    \centering
    \includegraphics[width = 0.49\textwidth]{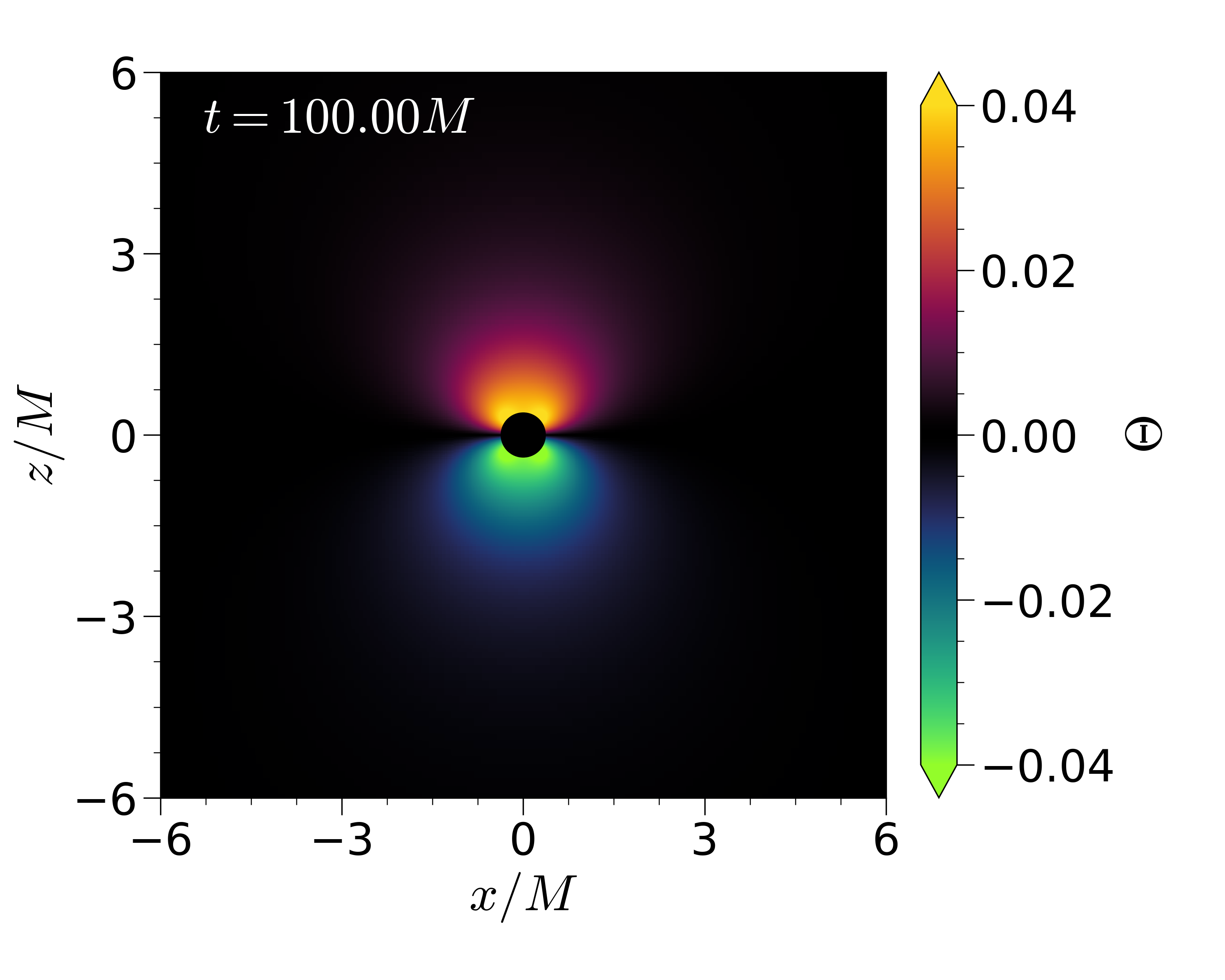}
    \includegraphics[width = 0.49\textwidth]{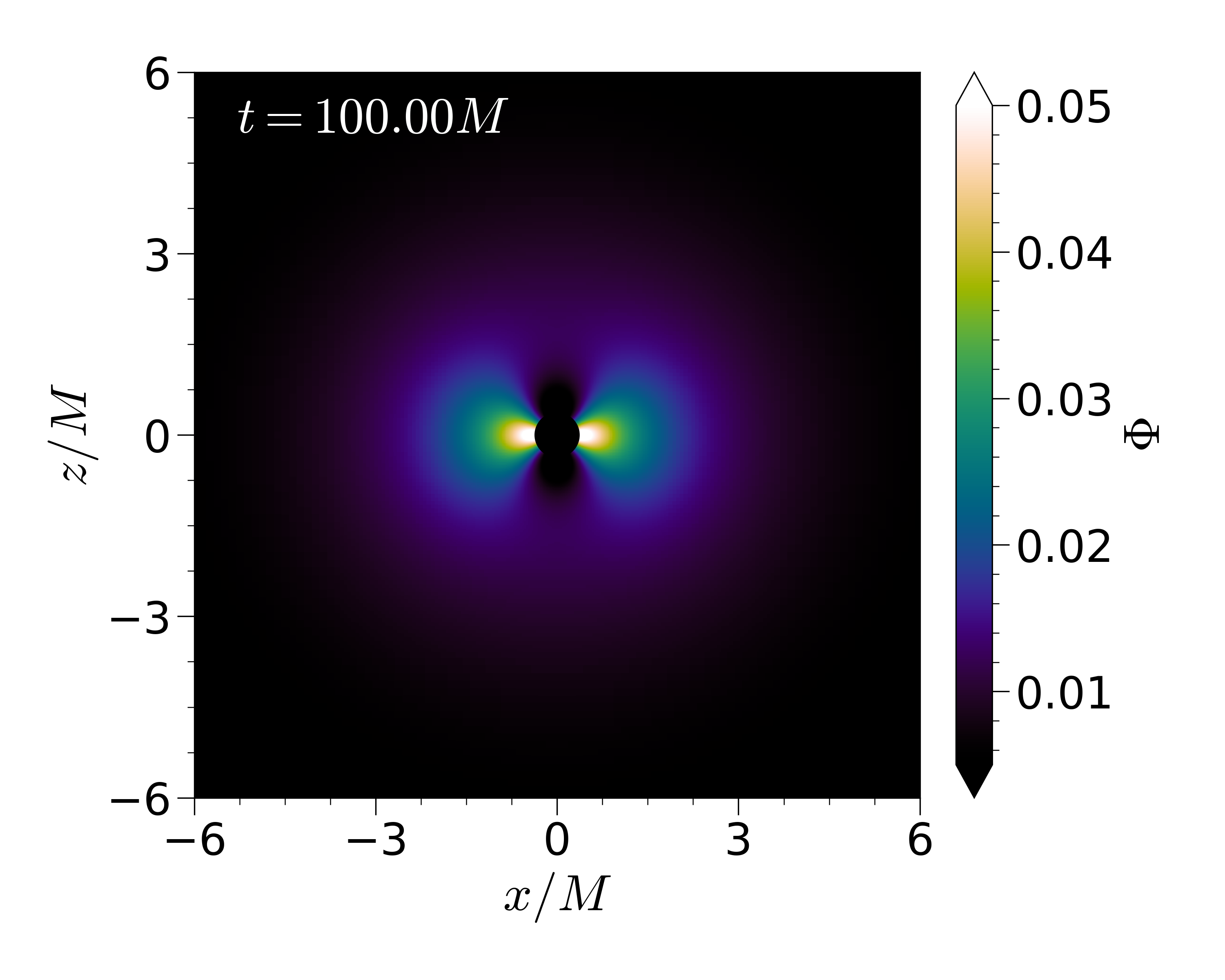}
    \caption{Same as Fig.~\ref{fig:2D_SBH_apt3_snaps} but for a \bh{} of dimensionless spin $\chi = 0.9$ and \bh{} 
    radius of $r_{h} \sim 0.36 M$ .}
    \label{fig:2D_SBH_apt9_snaps}
\end{figure*}

In this section, we investigate how \bh{} spin affects the growth, evolution, and end state of the axion and dilaton hairs.
We concentrate on simulations with initially vanishing axion and dilaton fields.
We show additional robustness checks using the approximate analytical initial data at the end of the section.
We present the fields' time evolution  in Figs.~\ref{fig:aSpt1_Theta10_ID0_comp_spins} and \ref{fig:aSpt1_Phi00_ID0_comp_spins}
and two-dimensional snapshots of their late-time profile in Figs.~\ref{fig:2D_SBH_apt3_snaps} and \ref{fig:2D_SBH_apt9_snaps}.

In Figs.~\ref{fig:aSpt1_Theta10_ID0_comp_spins} and \ref{fig:aSpt1_Phi00_ID0_comp_spins},
we show the evolution of the leading-order multipoles of the axion and dilaton
hairs, $\Theta_{10}$ and $\Phi_{00}$, in the left panels and the next-to-leading-order modes, $\Theta_{30}$ and $\Phi_{20}$, in the right panels.
We also evolved multipoles up to $l = 4$, but they are subdominant by several orders of magnitude, so we omit presenting them here. 
We fix the coupling strength $\ha = 0.1$
and show results for a range of \bh{} spin parameters $\chi \in \{0.1, 0.3, 0.5, 0.7, 0.9\}$ .
The multipoles are extracted at $r_{ex} = 20M$, 
and we shift time by $r_{ex}$ 
so the origin coincides with the start of the growth of the hairs. 

Let us first focus on the growth of the $\Theta_{10}$ multipole in the left panel of Fig.~\ref{fig:aSpt1_Theta10_ID0_comp_spins}
where $\Theta_{10}$ has been rescaled by $r_{ex}^2 = (20M)^2$ 
to show the leading-order contribution to the axion hair independent of the extraction radius.
As the spin increases, the axion hair initially fluctuates more than in
simulations with smaller \bh{} spins because the initial data is further from the end state.
Additionally, as the \bh{} spin increases, the final axion hair increases, indicating 
a proportional relationship
between the axion hair and the \bh{} spin
that is consistent with the approximate analytical solution given by Eq.~\eqref{eq:IDHairMultipoleTheta10}.

Next we consider the dilaton field and focus on the evolution of the $\Phi_{00}$ multipole in the left panel of Fig.~\ref{fig:aSpt1_Phi00_ID0_comp_spins}.
We rescale $\Phi_{00}$ 
by $r_{ex} = 20M$ 
to present the leading-order contribution to the dilaton hair independent of the extraction radius.
After the initial adjustment, 
the dilaton hair settles inversely proportional to the \bh{} spin; as the \bh{} spin increases, the dilaton hair decreases.
This numerical result is consistent with the analytical approximation for the dilaton hair given by Eq.~\eqref{eq:IDHairMultipolePhi00} as spin appears at second order with a minus sign, indicating a decrease in dilaton hair with increasing spin. 

Summarizing the discussion of Figs.~\ref{fig:aSpt1_Theta10_ID0_comp_spins} and \ref{fig:aSpt1_Phi00_ID0_comp_spins},
we find that the axion and dilaton fields -- for sufficiently small \bh{} spins -- dynamically reach configurations that are consistent with the perturbative solutions.
For larger \bh{} spins, the fields dynamically adjust to a new end state
that captures next-to-leading-order effects
in both the spin and coupling strength.

Up to this point, we have focused on the evolution of the dominant $\Theta_{10}$ and $\Phi_{00}$ multipoles.
Let us now consider the evolution of the next-to-leading-order modes, $\Theta_{30}$ and $\Phi_{20}$, in the right panels of Figs.~\ref{fig:aSpt1_Theta10_ID0_comp_spins} and \ref{fig:aSpt1_Phi00_ID0_comp_spins}.

In the right panel of Fig.~\ref{fig:aSpt1_Theta10_ID0_comp_spins}, we see that, rather than briefly adjusting 
before settling close to the final value, the $\Theta_{30}$ mode rings down before reaching its final state.
After the initial decay,
we find that 
the magnitude of $\Theta_{30}$ increases with \bh{} spin and roughly settles to
$\rex^2 \Theta_{30} \lesssim 10^{-4}$
for all \bh{} spins.
Thus, the $\Theta_{30}$ multipole is subdominant to the $\Theta_{10}$ multipole,
and their ratio is consistent with 
Eq.~\eqref{eq:IDHairMultipoleTheta30} as $\Theta_{30}/\Theta_{10} \propto \chi^2M^2/\rBL^2 \lesssim 10^{-3}$.

Observing the evolution of $\Phi_{20}$ in the right panel of Fig.~\ref{fig:aSpt1_Phi00_ID0_comp_spins}, we find that the magnitude of $\Phi_{20}$ increases with increasing \bh{} spin.
This behavior differs from the leading-order contribution to the dilaton field as $|\Phi_{00}|$ decreases with increasing spin as seen in the left panel of Fig.~\ref{fig:aSpt1_Phi00_ID0_comp_spins}.
This behavior is consistent with the small-spin, small-coupling approximation in Eqs.~\eqref{eq:IDHairMultipolePhi00} and~\eqref{eq:IDHairMultipolePhi20}, respectively,
but we here also capture the evolution for high \bh{} spins.
Since $|\Phi_{00}|$ decreases and $|\Phi_{20}|$ increases as \bh{} spin increases, the higher mode $|\Phi_{20}|$ contributes more to the final dilaton hair
of a highly rotating \bh{}
than that of a slowly rotating \bh{}.
Once the modes have settled, we compare the $|\Phi_{00}|$ mode with the $|\Phi_{20}|$ mode at $t = 150M$ and find
$\Phi_{00}/\Phi_{20} = 1.08\times 10^{5}$ for \bh{} spin $\chi = 0.1$ and $\Phi_{00}/\Phi_{20} = 7.91\times 10^{2}$ for \bh{} spin $\chi = 0.9$.
This not only visibly changes the dilaton's profile as we will see in the two-dimensional  snapshots below,
but it may also have important implications for the dilaton radiation sourced by spinning \bbh{s}.
In particular, the $\Phi_{00}$ and $\Phi_{20}$ multipoles source the dipolar and $l=3$ radiative modes of the dilaton.
Our results suggest a strong dependence of their excitation on the \bh{s'} spins.

In the introduction, Sec.~\ref{sec:intro}, we wondered what profiles the axion and dilaton fields will evolve to; see the sketch in Fig.~\ref{fig:AxiDil_sketch}. 
We present two-dimensional snapshots from the numerical simulations in Figs.~\ref{fig:2D_SBH_apt3_snaps} and \ref{fig:2D_SBH_apt9_snaps} to illustrate the fields' final states.
The snapshots depict 
the axion (left panels) and dilaton (right panels)
in the $xz$-plane, i.e., along the \bh{'s} axis of rotation,
evolved around a \bh{} with spin $\chi = \{0.3, 0.9\}$ in Figs.~\ref{fig:2D_SBH_apt3_snaps} and \ref{fig:2D_SBH_apt9_snaps}, respectively.

For visualization purposes, we add a black circle
indicating the \bh{,}
whose radius is given by the mean \bh{} apparent horizon $r_h = \{0.49M, 0.36M\}$
(corresponding to $\chi = \{0.3, 0.9\}$).
Comparing the left panels of 
Figs.~\ref{fig:2D_SBH_apt3_snaps} and \ref{fig:2D_SBH_apt9_snaps}, we see that the magnitude of the axion field, given by the green-orange colorbar, increases with increasing \bh{} spin.
The dipole shape along the axis of rotation remains,
indicating the dominance of the $l = 1, m = 0$ mode in the axion's final profile in the $xz$-plane.

We now compare the dilaton profiles in the xz-plane in the right panels of Figs.~\ref{fig:2D_SBH_apt3_snaps} and \ref{fig:2D_SBH_apt9_snaps}. 
As the spin increases, the magnitude of the dilaton field, given by the black-white colorbar, decreases,
consistent with the results presented in Fig.~\ref{fig:aSpt1_Phi00_ID0_comp_spins}.
The field's shape also changes with increasing \bh{} spin
due to the relative magnitude of the subdominant $\Phi_{20}$ increasing with \bh{} spin;
see also our discussion of Fig.~\ref{fig:aSpt1_Phi00_ID0_comp_spins}.
For smaller \bh{} spins such as $\chi = 0.3$, the dilaton field appears primarily as a monopole, i.e., dominated by the $l = m = 0$ mode.
For higher \bh{} spins, this monopole appears deformed as 
the $l = 2, m = 0$ mode becomes more excited and leaves an imprint in the dilaton field along the axis of rotation; for reference, the $l = 2, m= 0$ mode appears as a dipole along the axis of rotation with a donut in the equatorial plane.

To test the robustness of our results, we initialize a set of simulations with the approximate analytical solution given by initial data Type II  in Sec.~\ref{ssec:ScalarInitialData} and compare the results to those with zero initial data (Type I).
We present the results in
Figs.~\ref{fig:Theta10_comp_spinASpt1} and~\ref{fig:Phi00_comp_spinASpt1} for a range of \bh{} spins $\chi \in \{ 0.1, 0.3, 0.5, 0.7, 0.9\}$.
Note, that 
Figs.~\ref{fig:Theta10_comp_spin} and~\ref{fig:Phi00_comp_spin} also show results for smaller coupling constants that are discussed in the next section; for the comparison done here, we focus on the coupling strength of $\ha=0.1$.

Compared to the simulations where the fields are initialized as zero (Type I),
those with the approximate, analytic initial data (Type~II)
exhibit less adjustment before settling to their final state.
Despite the differences in their early-time behavior, the end state for both types of initial data fall within $0.6\%$
of each other, comparable to our numerical error.
This 
indicates
that our results are consistent
and that the final axion and dilaton hairs are independent of the choice of initial data.

In summary,
the results presented in this section demonstrate that 
\bh{s} in axi-dilaton gravity do indeed acquire two scalar charges in the form of an axion and a dilaton hair. 
We show that the two hairs form dynamically even when they are absent in the initial data.
Investigating the hairs' dependence on the \bh{} spin, we find that
both the leading-order and next-to-leading-order contributions to the axion hair, $\Theta_{10}$ and $\Theta_{30}$, increase with increasing \bh{} spin.
In the case of the dilaton field, we find that the
dilaton charge, $|\Phi_{00}|$, decreases as \bh{} spin increases while the next-to-leading-order contribution, $|\Phi_{20}|$, increases with the \bh{} spin.
This inverse relationship between $\Phi_{00}$ and $\Phi_{20}$ becomes most apparent for higher \bh{} spins, as shown by the deformed ``monopole'' profile in the right panel of Fig.~\ref{fig:2D_SBH_apt9_snaps}.

\subsection{Effect of coupling strength on final hairs}\label{ss:NumResCoupl}

\begin{figure}[t!]
    \centering
    \includegraphics[width = 0.48\textwidth]{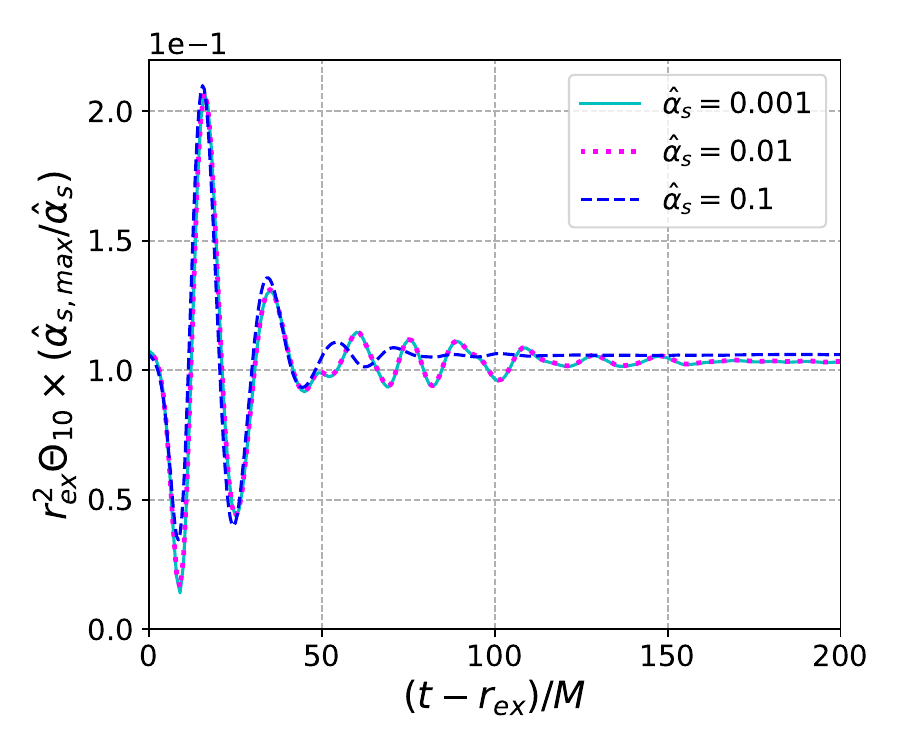}
    \caption{Evolution of the $\Theta_{10}$ multipole in the background of a \bh{} with spin $\chi = 0.9$. $\Theta_{10}$ is rescaled by $r_{ex}^2 = (20M)^2$ as well as the ratio between the maximum coupling strength $\ha= 0.1$ and the coupling strength used in the simulation, $\ha = 0.001$ (solid cyan), $\ha = 0.01$ (dotted magenta), and $\ha = 0.1$ (dashed blue). }
    \label{fig:Theta10_comp_aS}
\end{figure}

In this section, we consider the effect of the dimensionless coupling strength, $\ha$, on the evolution of the axion and dilaton fields.
We perform three sets of simulations in the range $\ha = \{0.001,0.01,0.1\}$,
and vary the \bh{'s} spin parameter $\chi\in\{0.1,0.3,0.5,0.7,0.9 \}$ for each value of the coupling strength.
We initialize the axion and dilaton fields with the analytic result in the small-spin, small-coupling approximation, corresponding to Type~II initial data in Sec.~\ref{ssec:ScalarInitialData}.
The analytic expressions given in Eqs.~\eqref{eq:IDAxionHair} and Eq.~\eqref{eq:IDDilatonHair} scale linearly with the coupling strength. 
In the remainder of this section, we assess the significance of higher-order terms in coupling.

\begin{figure}[t!]
    \centering
    \includegraphics[width = 0.48\textwidth]{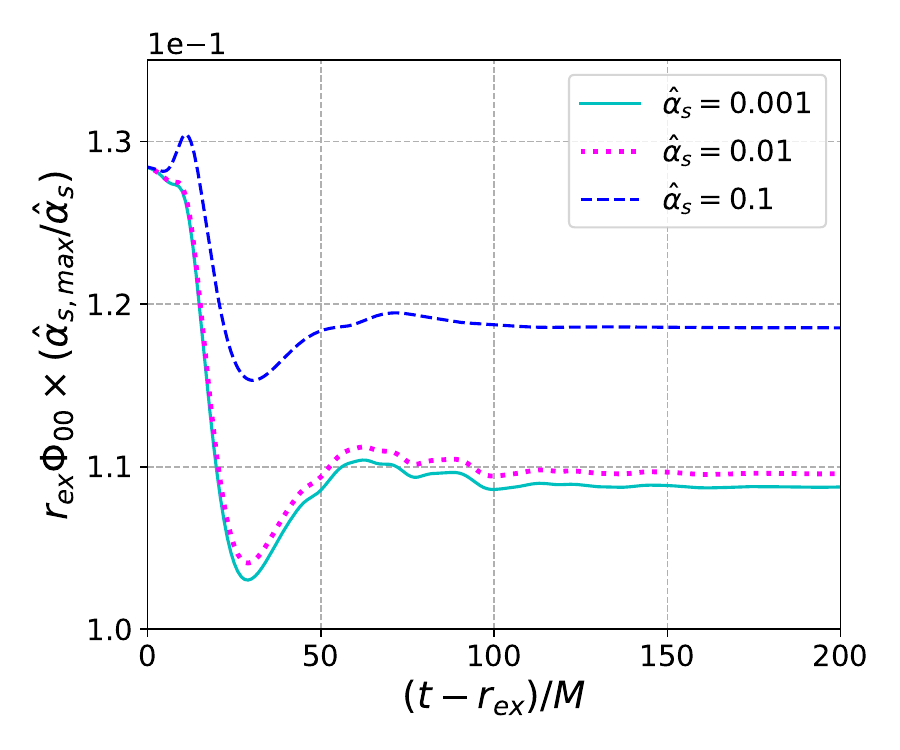}
    \caption{Same as Fig.~\ref{fig:Theta10_comp_aS} but for $\Phi_{00}$ rescaled by $r_{ex} = 20M$.}
    \label{fig:Phi00_comp_aS}
\end{figure}

\begin{figure*}[t!]
\centering
\subfloat[$\ha = 0.001$]{\includegraphics[width=0.325\textwidth]{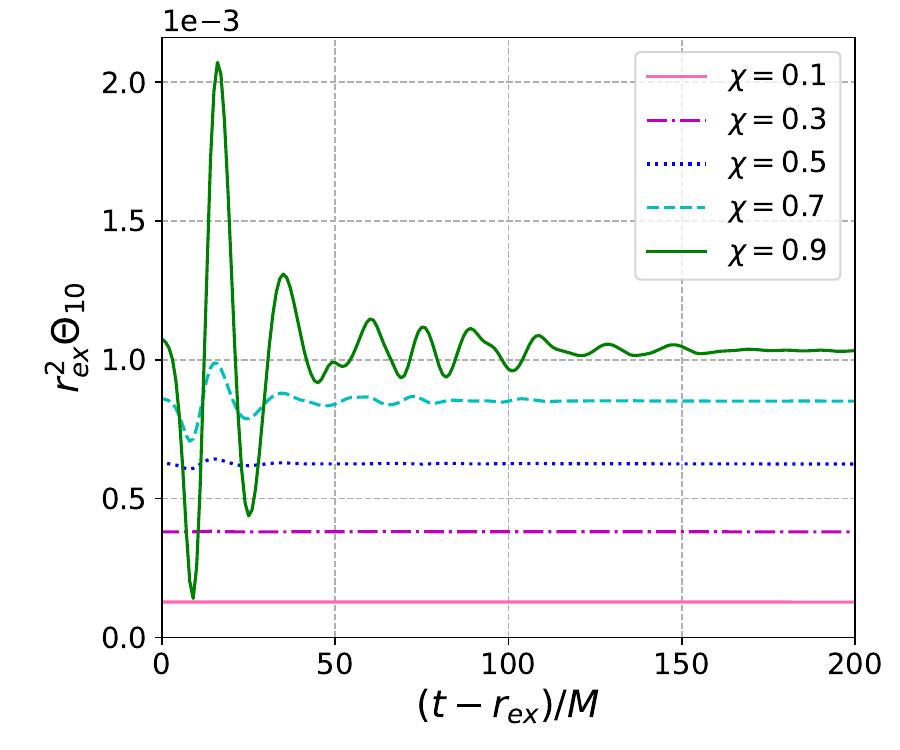}\label{fig:Theta10_comp_spinASpt001}}
\subfloat[$\ha = 0.01$]{\includegraphics[width=0.325\textwidth]{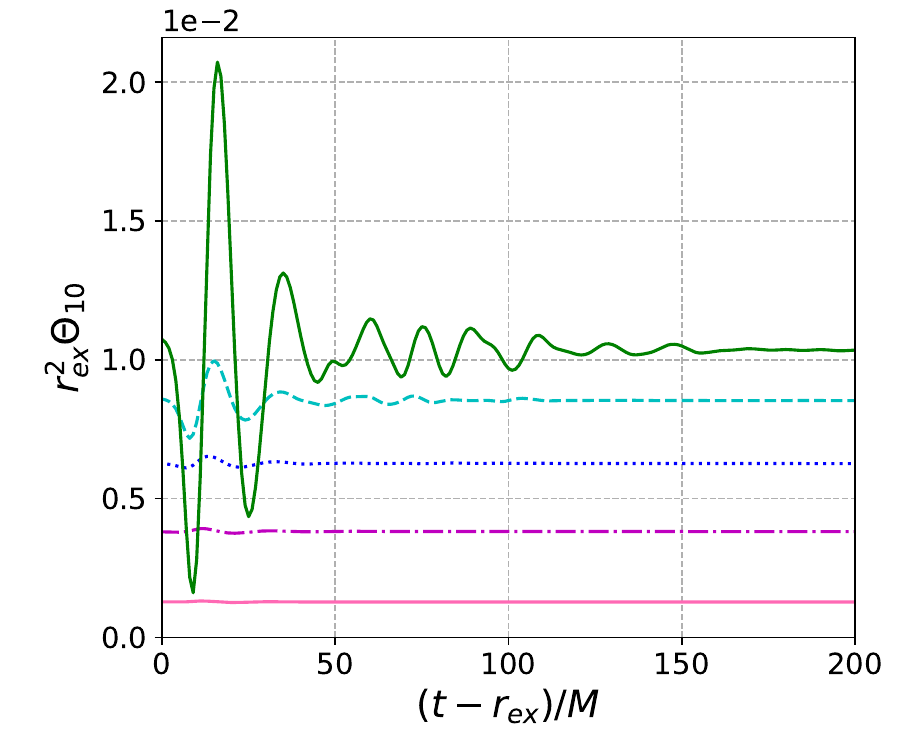}\label{fig:Theta10_comp_spinASpt01}}
\subfloat[$\ha = 0.1$]{\includegraphics[width=0.325\textwidth]{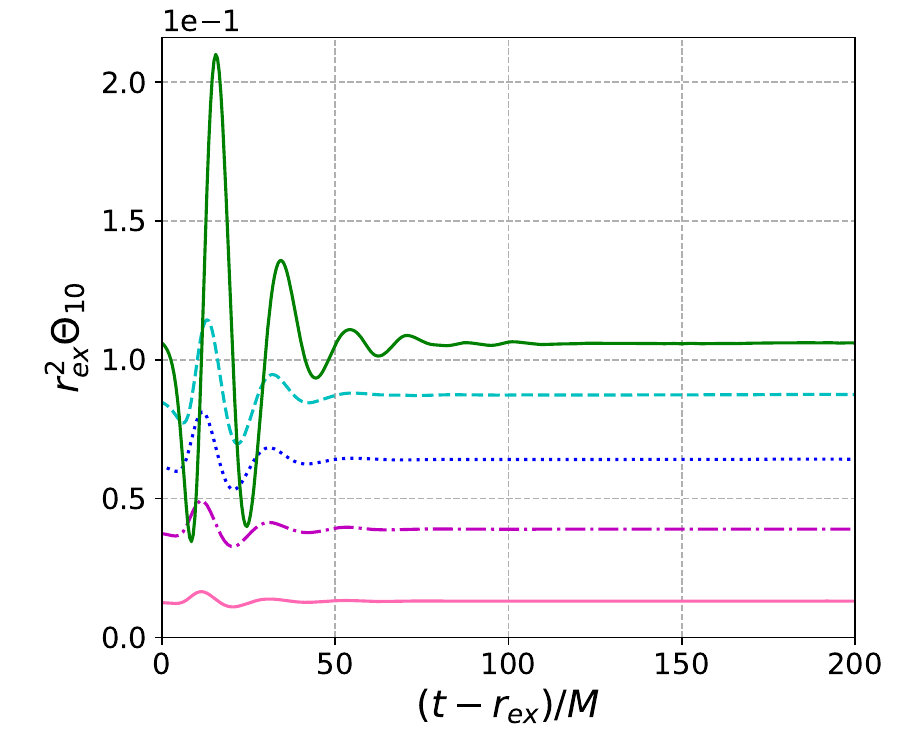}\label{fig:Theta10_comp_spinASpt1}}
    \caption{Evolution of the $l = 1, m = 0$ multipole of the axion field $\Theta$ extracted at $r_{ex}= 20M$ with coupling strength $\ha = 0.001$ (left panel), $\ha = 0.01$ (middle panel), $\ha = 0.1$ (right panel) in the background of a rotating \bh{}  with spin parameter $\chi=0.1$ (solid pink), $\chi=0.3$ (dash-dotted magenta), $\chi=0.5$ (dotted blue), $\chi=0.7$ (dashed cyan), $\chi=0.9$ (solid green).
    The axion is initialized by the small-coupling, small-spin solution, i.e., Type~II initial data in Sec.~\ref{ssec:ScalarInitialData}.
    }
\label{fig:Theta10_comp_spin}
\end{figure*}

\begin{figure*}[t!]
\centering
\subfloat[$\ha = 0.001$]{\includegraphics[width=0.325\textwidth]{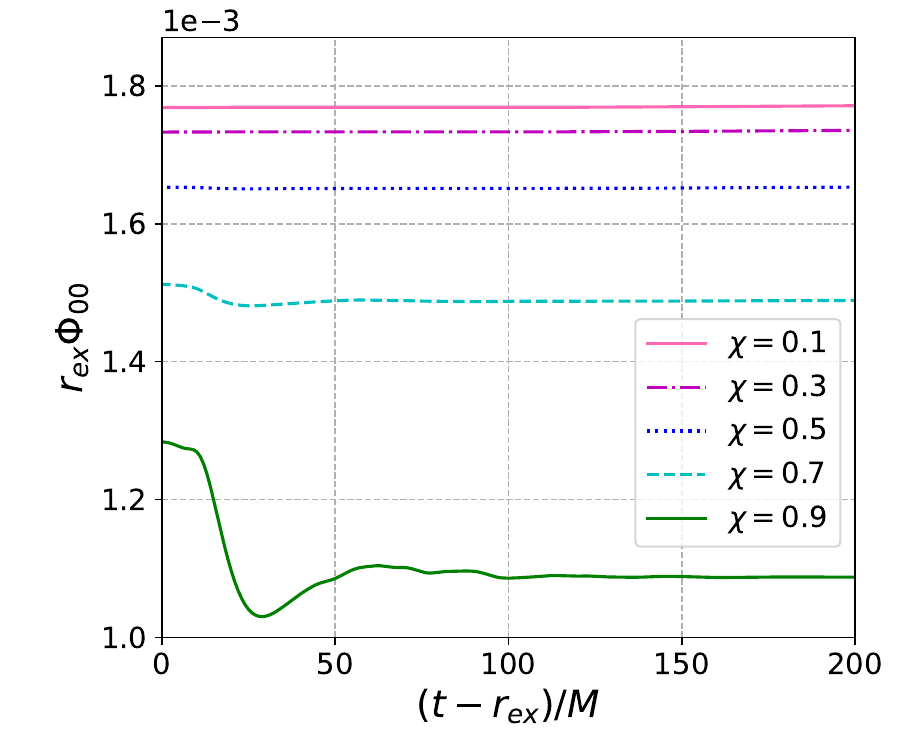}\label{fig:Phi00_comp_spinASpt001}}
\subfloat[$\ha = 0.01$]{\includegraphics[width=0.325\textwidth]{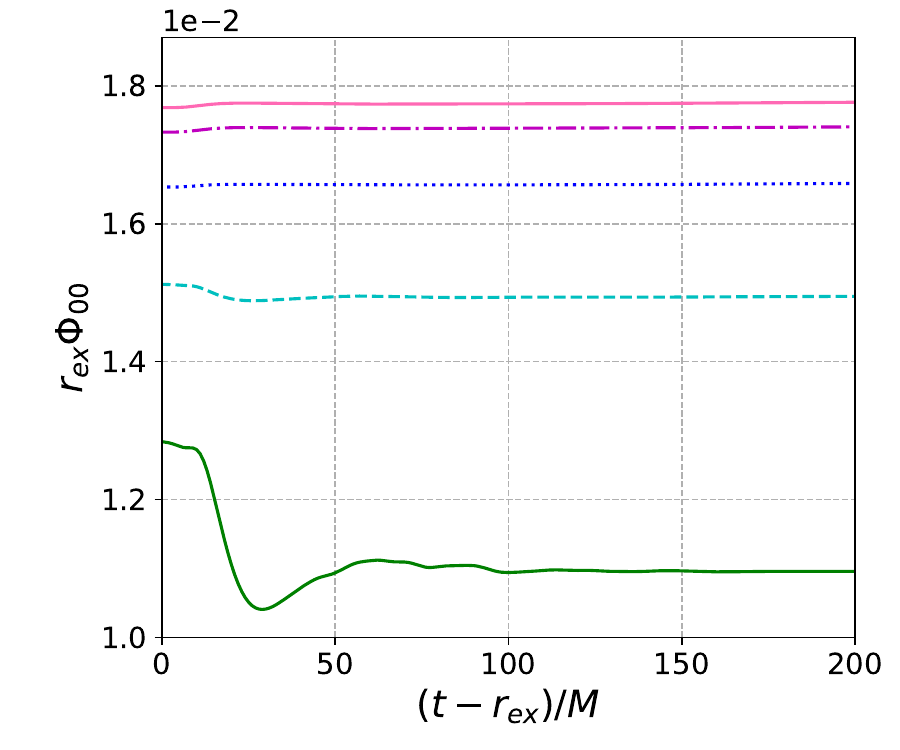}\label{fig:Phi00_comp_spinASpt01}}
\subfloat[$\ha = 0.1$]{\includegraphics[width=0.325\textwidth]{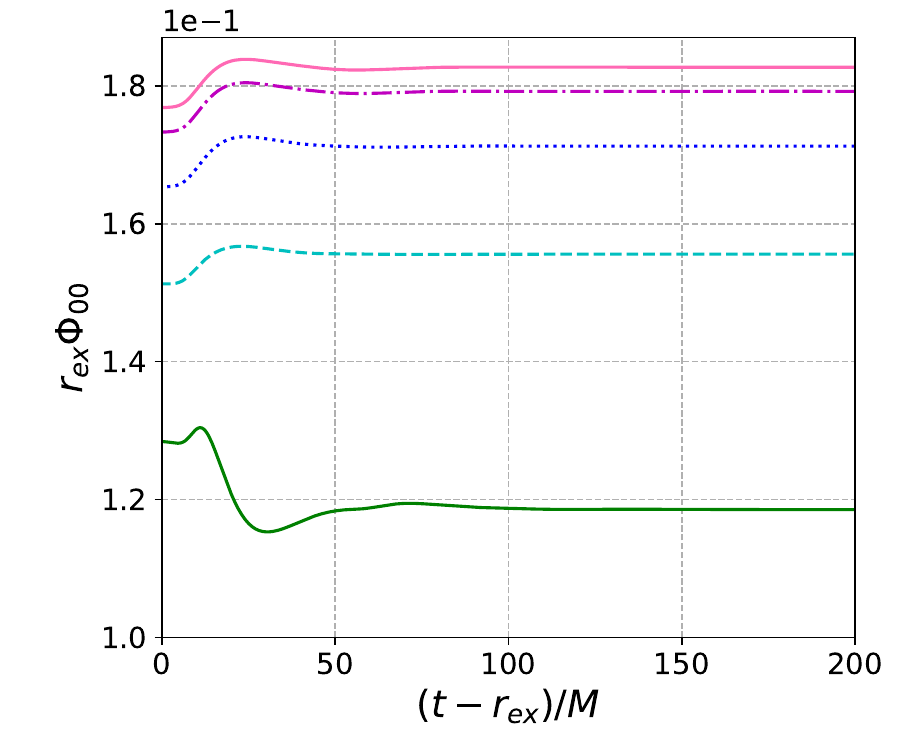}\label{fig:Phi00_comp_spinASpt1}}
    \caption{Same as Fig.~\ref{fig:Theta10_comp_spin} but for the evolution of the $l = m = 0$ multipole of dilaton field $\Phi$.
    The dilaton field is initialized by Type~II initial data in Sec.~\ref{ssec:ScalarInitialData}.
    }
\label{fig:Phi00_comp_spin}
\end{figure*}

In Figs.~\ref{fig:Theta10_comp_aS}
and~\ref{fig:Phi00_comp_aS}, we illustrate the dependence of the axion and dilaton hairs on the coupling strength,
focusing on simulations with a \bh{} spin of $\chi=0.9$.
Then in Figs.~\ref{fig:Theta10_comp_spin} and~\ref{fig:Phi00_comp_spin} we present the hairs' dependence on both the coupling strength and the \bh{} spin.

Fig.~\ref{fig:Theta10_comp_aS} shows the evolution of $\Theta_{10}$, rescaled by $\hat{\alpha}_{\rm{s,max}}/\ha$
with $\hat{\alpha}_{\rm{s,max}} = 0.1$, for all three coupling strengths.
For the simulations with small couplings, $\ha \in \{ 0.001, 0.01\}$,
we see that the final axion hairs overlap nearly perfectly.
This is consistent with the 
linear scaling of the small-coupling approximation.
However, the simulation with the largest coupling strength, $\ha = 0.1$, 
differs by approximately $2\%$.
While this difference is small, it is larger than the expected numerical error (see left panel of Fig.~\ref{appfig:percenterror})
and, thus, 
captures next-to-leading-order effects in the coupling strength, 
$\mathcal{O}(\ha^{2})$.
A more pronounced deviation from linear scaling with the coupling strength
is seen in the dilaton
shown in Fig.~\ref{fig:Phi00_comp_aS}.
Here we present $\Phi_{00}$,
rescaled by $\hat{\alpha}_{\rm{s,max}}/\ha$.
Comparing the simulations with small coupling strengths $\ha = 0.001$ and $\ha = 0.01$, we find that the final dilaton hairs differ by less than a percent.
Then, comparing the simulations with coupling strengths $\ha = 0.01$ and $\ha = 0.1$, we find that $\Phi_{00}$ differs by approximately $8\%$.
This difference indicates the significance of the  next-to-leading-order effects in the coupling.
Our results also indicate that effects of the order 
$\mathcal{O}(\ha^{2})$
are more significant for the dilaton than for the axion.

We now turn to Figs.~\ref{fig:Theta10_comp_spin} and~\ref{fig:Phi00_comp_spin},
where we show the dominant axion and dilaton multipoles
simulated with
$\ha\in \{ 0.001, 0.01, 0.1\}$ and varying the \bh{} spin.
We see that both the axion and dilaton hairs increase with the coupling strength as is expected from Eqs.~\eqref{eq:IDAxionHair} and~\eqref{eq:IDDilatonHair},
which are linear in the coupling.
Moreover, in Figs.~\ref{fig:Theta10_comp_spin} and~\ref{fig:Phi00_comp_spin} we can 
measure the true value of the scalar hair and quantify the effect of higher-order terms 
by comparing the end state to the approximate analytic solution.

To isolate the effect of high coupling, we do this comparison for simulations with small \bh{} spin $\chi = 0.1$.
In the case of the axion, we find the deviation between the final hair and the approximate analytic solution to be
$\sim 0.2\%$, $0.02\%$, and $2.4\%$ for coupling strengths $\ha = 0.001$, $0.01$, and $0.1$, 
with the approximation 
overestimating the scalar charge for $\ha = 0.001$ and 
underestimating the scalar charge for $\ha = \{0.01,0.1\}$. 
Similarly for the dilaton, we find the deviation to be 
$\sim 0.2\%$, $0.5\%$, and $3.3\%$ for coupling strengths $\ha = 0.001$, $0.01$, and $0.1$,
with the approximation underestimating the true value for all cases.
The differences at small coupling are comparable to numerical error, as expected, while the differences at high coupling capture higher-order effects.

Since the approximate analytic solution we use for our initial data 
employs both the small-spin and small-coupling approximation, we also study the interplay between the coupling and the spin.
As shown in Figs.~\ref{fig:Theta10_comp_spin} and~\ref{fig:Phi00_comp_spin}, in the highest spin case ($\chi = 0.9$), we note that for both the axion and dilaton, the final hairs differ from the initial approximations more for small-coupling simulations ($\ha \leq 0.01$) than for the high-coupling simulation ($\ha = 0.1$), counter to intuition.
Given that higher coupling pushes the final hair value up and higher spin values appear to push the final hair down relative to the analytic approximation, we can understand this trend to be the effects of higher-order spin and higher-order coupling corrections ``canceling'' each other out. 
This effect is even more apparent for the evolution of the dilaton hair in the $\chi = 0.7$ case, as seen in Fig.~\ref{fig:Phi00_comp_spin}. 
It can be seen that the approximate initial data is overestimating the dilaton hair for the first two panels and is underestimating it for the rightmost panel.
This shows the higher-order spin terms dominating in the first two panels and the higher-order coupling terms dominating in the rightmost panel.

Combining our results for $\Theta_{10}$ and $\Phi_{00}$, we find that the axion and dilaton hairs form dynamically for different coupling strengths.
We also show that the linear scaling in coupling strength fails for large coupling, indicating the importance of higher-order coupling terms.
Finally, we note the interplay of higher-order spin and coupling terms and anticipate that they enter with opposite signs.

\subsection{Effect of kinetic coupling on final hairs}
\label{ssec:NumResCoupBTWFields}

\begin{figure}[t!]
    \centering
    \includegraphics[width=0.48\textwidth]{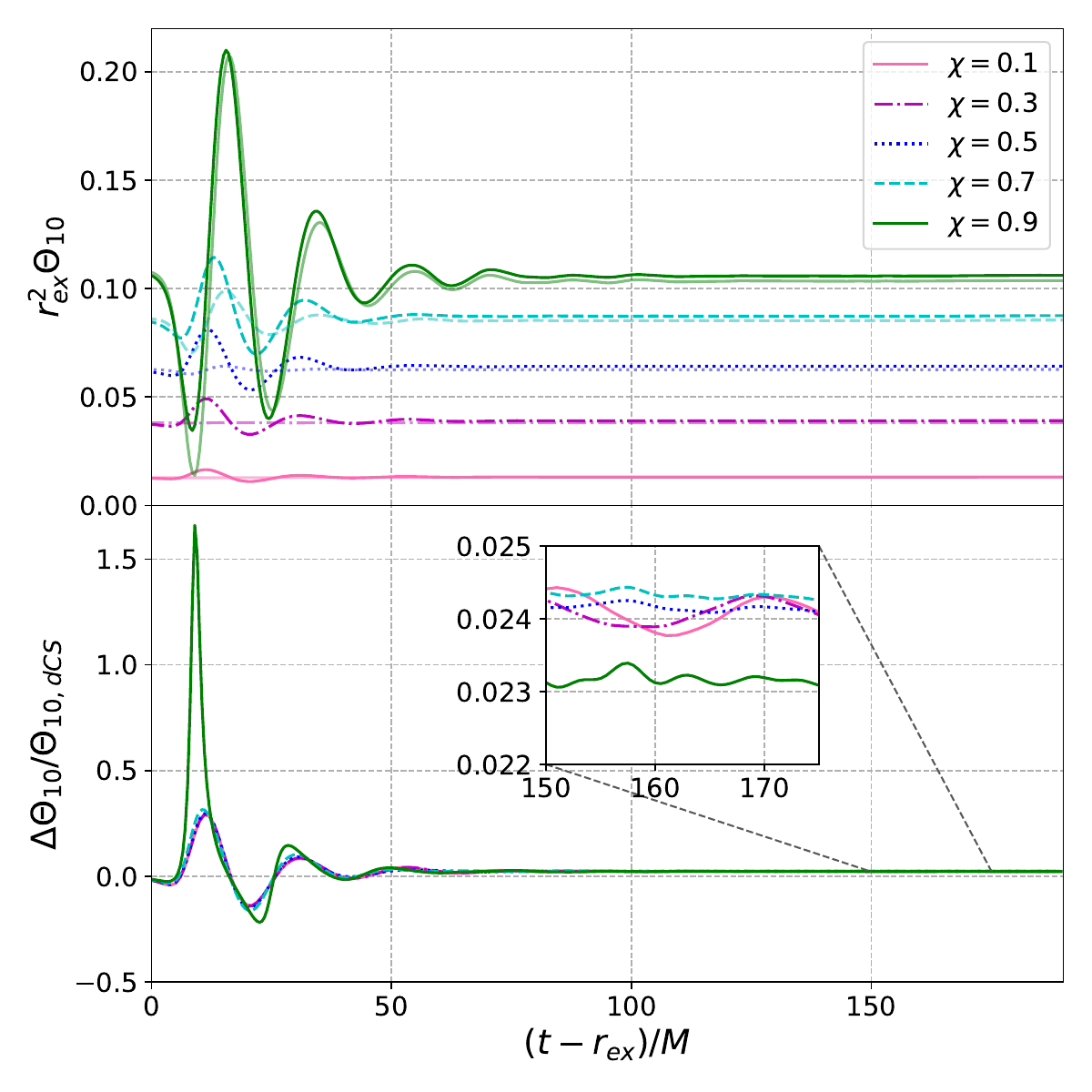}
    \caption{Top panel: Evolution of $\Theta_{10}$ rescaled by $\rex^2 = (20M)^2$ 
    in axi-dilaton gravity
    (coupled to the dilaton; Model~E in Sec.~\ref{sec:GravityModels}, opaque lines)
    and 
    in \dCS gravity (no dilaton; Model~B in Sec.~\ref{sec:GravityModels}, faded lines)
    for coupling strength $\ha = 0.1$ and a range of \bh{} spins.
    Bottom panel: Relative difference between $\Theta_{10}$ evolved in axi-dilaton gravity and in \dCS gravity. 
    \vspace{-0.2cm}
    }
    \label{fig:Theta10_AD_vs_dCS}
\end{figure}

\begin{figure}[t!]
    \centering
    \includegraphics[width=0.48\textwidth]{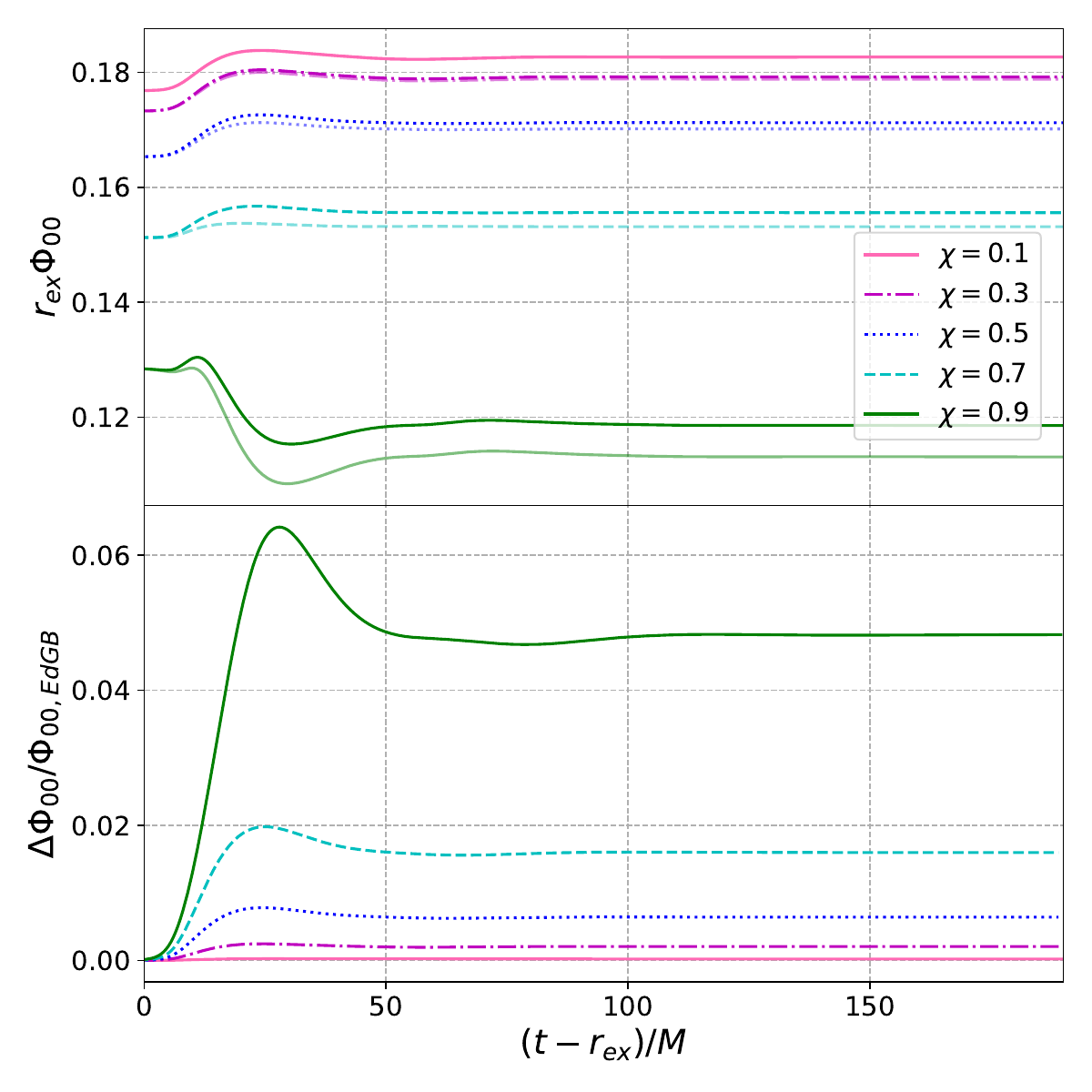}
    \caption{
    Top panel: Evolution of $\Phi_{00}$ rescaled by $\rex = 20M$ 
    in axi-dilaton gravity
    (coupled to the axion; Model~E in Sec.~\ref{sec:GravityModels}, opaque lines)
    and in \EdGB gravity 
     (no axion; Model~C in Sec.~\ref{sec:GravityModels} with Eq.~\eqref{eq:sGB_DilatonCoupling}, faded lines)
    for coupling strength $\ha = 0.1$ 
    and a range of \bh{} spins.
    Bottom panel: Relative difference between $\Phi_{00}$ evolved in axi-dilaton gravity and in \EdGB gravity.
    \vspace{-0.1cm}
    }
    \label{fig:Phi00_AD_vs_EdGB}
\end{figure}

\begin{figure}
    \centering
    \includegraphics[width=\columnwidth]{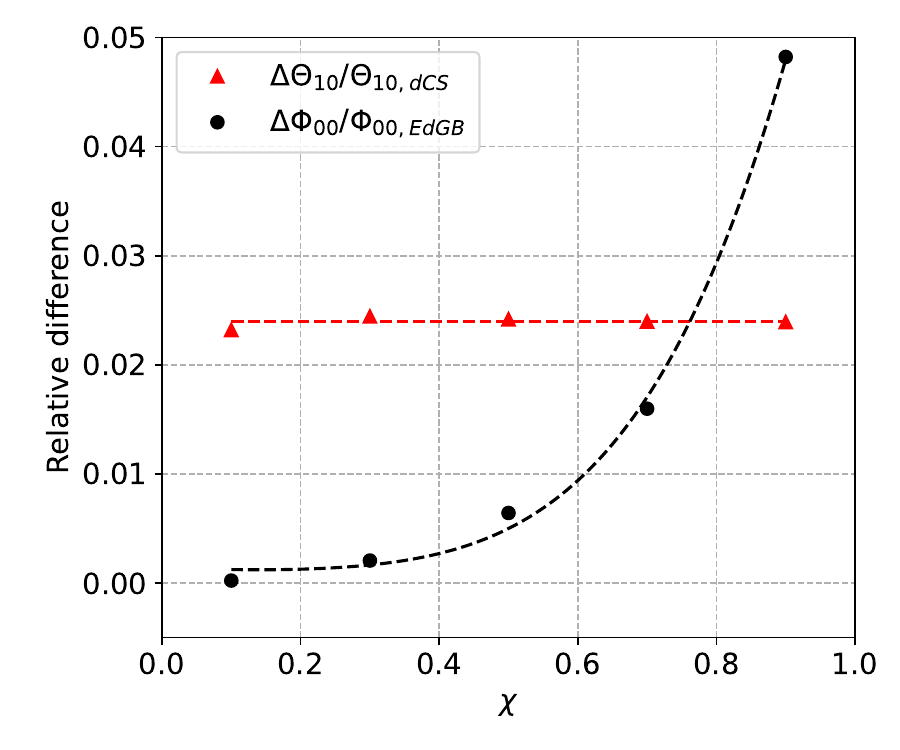}
    \caption{Relative difference at late times ($t = 150$M) between $\Theta_{10}$ evolved in axi-dilaton and \dCS gravity (red triangles)
    and $\Phi_{00}$ evolved in axi-dilaton and \EdGB gravity (black circles) as a function of \bh{} spin $\chi$ with coupling strength $\ha = 0.1$.
    We give lines of fit for the axion field (dashed red) and dilaton field (dashed black).
    }
    \label{fig:RelDiffVsSpin}
\end{figure}

To gain further insight into the effect of the kinetic coupling between the axion and dilaton fields on the final hairs,
we compare simulations in axi-dilaton gravity to simulations in which the fields are not coupled to each other.
The latter correspond to the axion determined by \dCS gravity and the dilaton determined by \EdGB gravity.
The results are shown in Figs.~\ref{fig:Theta10_AD_vs_dCS} and \ref{fig:Phi00_AD_vs_EdGB}.
All simulations presented in this section were performed using the \CanudaAD thorns with suitable parameter choices; see App.~\ref{appssec:ModelSelection}.
For cases in which the axion and dilaton evolve separately according to \dCS and \EdGB gravity,
we simulate them in the same run 
and realize the decoupling through the appropriate parameter choice.

In the top panels of Figs.~\ref{fig:Theta10_AD_vs_dCS} and \ref{fig:Phi00_AD_vs_EdGB}, we show the evolution of $r_{ex}^2\Theta_{10}$ and $r_{ex}\Phi_{00}$
in axi-dilaton gravity (opaque lines),
together with their evolution in \dCS and \EdGB gravity (faded lines), respectively.
We set the coupling strength to $\ha=0.1$ and consider a range of \bh{} spins $\chi \in \{ 0.1, 0.3, 0.5, 0.7, 0.9\}$.
We initialize the axion and dilaton as the approximate analytical solution,
i.e., Type~II initial data in Sec.~\ref{ssec:ScalarInitialData}.
At this order in 
$\ha$, the initial data given by Eq.~\eqref{eq:IDAxionHair} is the same for the axion in \dCS and in axi-dilaton gravity,
and 
Eq.~\eqref{eq:IDDilatonHair} is the same for the dilaton in \EdGB and in axi-dilaton gravity.

For both the axion and the dilaton, we observe that the magnitude of the final hair in axi-dilaton gravity is larger than that in \dCS and \EdGB gravity individually.
Since the interaction between the axion and dilaton fields is the only alteration between the two 
sets of simulations,
this difference in magnitude highlights the effect of the kinetic coupling between the fields.
In particular, this makes it clear that the kinetic coupling between the axion and dilaton fields comprises some of the higher-order coupling effects observed in Sec.~\ref{ss:NumResCoupl}.

To quantify the effect of the kinetic coupling between the fields on the axion hair,
we present the 
relative difference
between the axion in axi-dilaton and \dCS gravity, 
$\left(\Theta_{10,\rm{AxiDil}}-\Theta_{10,\rm{dCS}}\right)/\Theta_{10,\rm{dCS}}$, 
in the bottom panel of Fig.~\ref{fig:Theta10_AD_vs_dCS}.
After the initial perturbation decays,
we see that the relative difference, and thus the impact of the coupling between the fields, 
is the same for all \bh{} spins.

Turning to the effect of the fields' kinetic coupling 
on the dilaton hair, we show the relative difference between the dilaton in axi-dilaton and \EdGB gravity,
$\left(\Phi_{00,\rm{AxiDil}}-\Phi_{00,\rm{EdGB}}\right)/\Phi_{10,\rm{EdGB}}$
in the bottom panel of Fig.~\ref{fig:Phi00_AD_vs_EdGB}. 
Here, we see a clear dependence on the \bh{} spin.

We quantify the relationship between this relative difference 
and the \bh{} spin in  Fig.~\ref{fig:RelDiffVsSpin}.
As mentioned above,
$\Delta\Theta_{10}/\Theta_{10, \textrm{dCS}}$ does not depend on \bh{} spin,
so we fit it as a constant and find $\Delta\Theta_{10}/\Theta_{10, \textrm{dCS}} = 0.024$.
In the case of the dilaton, we fit $\Delta\Phi_{00}/\Phi_{00, \textrm{EdGB}}$ with 
\begin{equation}
    \Delta\Phi_{00}/\Phi_{00, \textrm{EdGB}} = a(\chi + b)^2 + c(\chi+b)^4 + d \,,\nonumber 
\end{equation}
and find $a = 0.014$, $b = -0.139$, $c = 0.115$, and $d = 0.001$. 
While not presented here, we note that for small \bh{} spins $\chi = \{0.1, 0.3,0.5\}$, 
$\Delta\Phi_{00}/\Phi_{00, \textrm{EdGB}}$ 
fits extremely well with 
only a quadratic function (i.e. $c=0$)
while the $\chi^4$ term is necessary to 
accurately
fit high spins.

The spin dependence of the 
effect of the
kinetic coupling 
is understood by combining the fields' evolution equations with their approximate analytical solutions.
In Eqs.~\eqref{eq:AxiDil_eoms_with_ahat},
the axion and dilaton are 
determined in part by the kinetic coupling terms 
$\Box \Theta \propto \nabla_i\Theta \nabla^i \Phi$ and 
$\Box \Phi \propto \nabla_i\Theta \nabla^i\Theta$.
Inserting the leading-order contributions to the approximate analytical solutions 
$\Theta \propto \ha \chi$ (see Eq.~\eqref{eq:IDAxionHair}) and
$\Phi \propto \ha$ (see Eq.~\eqref{eq:IDDilatonHair}), 
into these pseudo-equations,
we recognize that 
for the axion, the spin cancels on both sides
while
the dilaton picks up a factor of $\chi^{2}$ 
from the axion source
on the right-hand-side.

In summary, we find that the effect of the kinetic coupling between the axion and dilaton fields is 
comparable for all \bh{} spins for the final axion hair
and
increases 
(at least) quadratically with
\bh{} spin for the dilaton hair. 
For both the axion and dilaton hairs, the effect of the kinetic coupling leads to larger end states in comparison to the hairs evolved only in \dCS or \EdGB gravity.
Extending to \bbh mergers, we expect 
the effect of the kinetic coupling to give stronger deviations from \gr \bh{s} than in \dCS and \EdGB gravity.

\section{Axion and dilaton evolution around binary black holes}
\label{sec:BBHSimulations}

In this section, we present first results of a \bbh{} merger in axi-dilaton gravity, in the decoupling approximation.
We perform
a short, proof-of-principle simulation
to
(i) study the 
formation of the axion and dilaton hairs around the binary, 
and
(ii) compute the axion's and dilaton's radiative multipoles produced by the inspiraling hairs that represent orbiting scalar charges.

The background spacetime, described by \gr{,} consists of an equal-mass, nonspinning \bbh{} system with a total \bh{} mass of $\textrm{M}=m_{1}+m_{2}=1$,
where $m_1$ and $m_2$ are the ADM masses of the individual punctures, as defined in ~\cite{Ansorg:2004ds}.
The \bh{s} have an initial coordinate separation of $d=6$M. 
They inspiral for about
two 
orbits before merging at $t\sim108$M
(determined by the formation of the common apparent horizon).
The newly formed \bh{} rings down to a rotating remnant \bh{}
with dimensionless spin parameter $\chi = 0.688$.

\begin{figure*}[th!]
\centering
\subfloat[Top panel: evolution of the \bbh{'s} coordinate separation.
Middle panel: evolution of dominant axion multipole, $\Theta_{10}$. 
Bottom panel: evolution of dominant dilaton multipole, $\Phi_{00}$.]{\includegraphics[width=0.48\textwidth]{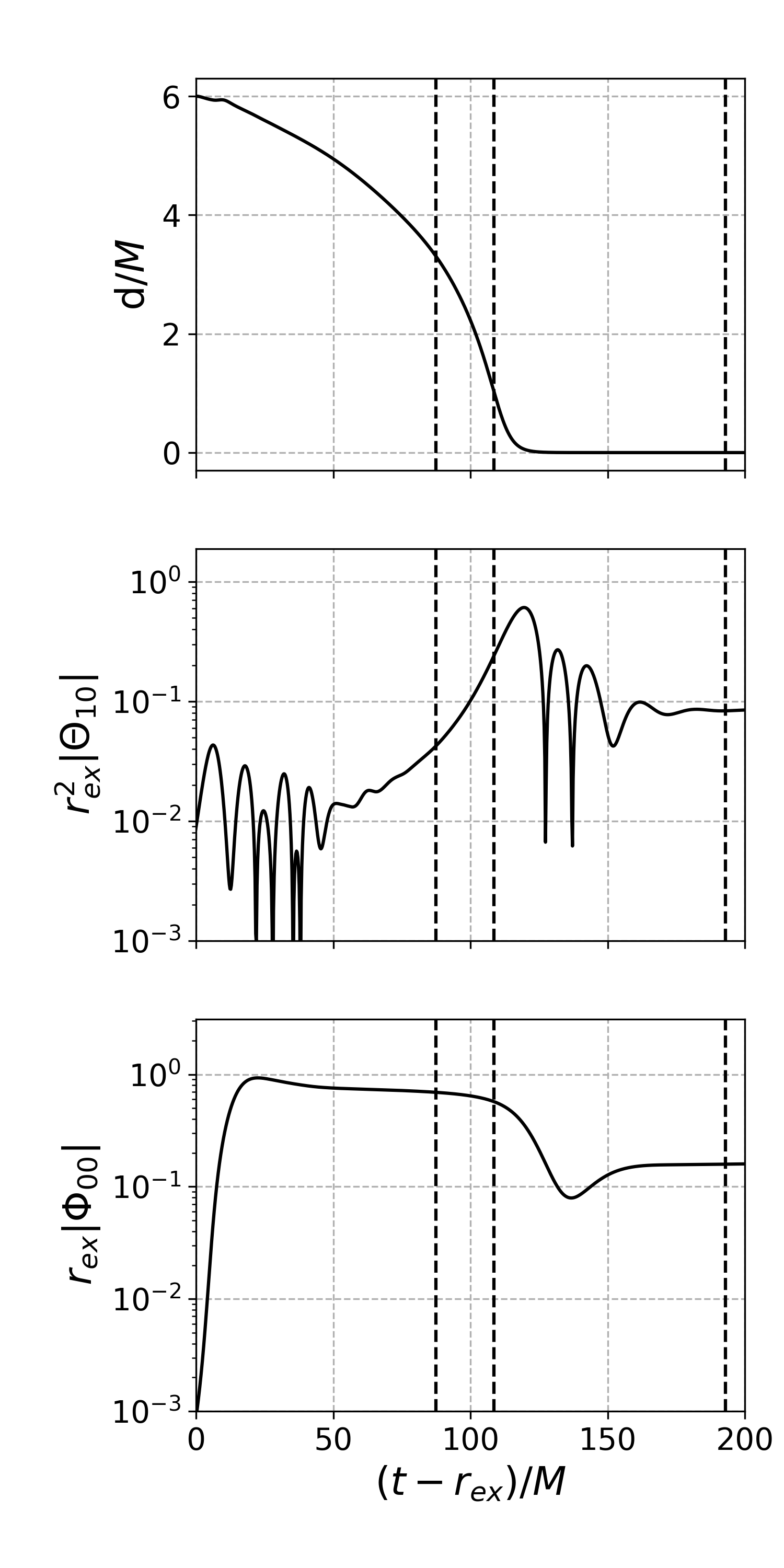}\label{fig:binary_scalar_hair}}
\hfill
\subfloat[Top panel: dominant $\l=2$, $m=2$ mode of the gravitational waveform, $\Psi_{4,22}$ of the \bbh background. 
Middle panel: dominant radiative multipole,  $\Theta_{32}$, of the axion.
Bottom panel: dominant radiative multipole, $\Phi_{22}$, of the dilaton.
]{\includegraphics[width=0.48\textwidth]{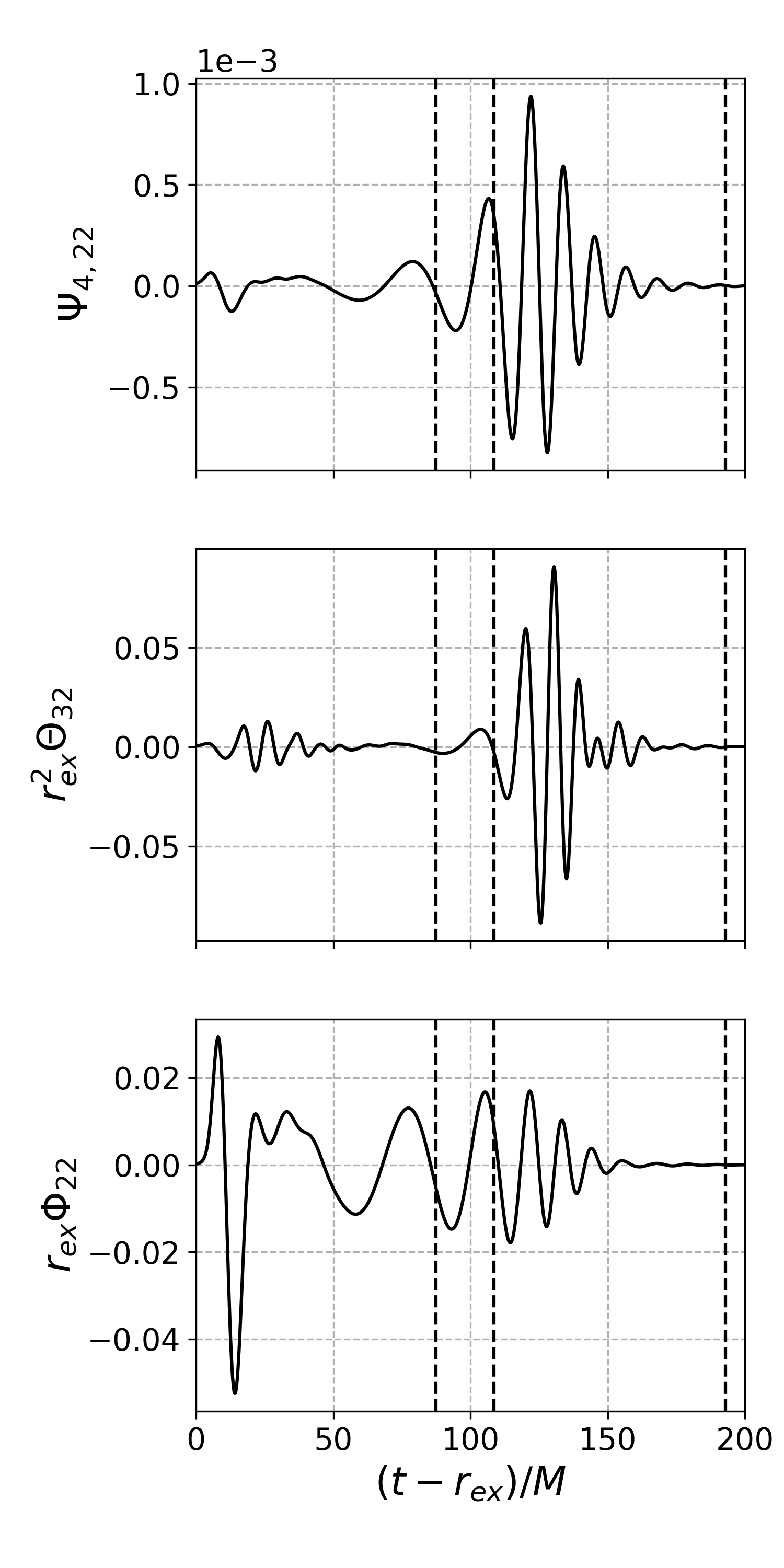}\label{fig:binary_radiative_modes}}
\caption{Evolution of axion and dilaton fields in the background of an equal mass, nonspinning \bbh{} merger.
The shown multipoles are extracted at a radius of $r_{ex}=80$M. 
The coupling strength is $\ha = 0.1$, and the fields are set to zero at the beginning of the simulation. 
The remnant \bh{} has a spin of $\chi=0.688$. 
The dashed, vertical lines denote each of the timestamps shown in Fig.~\ref{fig:binary_2d}, the middle of which is the first appearance of the common horizon at $t = 108$M.
\vspace{-0.2cm}
}
    \label{fig:binary_modes}
\end{figure*}

\begin{figure}[t!]
    \centering
    \includegraphics[width = 0.5\textwidth]{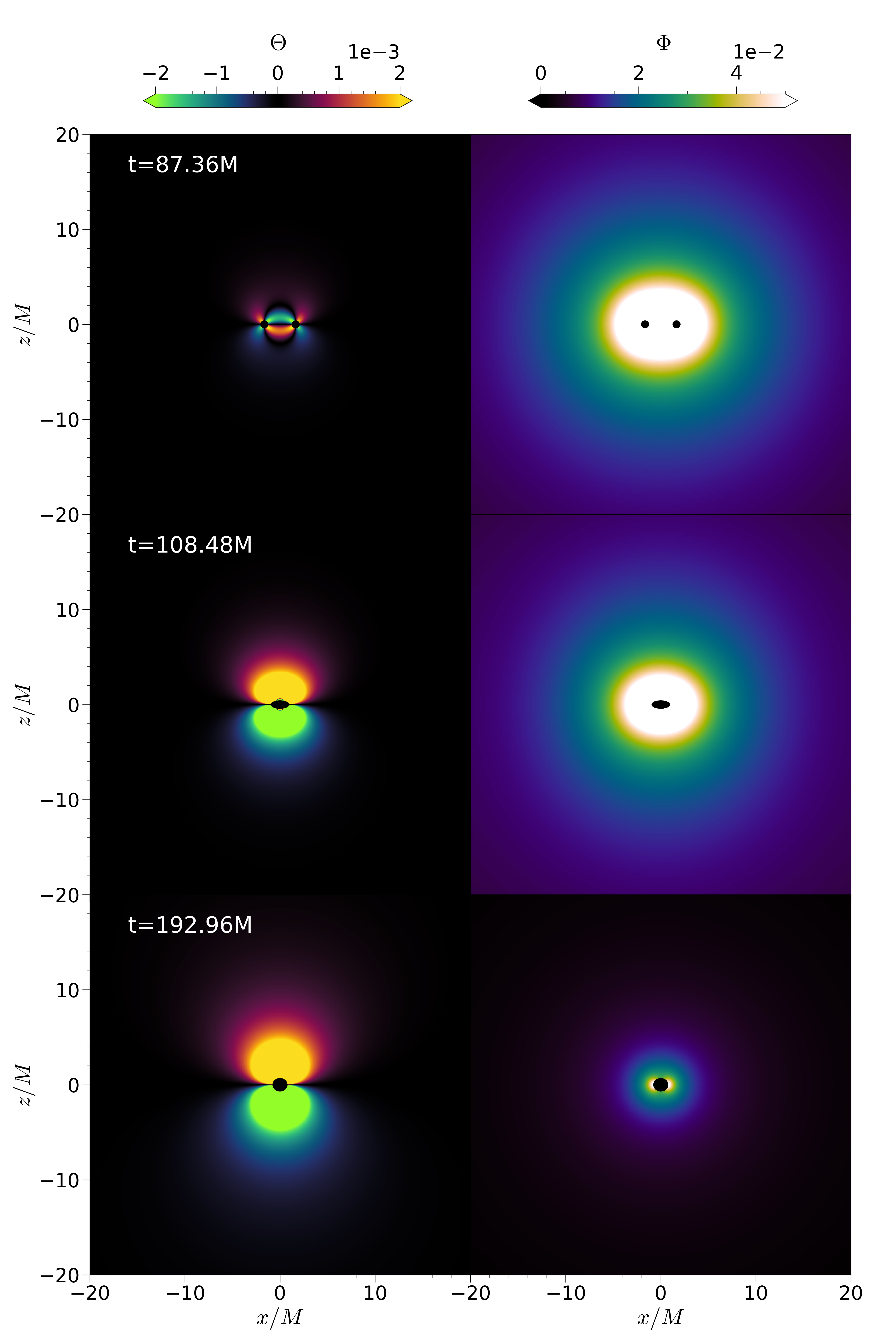}
    \caption{Axion (left) and dilaton (right) fields in the xz-plane (a slice perpendicular to the orbital plane) during the inspiral at $t=87.36M$ (top), the merger at $t=108.48M$ (middle), and post-merger at $t=192.96M$ (bottom) for a \bbh{} system with equal mass, nonspinning initial \bh{s}. The coupling strength is $\ha = 0.1$, and the fields are set to zero at the beginning of the simulation.} 
    \label{fig:binary_2d}
\end{figure}

We illustrate the evolution of the \bh{s} in the top panels of Fig.~\ref{fig:binary_modes}.
In Fig.~\ref{fig:binary_scalar_hair} we show the evolution of the \bh{s'} coordinate separation
to guide our understanding of
the axion's and dilaton's 
behavior
throughout the coalescence.
The dashed vertical lines indicate times during the late inspiral,
the formation of the common apparent horizon that signifies the merger, and after the ringdown.
They correspond to the times of the two-dimensional snapshots
of the axion and dilaton
presented later in Fig.~\ref{fig:binary_2d}.
In the top panel of Fig.~\ref{fig:binary_radiative_modes} we present the dominant gravitational waveform, i.e., the quadrupole of the Newman-Penrose scalar $\Psi_{4,22}$.
After a short burst of unphysical ``junk'' radiation, the waveform exhibits
the characteristic morphology of the late inspiral, merger peak, and exponentially decaying ringdown.

We now turn to the evolution of the axion and dilaton
whose dynamics are driven by the binary.
They are initialized as
zero according to initial data Type~I in Sec.~\ref{ssec:ScalarInitialData}.
We set the dimensionless coupling strength $\ha = 0.1$.
For \bbh simulations we define the dimensionless coupling parameter 
$\ha=\as/(m_1+m_2)^2$.

We illustrate the evolution of the axion's and dilaton's  
multipoles in Fig.~\ref{fig:binary_modes}. 
In the left panels, we show the dominant multipoles 
making up the hair
of each field, and in the right panels, we show their radiative modes.
We extract our data at $r_{ex}=80$M with the origin placed at the center of mass of the \bbh{}, thus capturing the 
behavior of the fields around both \bh{s.} 

In the bottom panel of Fig.~\ref{fig:binary_scalar_hair}, we show the evolution of the dominant mode of the dilaton hair,
$\rex \Phi_{00}$.
The dilaton hair initially grows and settles to a nearly constant value as the \bh{s} 
coalesce.
After the \bh{s} merge,
the dilaton charge decreases and settles to a value lower than that during the inspiral.
This is predominantly due to the remnant \bh{'s} larger mass, i.e. lower curvature, as well as due to the remnant's higher spin; 
see Eq.~\eqref{eq:IDHairMultipolePhi00}.

In the middle panel of Fig.~\ref{fig:binary_scalar_hair}, we present the evolution of the leading-order multipole of the axion hair, $\rex^{2}\Theta_{10}$.
At the beginning of the simulation, the axion hair oscillates and settles close to zero.
This behavior is expected during the  inspiral of {\emph{nonspinning}} \bh{s} as the axion hair scales with the \bh{} spin; 
see Eq.~\eqref{eq:IDAxionHair}.
We observe, however, that the axion starts to grow
in the late inspiral, but before the final, rotating \bh{} is born.
The origin of this growth around the nonspinning \bh{s} may be two-fold:
First, the orbital angular momentum in the late inspiral may generate a nonzero Pontryagin density that sources the axion, see Eq.~\eqref{eq:AxiDil_gen_axioneom}.
This is consistent with results of the axion's dynamics in \dCS gravity~\cite{Okounkova:2017yby}, i.e., in the absence of the dilaton.
Second, the kinetic coupling to the dilaton
may enhance the growth of the axion hair 
(see
the last term in Eq.~\eqref{eq:AxiDil_gen_axioneom}).
The axion hair peaks as the \bh{s} merge.
As the newly formed \bh{} settles to a Kerr \bh{} with a spin of $\chi\sim0.688$,
the axion rings down and settles to its final hair profile.
A detailed analysis of  the pre-merger axion growth 
is forthcoming.

Since each \bh{} carries a dilaton and, in the late inspiral, axion charge, their inspiral generates axion and dilaton radiation.
In Fig.~\ref{fig:binary_radiative_modes} we present the dominant radiative modes, $\Theta_{32}$ and $\Phi_{22}$, respectively.
In the first $t\sim50$M, both fields show an initial oscillation as they adjust from the zero initial data to the binary configuration.
The axion's radiation, $\Theta_{32}$, shown in the middle panel of Fig.~\ref{fig:binary_radiative_modes}, remains close to zero during the inspiral 
which is consistent with the zero axion hair.
Only when the latter starts growing in the late inspiral does it generate axion radiation that is increasing in amplitude and frequency. 
It peaks during the \bh{s'} merger, and then decays exponentially.
The dilaton's radiation, $\Phi_{22}$, shown in the bottom panel of Fig.~\ref{fig:binary_radiative_modes},
oscillates with increasing amplitude and frequency until merger.
After the \bh{s'} merger
the dilaton radiation decays to zero.

In Fig.~\ref{fig:binary_2d}, we show two-dimensional snapshots of the axion and dilaton fields at $t=87.36$M (during the inspiral after junk has settled), $t=108.48$M (just after the common apparent horizon is found), and $t=192.96$M (after the final \bh{} has settled).
We see that the strength of the axion field is significantly higher after the merger, consistent with the nonzero spin of the remnant \bh{} which leads to growth of axion hair. 
The dilaton field changes in both magnitude and shape between the inspiral and the post-merger. 
Due to the larger mass of the remnant \bh{} relative to that of the initial, individual \bh{s}, the dilaton's magnitude decreases after the merger.
This is compounded by the nonzero spin of the remnant which also drives down the overall amplitude of the dilaton field.
Since the dilaton's next-to-leading-order multipole increases with spin, its relative amplitude as compared to the dominant mode increases, and we observe a deviation from the monopole structure of the dilaton field around the \bh{}.
This behavior is consistent with the results 
shown in Fig.~\ref{fig:2D_SBH_apt9_snaps} for the highly spinning single \bh{}.

In summary, using \CanudaAD, we evolve axion and dilaton fields coupled to each other in a \bbh{} background, showing how their profiles change throughout the evolution.
We present the growth of axion hair during the inspiral before the formation of a spinning remnant, an effect that has only been captured by numerical relativity simulations
and has never before been shown in axi-dilaton gravity.
This growth may come from several sources including a possible nonzero Pontryagin density caused by the orbital angular momentum or the coupling with the dilaton field.
We also show the presence of radiative modes for both the axion and dilaton fields which  
carry energy away from the system.
Such energy loss leads to changes in the \GW{s} which may be observable with future \GW{} observatories.
As a consequence of the increased axion and dilaton hairs observed for axi-dilaton gravity versus \dCS and \EdGB gravity, it is likely that this radiation, and thus energy loss, will be larger for axi-dilaton gravity as well.
In future work, we will quantify this energy loss and expand this analysis to include more coverage of the \bbh{} parameter space. 

\section{Conclusions}\label{sec:Summary}

In this paper, we probed the dynamical 
evolution of
an axion and a dilaton, kinetically coupled to each other,
around a single and binary \bh{s} in axi-dilaton gravity.
To accomplish this, we created \CanudaAD, the first open-source, parameterized numerical relativity code to perform simulations in 
(bi-)scalar-tensor theories of gravity.
Through this parameterization the code has the flexibility
to study \bh{s} in variety of theories beyond \gr{,}
including axi-dilaton gravity as well as popular models such as \dCS and \sGB gravity.

Using our new 
code, we performed a series of simulations to investigate how axion and dilaton fields evolve in axi-dilaton gravity in the background of \bh{s}.
Our results present the first simulations of the 
dynamical formation 
of axion and dilaton
hair.
In the case of simulations with high \bh{} spin or high coupling strength, we captured a new end-state solution of the axion and dilaton hairs.

We then addressed three questions: i) how do the axion and dilaton hairs depend on the \bh{} spin, ii) how does the strength of the coupling to curvature affect the axion and dilaton hairs, and iii) how does the kinetic coupling between the axion and dilaton fields impact these hairs?

We showed the dynamical growth of the axion and dilaton hairs from zero initial data for a variety of \bh{} spins.
For backgrounds with small \bh{} spin, we confirmed consistency with the approximate analytical solution
valid in the small-spin, small-coupling limit.
For larger \bh{} spins, we captured a new end-state solution for the axion and dilaton hairs.
By varying the spin of the \bh{}, we found that the final axion hair increases with increasing spin. 
This holds true for both the leading-order and next-to-leading-order multipoles. 
For the dilaton hair, we found that the magnitude of the dominant mode decreases with increasing spin while the next-to-leading-order multipole increases with increasing spin. 
For high spins, this 
deformation away from the monopole
is visible in the dilaton profile on a slice perpendicular to the equatorial plane.

To continue our analysis, we varied the strength of the fields' coupling to their 
respective curvature invariants. 
For small coupling strengths, the magnitude of the axion and dilaton hairs increase linearly 
with the coupling parameter,
which is consistent with the 
approximate analytical solution.
In simulations with large coupling strengths, however,
the increase in the fields' magnitude
exceeds a linear scaling,
thus indicating the importance of higher-order effects in the coupling.

We investigated one such higher-coupling effect in detail, namely the kinetic coupling between the axion and dilaton fields. 
We found that the kinetic coupling yields
axion and dilaton end states with a larger magnitude 
in comparison to simulations in which the fields are not coupled to each other.
To quantify the impact of the kinetic coupling 
on the final axion and dilaton hairs, we performed simulations where the coupling between the fields was turned on ($\gph = e^{-\Phi}$) and simulations where it was turned off ($\gph = \text{constant}$). 
For the axion field, 
switching on the kinetic coupling increased the final axion hair by $\approx 2.4\%$,
independent of the spin of the \bh{}.
The kinetic coupling between the fields also caused an increase in the magnitude of the dilaton hair.
We observed that, in this case, the increase depends on the \bh{} spin,
and found a quadratic scaling 
for small to moderate spins, while for high spins quartic terms also become important.
For a \bh{} of spin $\chi=0.9$, the dilaton hair was $\approx 5\%$ larger in simulations
with the kinetic coupling switched on
as compared to simulations without it.
For both, the axion and the dilaton, the difference due to the kinetic coupling between the fields
accounts, in part, for 
the higher-order coupling effects observed earlier. 

We also performed a simulation of the 
axion and dilaton fields evolving around a \bbh{} system in axi-dilaton gravity,
which nicely captured several of the trends observed in the single \bh{} simulations.
We began with two nonspinning \bh{s} with both the axion and dilaton fields set to zero and observed their evolution.

Although the initial \bh{s} are nonspinning, and therefore their Pontryagin density vanishes initially, we observed a growth of the axion hair in the late inspiral before it settled to a static hair around the spinning remnant \bh{.}
The growth of the axion hair before the merger could be indicative of a nonzero Pontryagin density caused by the orbital angular momentum, or it could be the result of coupling with the dilaton field which is present throughout the inspiral.
We note that the pre-merger growth of the axion has only been captured with numerical relativity simulations and we plan to investigate this effect further in upcoming work.

The opposite trend could be seen for the dilaton field, which began at a larger value around the nonspinning initial \bh{s} and decreased in magnitude post merger due to the higher mass and, in part, due to spin of the remnant \bh{}.
The final dilaton hair profile again displayed a slight deviation away from a monopole profile due to higher multipoles 
being present in the stationary solution around the spinning remnant \bh{.}

Radiative modes were present for both the axion and 
dilaton fields, carrying away energy from the binary.
Given the effect of the kinetic coupling increasing the axion and dilaton hairs, we anticipate that this radiation will be larger for axi-dilaton gravity than for \dCS or \EdGB gravity alone.
Such energy dissipation would impact the \bbh{} system itself as well as its gravitational radiation, effects which we will study in future works.

\section*{Acknowledgments}
We thank 
Roland Haas,
Leah Jenks,
Paolo Pani,
Hector O. Silva,
Kent Yagi, and
Nico Yunes
for insightful discussions and comments.
The authors acknowledge support provided by the National Science Foundation under NSF Award No.~OAC-2004879, No.~PHY-2110416,
No.~OAC-2411068 and No.~PHY-2409726.
CBR acknowledges support provided by the ICASU-Physics Fellowship. 
A.D. acknowledges support provided by the MUR PRIN Grant 2020KR4KN2 ``String Theory as a bridge between Gauge Theories and Quantum Gravity''; by the FARE program (GW-NEXT, CUP:~B84I20000100001), and by the INFN TEONGRAV initiative.
We acknowledge the Texas Advanced Computing Center (TACC) at the University of Texas at Austin for providing HPC resources on Frontera via allocations PHY22041.
This research used resources provided by the Delta research computing project, which is supported by the NSF Award No. OCI-2005572 and the State of Illinois.
This research was supported in part by the Illinois Computes project which is supported by the University of Illinois Urbana-Champaign and the University of Illinois System.
This work used the open-source softwares
\textsc{xTensor}~\cite{xact_url,Brizuela:2008ra},
the \ETK~\cite{Loffler:2011ay,EinsteinToolkit:2022_11},
\Canuda~\cite{witek_helvi_2023_7791842}, and \kuibit~\cite{kuibit}.
The \CanudaAD{} code developed to conduct the simulations in this work is open source and available in a git repository~\cite{CanudaAxiDil_repo}.
A YouTube playlist with two-dimensional animations rendered from data produced with our simulations is available at \href{https://www.youtube.com/@canudanumericalrelativity1634}{www.youtube.com/@canudanumericalrelativity1634}.

\appendix 

\begin{table*}[t!]
    \centering
    \begin{tabular}{|c|c|c|c|c|c|c|c|}
    \hline
    Model of gravity &
    \texttt{axi\_coupling} &
    \texttt{aCS} &
    \texttt{dil\_coupling} &
    \texttt{aGB} &
    \texttt{dil\_lambda} &
    \texttt{AD\_coupling} &
    \texttt{AD\_lambda} \\
    \hline
    \hline
    Minimally coupled scalar & none & $0.0$ & none & $0.0$ & -- & none & -- \\
    Shift-symmetric \dCS & linear & $10^{-3}-10^{-1}$ & none & $0.0$ & -- & none & -- \\
    Cubic \dCS & cubic & $10^{-3}-10^{-1}$ & none & $0.0$ & -- & none & -- \\
    \EdGB & none & $0.0$ & exponential & $10^{-3}-10^{-1}$ & $1.0$ & none & -- \\
    Shift-symmetric \sGB & none & $0.0$ & linear & $10^{-3}-10^{-1}$ & -- & none & -- \\
    Quadratic \sGB & none & $0.0$ & quadratic & $10^{-3}-10^{-1}$ & -- & none & -- \\
    Minimally coupled bi-scalar  & none & $0.0$ & none & $0.0$ & -- & exponential & any value \\
    Axi-dilaton & linear & $10^{-3}-10^{-1}$ & exponential & $10^{-3}-10^{-1}$ & $1.0$ & exponential & $1.0$ \\
    \hline
    \end{tabular}
    \caption{Parameter choices in the \CanudaAD code to select different models of gravity
    as outlined in Sec.~\ref{sec:GravityModels}.
    \texttt{axi\_coupling} and \texttt{aCS} describe the coupling function $\fth$ and coupling strength
    between the axion and the Pontryagin density.
    \texttt{dil\_coupling} and \texttt{aGB} describe the coupling function $\fph$ and coupling strength
    between the dilaton and the Gauss-Bonnet invariant.
    \texttt{dil\_lambda} is the dimensionless constant $\lambda_{\text{GB}}$ in the coupling function $\fph = e^{\lambda_{\text{GB}} \Phi}$.
    \texttt{AD\_coupling} determines the coupling function $\gph$ between the fields.
    \texttt{AD\_lambda} is the dimensionless constant $\lambda_{\text{AD}}$ in the coupling function $\gph = e^{-\lambda_{\text{GB}} \Phi}$.}
    \label{tab:parameter choices}
\end{table*}

\section{\CanudaAD code description}\label{appsec:code_description}

In this section, we complete the description of the \CanudaAD code~\cite{CanudaAxiDil_repo}.
\CanudaAD{} is a parameterized numerical relativity code for
(bi-)scalar-tensor theories of gravity
in which the scalars can be 
coupled to each other, and they can be
minimally or nonminimally coupled to gravity.
This includes scalar fields in \gr{} as well as
popular quadratic gravity models such as \sGB, \dCS and axi-dilaton gravity.
In Sec.~\ref{appssec:ModelSelection}, we describe the 
parameter choices in \CanudaAD{} that need to be set 
to select the models presented in Sec.~\ref{sec:GravityModels}.
In Sec.~\ref{appssec:SFID}, we summarize the choices of axion and dilaton initial data that are available in the \CanudaAD code.

\subsection{Model selection  parameters}\label{appssec:ModelSelection}

Table ~\ref{tab:parameter choices} lists the parameters, and their appropriate settings, to select the models described in Sec.~\ref{sec:GravityModels}. 
Note that the dimensionless coupling parameter $\ha$ introduced in Sec.~\ref{ssec:4dstring} is given by the (dimensionless) parameters \texttt{aGB} and \texttt{aCS} in the code. 
That is, the user may choose a coupling such that \texttt{aGB} $\neq$ \texttt{aCS}
(e.g., when \sGB and \dCS gravity are simulated in the same run). 
This is not to be confused with the dimensionful coupling constants $\aGB$ and $\aCS$ used throughout the paper, all of which are related via \texttt{aGB}\,=\,$\haGB = \aGB/\textrm{M}^2$ and \texttt{aCS}\,=\,$\haCS=\aCS/\textrm{M}^2$,
where  $\textrm{M}$ is a characteristic mass scale of the considered problem.

The parameters \texttt{axi\_coupling} and \texttt{dil\_coupling} 
specify the coupling functions, $\fth$ and $\fph$, of the  axion and dilaton fields.
The parameters \texttt{dil\_lambda} and \texttt{AD\_lambda} refer to 
$\lambda_{\rm{GB}}$ 
and $\lambda_{\rm{AD}}$; 
see Eqs.~\eqref{eq:sGB_DilatonCoupling} and~\eqref{eq:axidilmodel}.
Finally, the parameter \texttt{AD\_coupling} refers to the function $\gph$ and determines the coupling function connecting the axion and dilaton fields.
Setting it to ``none'' decouples the fields and allows the user to 
individually
evolve \sGB and \dCS gravity
with the same simulation.

\subsection{Scalar field initial data, continued}\label{appssec:SFID}

The \CanudaAD code offers a suite of initial data choices for the axion and dilaton fields.
In Sec.~\ref{ssec:ScalarInitialData} we presented zero initial data and the approximate analytic solution in axi-dilaton gravity around a single \bh{.}
The latter coincides with the small-spin, small-coupling solutions in \dCS and \sGB gravity.
The initial data thorn also provides 
axion and dilaton profiles around \bbh{s}
through a superposition of the single \bh{} solutions.~\footnote{
Note that all scalar initial data are currently treated in the decoupling approximation, i.e., do not back-react onto the metric.
The construction of consistent metric--(bi-) scalar initial data
is beyond the scope of this paper.
\Canuda{} provides constraint-satisfying initial data of the Einstein--Klein-Gordon equations
in its {\textsc{Scalar}} module.
}
Here we list additional options and their applicability.

\noindent{\bf{Type III:} Gaussian}
We provide the option to initialize the axion and dilaton fields as Gaussian profiles centered around
$r=r_0$ as,
\begin{subequations}
\label{eq:IDThetaGauss}
\begin{align}
\Theta|_{t=0} & = \Theta_0 \,\mathrm{exp}\left( -\frac{(r-r_0)^2}{\sigma_{\Theta}^2}\right)\Sigma_{\Theta}(\theta,\phi)
\,, \\
\Ktheta|_{t=0}&=0
\,,
\end{align}
\end{subequations}
and 
\begin{subequations}
\label{eq:IDPhiGauss}
\begin{align}
\Phi|_{t=0}&=\Phi_0 \,\mathrm{exp}\left( -\frac{(r-r_0)^2}{\sigma_{\Phi}^2}\right)\Sigma_{\Phi}(\theta,\phi)
\,,\\
\Kphi|_{t=0}&=0
\,, 
\end{align}
\end{subequations}
where 
$\sigma_{\Theta}$ and $\sigma_{\Phi}$ are the Gaussians' width,
and $\Theta_{0}$ and $\Phi_{0}$ are their maximum amplitude.
The angular profiles $\Sigma_{\Theta}(\theta,\phi)$ and $\Sigma_{\Phi}(\theta,\phi)$
are given by a superposition of spherical harmonics, $Y_{lm}(\theta,\phi)$, that can be chosen by the user.
The code also offers a superposition of two Gaussians in $\Theta$ and $\Phi$
that can be used to set up field profiles around the individual \bh{s} of a binary.
This type of initial data is suitable for all models of gravity in the \CanudaAD{} code.

\begin{figure*}[t!]
    \centering
    \includegraphics[width = 0.48\textwidth]{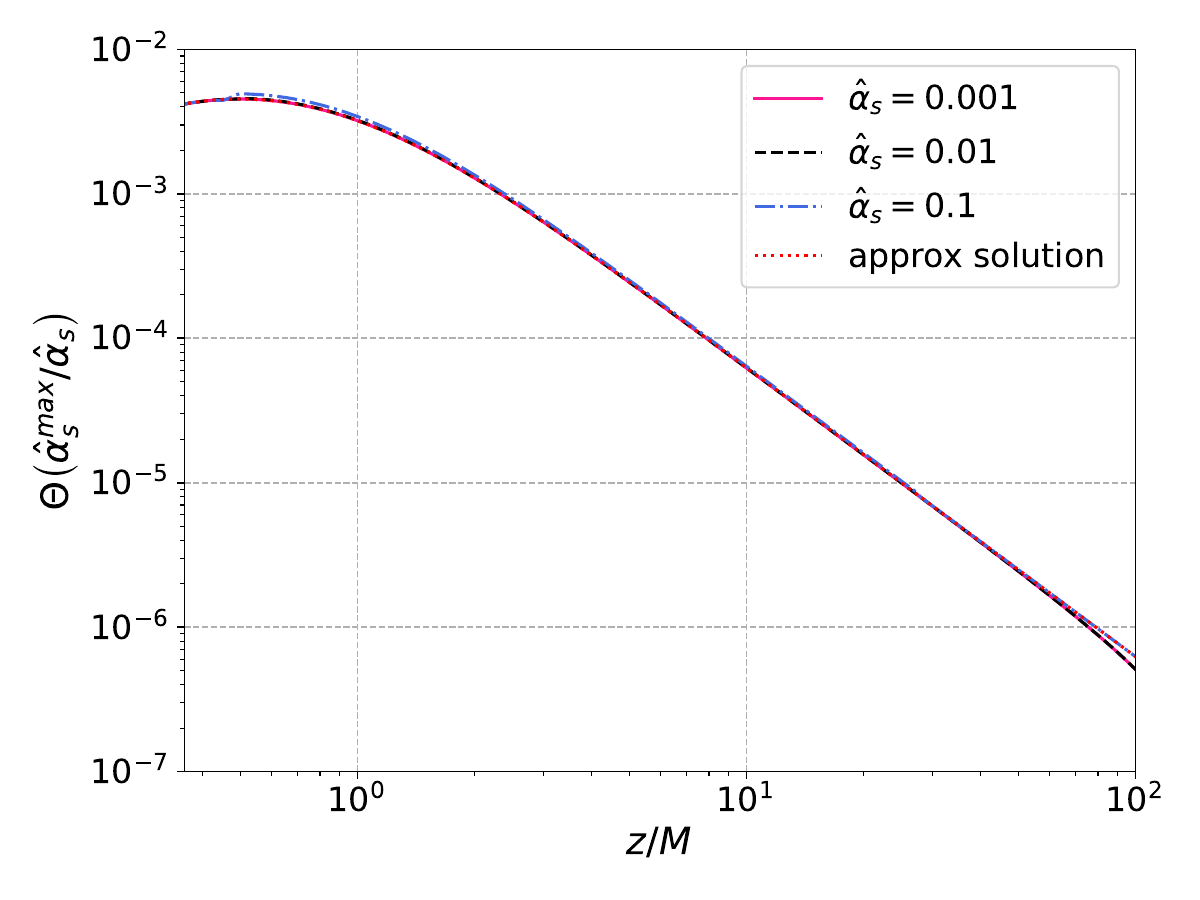}
    \includegraphics[width = 0.48\textwidth]{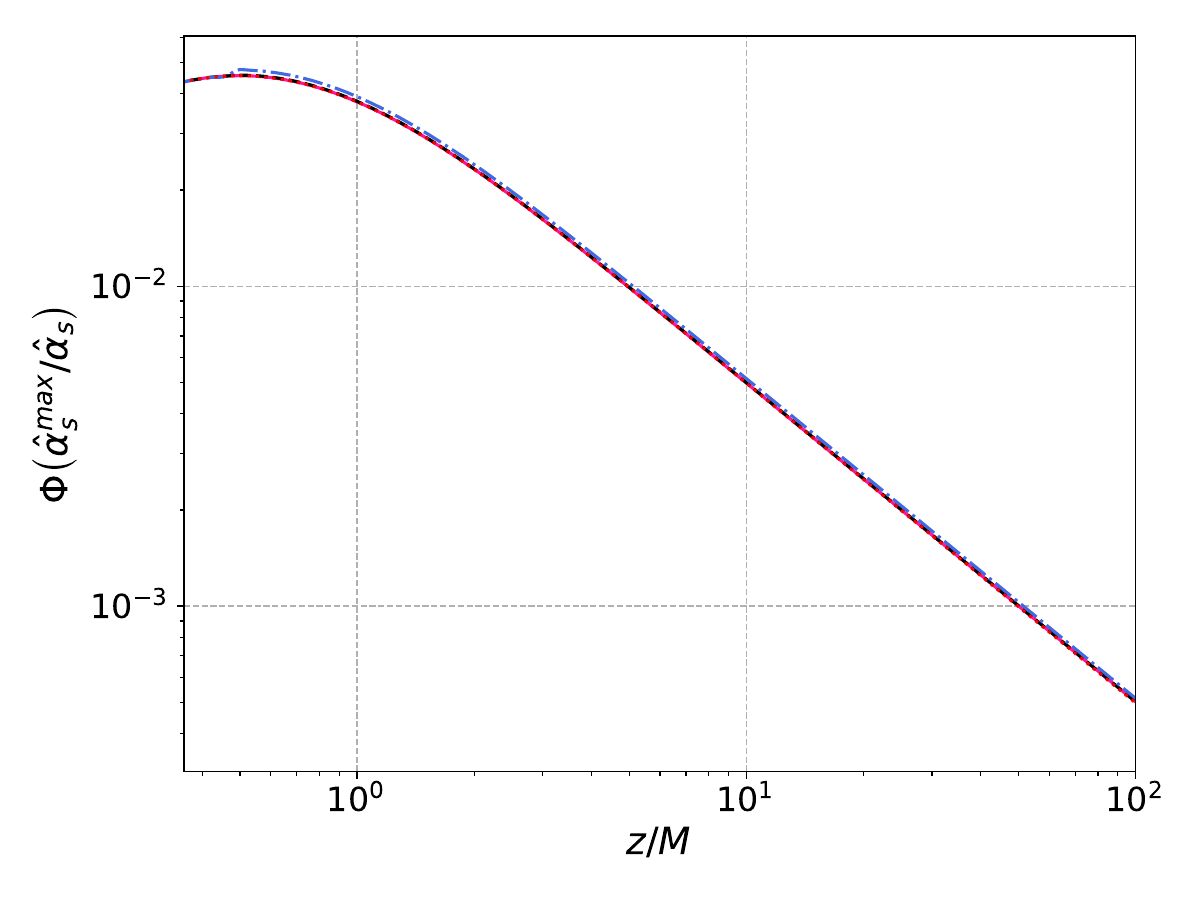}
    \caption{Profile of the axion (left panel) and dilaton (right panel) fields 
    along the z-axis,
    at $t=200$ M.
    The fields are evolved around a \bh{} with spin $\chi=0.1$.
    The numerical data is rescaled by $\hat{\alpha}_{\rm{s,max}}/\ha$, with $\hat{\alpha}_{\rm{s,max}} = 0.1$,
    for $\ha = 0.001$ (pink solid), $\ha = 0.01$ (black dashed), and $\ha = 0.1$ (blue dash-dotted).
    We validate the numerical data against the analytic solution in the small-spin, small-coupling approximation (red dotted line).}
\label{fig:ThetaPhi_vs_z_apt1_vary_aS}
\end{figure*}

\noindent{\bf{Type IV: Quasi-bound state of a massive scalar field in \gr{}
}}
We also implement the
quasi-bound state of a massive scalar field, with mass parameter $\mu_{\rm{S}}$, around a single \bh{} in \gr{.}
The implementation is available for both the axion and dilaton.
Here, we summarize the construction exemplarily for the axion.
The massive quasi-bound state is a solution of the Klein-Gordon equation
\begin{align}
\label{appeq:Klein-Gordon}
\left(\Box - \mu_{\rm{S}}^{2} \right)\Theta & = 0
\,,
\end{align}
with the d'Alembertian determined by the Kerr metric.
We implemented the solution found by Dolan~\cite{Dolan:2007mj} that is valid for any mass parameter $\mu_{\rm{S}}$ and  \bh{} spin $0\leq|\chi|<1$.
The full scalar field takes the form
\begin{align}
\Theta_{lm} & = \exp(-i\omega_{\Theta} t) \exp(i m \phi) \S_{lm}(\theta) \R_{lm}(\rBL)
\,,
\end{align}
where $\omega_{\Theta}$ is the field's frequency,
$\S_{lm}(\theta)$ are spheroidal harmonics
and  $\R_{lm}(\rBL)$ is the radial profile that we construct numerically.

The corresponding initial data is given by
\begin{subequations}
\begin{align}
\Theta_{lm}(t=0) & = \exp(i m \phi) \S_{lm}(\theta) \R_{lm}(\rBL)
\,,\\
\Ktheta{}_{lm}(t=0) & = i \omega_\Theta \Theta_{lm}(t=0)
\,.
\end{align}
\end{subequations}
We apply the same procedure to $\Phi$. 

\noindent{\bf{Type V: Dilaton bound state in quadratic \sGB gravity}}
The \CanudaAD{} code facilitates simulations of \bh{s} in different types of \sGB gravity that are determined by the coupling function $\fph$; see Eq.~\eqref{eq:sGB_CouplingFunctions}.

In particular, shift-symmetric \sGB or \EdGB yield unique \bh{s} with dilaton hair given by Type~II initial data in Sec.~\ref{ssec:ScalarInitialData}.
Instead, quadratic \sGB{,}
which is determined by the coupling function $\fph \propto \Phi^2$, 
has a richer phase-space of solutions that includes both the Kerr metric and scalarized solutions~\cite{Silva:2017uqg,Doneva:2017bvd}.
The latter refers to
a bound state of the scalar $\Phi$ around a single \bh{} located at $r=R$.
The scalar's profile is given by,
\begin{subequations}
\begin{align}
\Phi(t=0) & = \frac{M R}{\varsigma^2}\left(c_1+c_2\frac{M R}{\varsigma^2}+c_3\frac{(M R)^2}{\varsigma^4}\right)
\,,\\
\Kphi(t=0) & = 0
\,,
\end{align}
\end{subequations}
where $\varsigma = M + 2R$, and the fitting parameters are $c_1 = 3.68375$, $c_2 = 4.97242$, and $c_3 = 229.938$~\cite{Elley:2022ept}.
The \CanudaAD code also provides initial data around \bbh{s} through a superposition of the scalar bound states.

Recall that the initial data choices for the axion and dilaton fields can be chosen to be different from each other. 
Since the initial data here only describes the dilaton field and its momenta, the axion field and its respective momenta can be chosen to be any of the other initial data choices described in this paper.

\section{Numerical tests}

In this section, we present several numerical tests we performed to validate our results.
In Sec.~\ref{ssec:analytic_comaprison}, we compare our results with analytic approximations in the small-spin, small-coupling limit.
In Sec.~\ref{ssec:code_comparison}, we compare the results produced with our new \CanudaAD  module to
those obtained with our \CanudadCS~\cite{Richards:2023xsr} and \CanudaEdGB~\cite{Witek:2018dmd} codes, which have been thoroughly validated.
Finally, in Secs.~\ref{ssec:sbh_convergence} and ~\ref{ssec:bbh_convergence}, we present convergence tests for single and binary \bh{s}.

\begin{figure*}[t!]
    \centering
    \includegraphics[width = 0.49\textwidth]{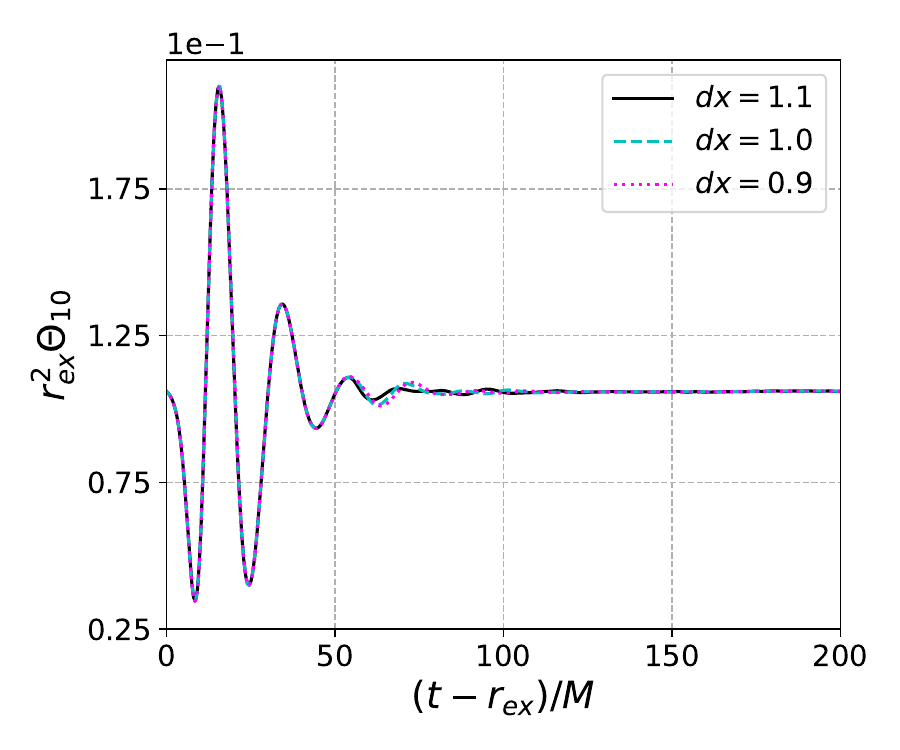}
    \includegraphics[width = 0.49\textwidth]{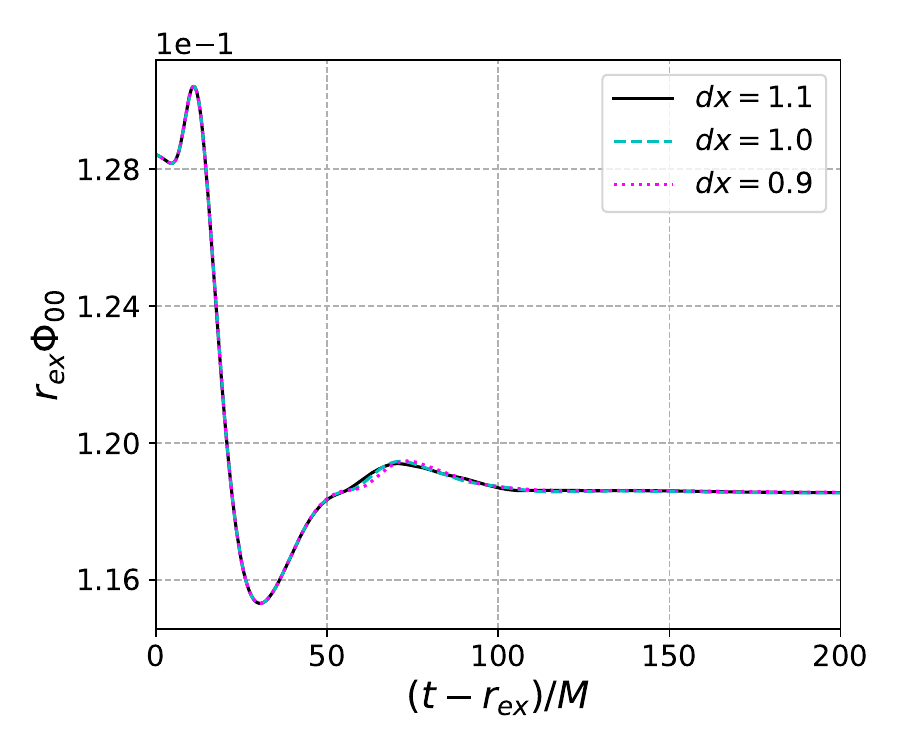}
    \caption{Evolution of the dominant multipoles of the axion (left panel) and dilaton (right panel) fields for different resolutions.}
    \label{appfig:sbh_resolutions}
\end{figure*}

\subsection{Comparison with approximate analytic solution}
\label{ssec:analytic_comaprison}

In Fig.~\ref{fig:ThetaPhi_vs_z_apt1_vary_aS}
we compare our numerical results against the approximate analytical solutions for the axion and dilaton hair (see Eqs.~\eqref{eq:IDAxionHair} and~\eqref{eq:IDDilatonHair}) and find excellent agreement.
Specifically, 
we show the profiles of the axion, $\Theta$, and dilaton, $\Phi$,
evolved in the background of slowly-rotating \bh{} with spin $\chi=0.1$
along the axis of rotation 
(z-axis)
at $t=200$M.
We rescale the fields by 
$\ha^{max}/\ha$
with $\ha^{max} = 0.1$ and
coupling strengths $\ha = \{0.001,0.01,0.1\}$.
Fig.~\ref{fig:ThetaPhi_vs_z_apt1_vary_aS} shows
excellent agreement between the numerical simulations around slowly-rotating \bh{s} and the small-spin analytic solutions.

\subsection{Code comparison}
\label{ssec:code_comparison}

To verify that our newly developed code, \CanudaAD, obtains physically sound results,
we benchmark it against our \CanudadCS~\cite{Richards:2023xsr} and \CanudaEdGB~\cite{Witek:2018dmd} codes.
They had been designed to simulate scalar fields in \dCS and \sGB gravity, respectively,
both of which are special cases in our new code.

We simulate the axion and dilaton fields around a single Kerr \bh{} with a moderate spin of $\chi=0.5$.
They are initialized as the approximate analytical solution (see Type~II in Sec.~\ref{ssec:ScalarInitialData})
which
also represents the small-spin, small-coupling solutions in \dCS and \sGB gravity. 

We run a simulation with \CanudaAD where the fields are ``decoupled", i.e., the axion and dilaton fields evolve at the same time but without any coupling between them, see Secs.~\ref{ssec:dCSgravity} and ~\ref{sssec:sGBgravity}. 
We compare the fields thus evolved with \CanudaAD to the analogous simulations of the axion with \CanudadCS and of the dilaton with \CanudaEdGB{.}
After the fields have settled to their final hairs, the decoupled \CanudaAD results
are in excellent agreement with those obtained from the 
\CanudadCS code for the axion hair and the \CanudaEdGB thorn for the dilaton hair.

\subsection{Single Black Hole Convergence Tests}
\label{ssec:sbh_convergence}

\begin{table}[]
    \centering
    \begin{tabular}{|c|c|c|}
    \hline
    Run & $dx/M$ & $h/M$ \\
    \hline
    \hline
    low & $1.1$  & $8.6 \times 10^{-3}$\\
    med & $1.0$  & $7.8 \times 10^{-3}$\\
    high & $0.9$ & $7.0 \times 10^{-3}$\\
    \hline
    \end{tabular}
    \caption{Simulations for convergence test around a single \bh{.}
    We ran the configuration
    IDII\_AD\_chi09
    in Table~\ref{tab:SBHSims}
    with three different resolutions.
    We list the resolution $\dif x/M$ on the outermost refinement level and the resolution $h=\dif x / 2^{n-1}$ (with $n=8$) on the innermost refinement level.}
    \label{tab:SBH_convergence}
\end{table}

In this section, we present convergence tests 
of the axion and dilaton fields evolved around a single \bh{}. 
For codes that use finite differencing, the numerical (i.e., discretized) data can be expressed in terms of the true values as
$\Theta_i = \Theta + c\, {dx}_{\rm i}^n$,
where $n$ is the convergence rate, $\rm i$ identifies a given resolution, $dx$ is the grid spacing, 
and $c$ is some function of time which does not depend on resolution.
Since we use mesh refinement, we use $dx$ to refer to the grid spacing on the coarsest grid. 
The residuals between different resolutions (low, medium, high) are then related by
\begin{equation}
\Theta_{\rm l} - \Theta_{\rm m} = Q_n \left(\Theta_{\rm m} - \Theta_{\rm h} \right)
\end{equation}
where the convergence factor is
\begin{equation}
    Q_n := \frac{|dx_{\rm l}^n-dx_{\rm m}^n|}{|dx_{\rm m}^n-dx_{\rm h}^n|}\,.
\end{equation}

\begin{figure*}[t!]
    \centering
    \includegraphics[width = 0.49\textwidth]{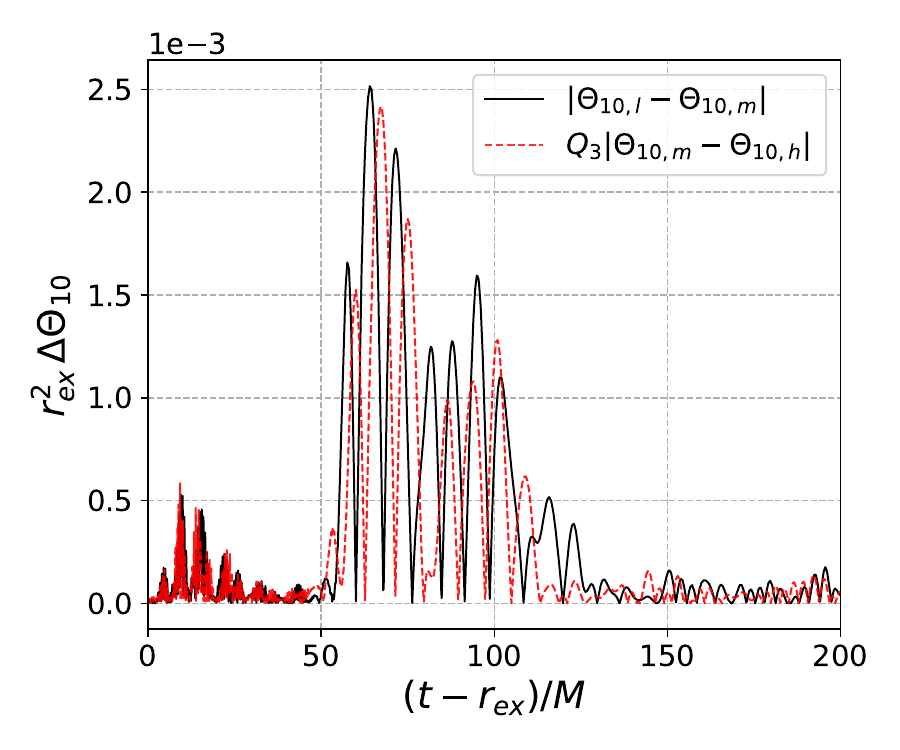}
    \includegraphics[width = 0.49\textwidth]{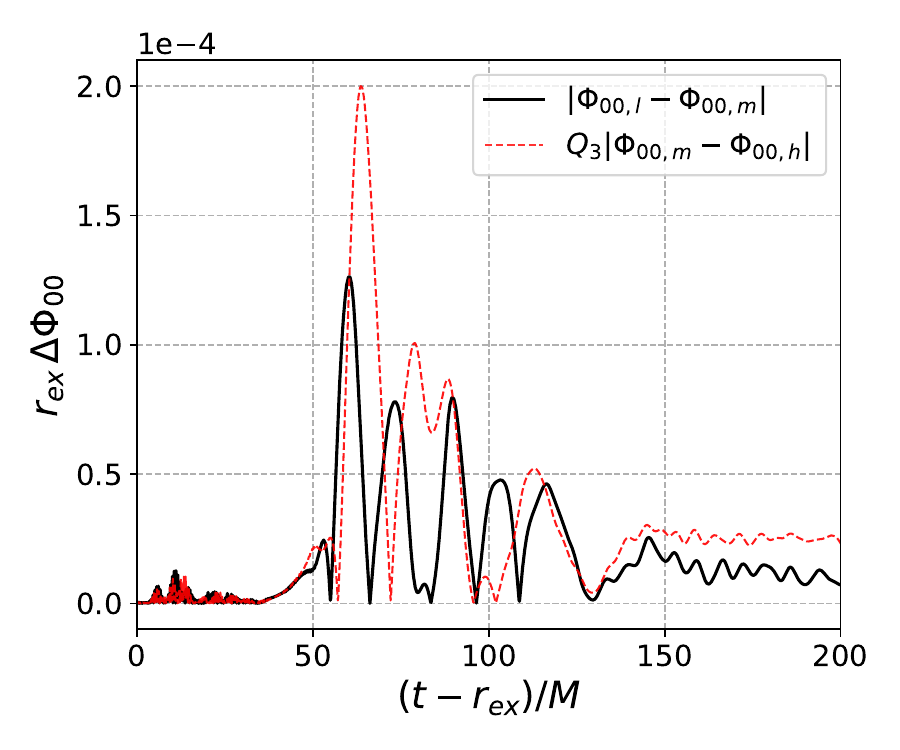}
    \caption{Convergence tests of the dominant multipoles of the axion (left panel) and dilaton (right panel) fields. 
    We show the residual between the low and medium resolution runs (black solid line), and the residual between the medium and high resolution runs rescaled by $Q_{3}$ (red dashed line).
    The latter indicates third order convergence. 
}
\label{appfig:sbh_convergence}
\end{figure*}

\begin{figure*}[t!]
    \centering
    \includegraphics[width = 0.49\textwidth]{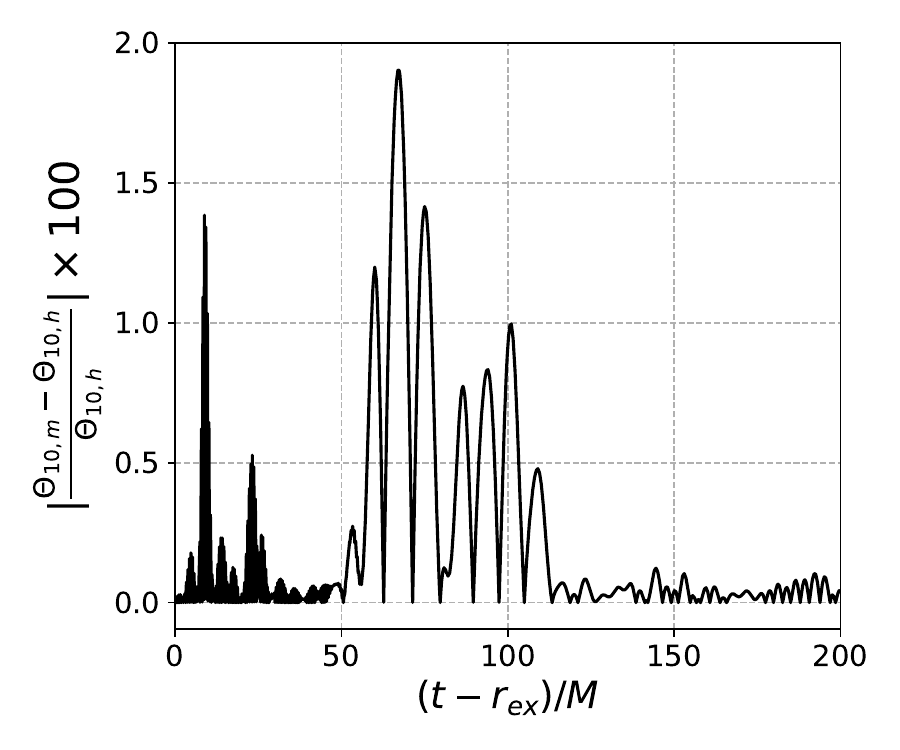}
    \includegraphics[width = 0.49\textwidth]{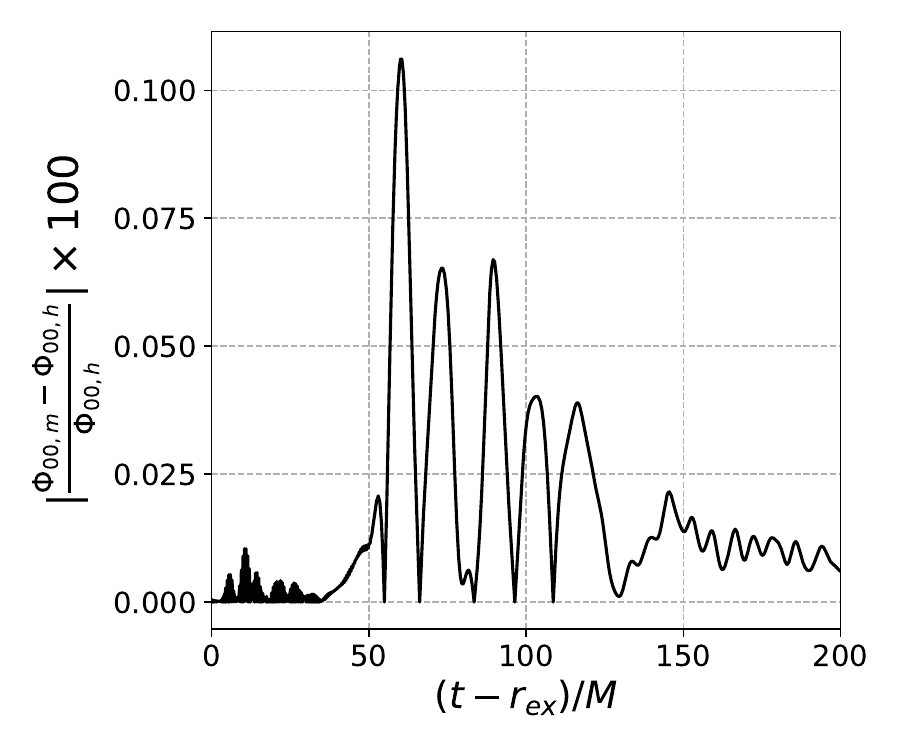}
    \caption{Percent error in the dominant modes of the axion (left) and dilaton (right) fields computed between the medium and high resolutions.}
    \label{appfig:percenterror}
\end{figure*}

We perform convergence tests with our most demanding parameters, \bh{} spin $\chi = 0.9$ and coupling strength $\ha = 0.1$ with the same grid set-up as described in Sec.~\ref{ssec:Simulations},
corresponding to run IDII\_AD\_chi09 
in Table~\ref{tab:SBHSims}
but with the resolutions listed in Table~\ref{tab:SBH_convergence}.
Fig.~\ref{appfig:sbh_resolutions} shows the dominant multipoles of the axion and dilaton fields for three different resolutions.
They are in excellent agreement, demonstrating that the data obtained with different resolutions is consistent.
Fig.~\ref{appfig:sbh_convergence} presents the convergence tests for the axion (left panel) and dilaton (right panel) fields.
Specifically, it shows the residuals between the low and medium resolutions, and the medium and high resolutions scaled by $Q_3 = 1.22$, 
indicating third order convergence.
Fig.~\ref{appfig:percenterror} shows the percent error between the medium and high resolutions for each of the fields.
At late times $t > 150M$, the fields settle to their final profiles and we find a numerical error of $\Delta \Theta_{10}/\Theta_{10,h}< 0.1\%$ (left panel) and $\Delta \Phi_{00}/\Phi_{00,h} < 0.02\%$ (right panel).

\vspace{-0.5cm}
\subsection{Binary Black Hole Convergence Tests}
\label{ssec:bbh_convergence}
\vspace{-0.2cm}

\begin{figure}[t!]
    \centering
    \includegraphics[width = 0.4\textwidth]{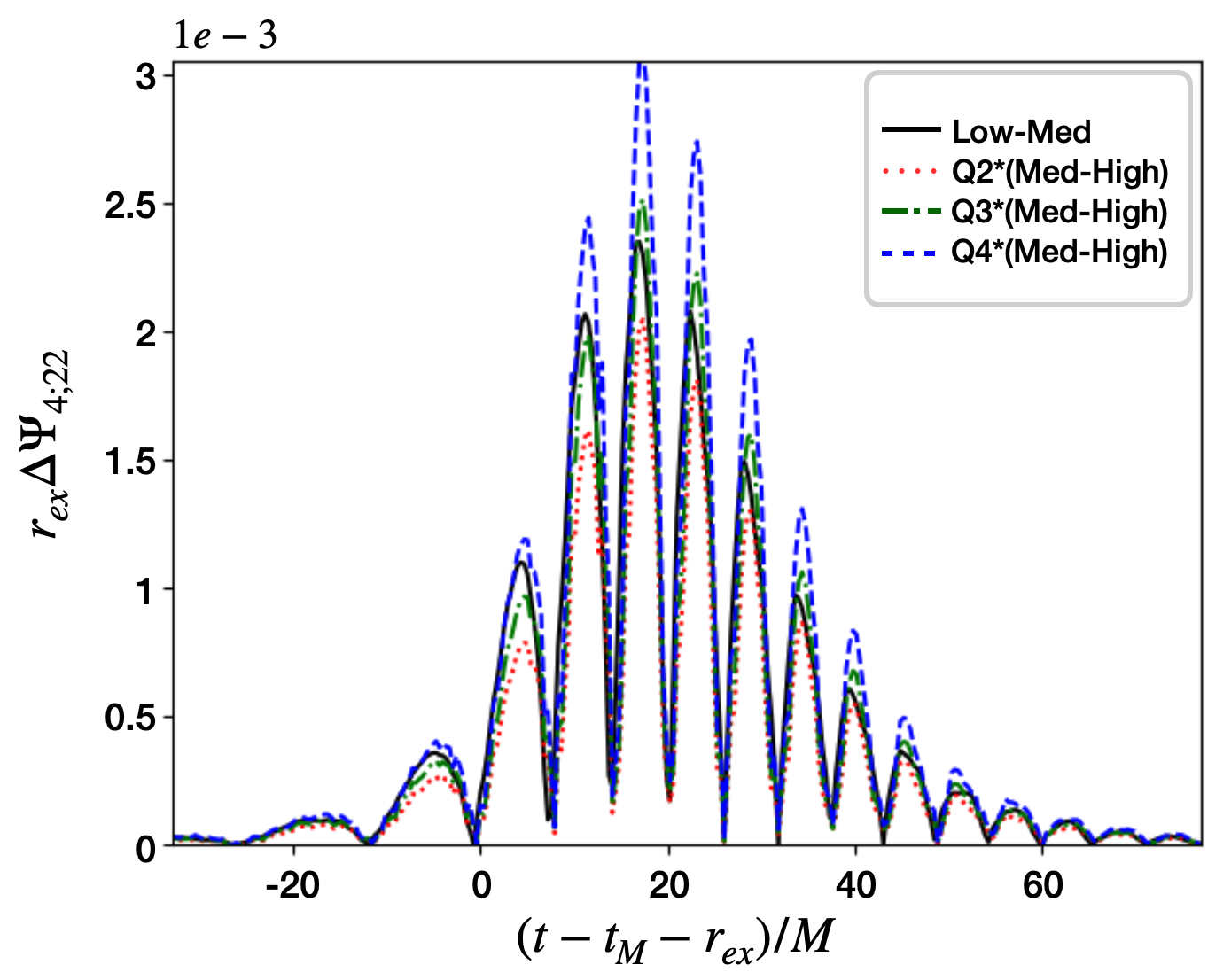}
    \caption{Convergence plots of $\Psi_{4;22}$. Binary simulation of equal-mass nonrotating BHs with coupling strength $\ha = 0.1$. 
    We compare the difference between the low and medium resolution run (black) with the medium and high resolution run.
    The latter is rescaled by $Q_2$ (red), $Q_3$ (green), and $Q_4$ (blue) indicating second, third, and fourth order convergence, respectively.}
    \label{fig:Psi422_conv_test}
\end{figure}

\begin{figure}[t!]
    \centering
    \includegraphics[width = 0.4\textwidth]{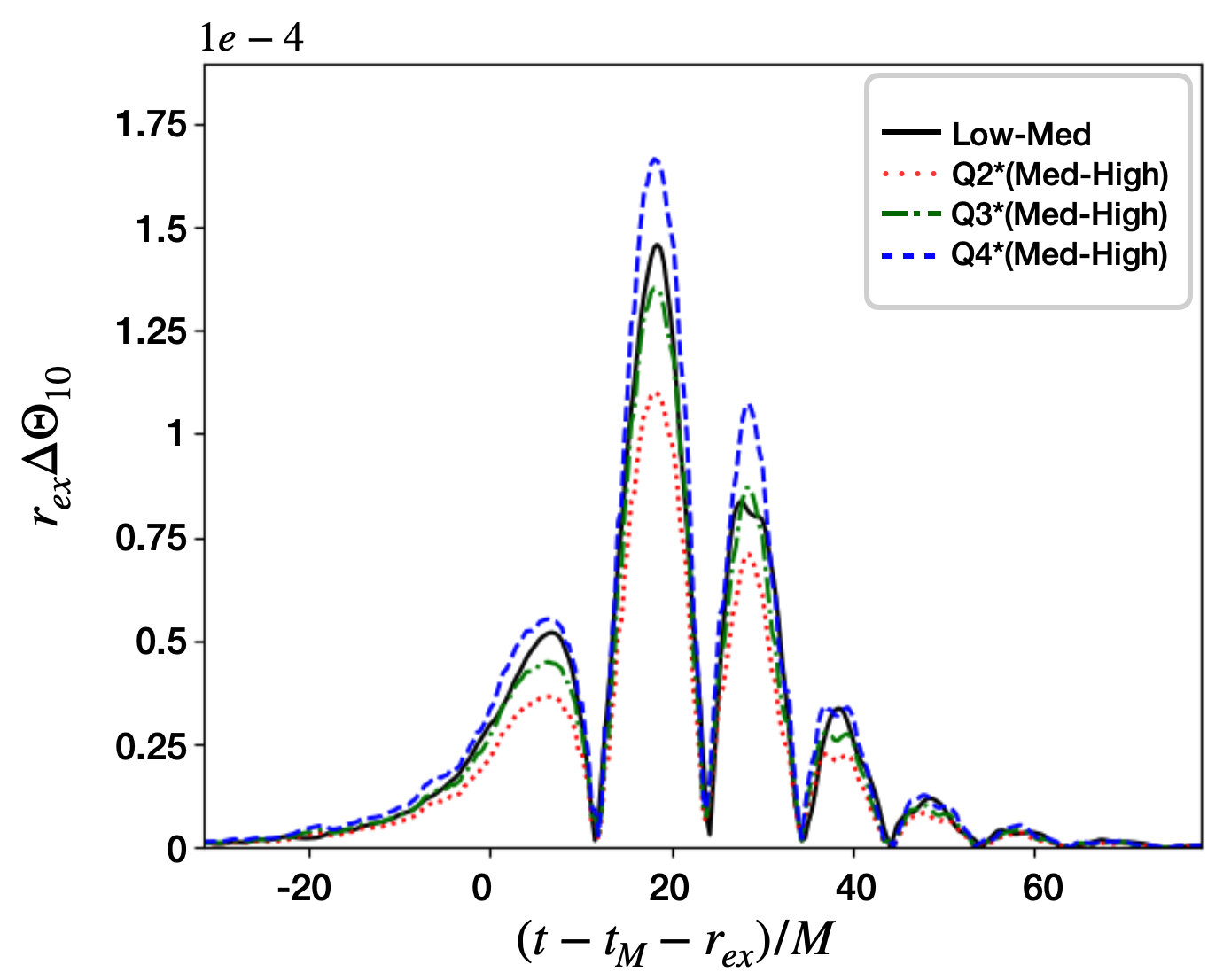}
    \caption{Same as Fig.~\ref{fig:Psi422_conv_test} but for $\Theta_{10}$.
    }
    \label{fig:Theta10_conv_test}
\end{figure}

\begin{figure}[t!]
    \centering
    \includegraphics[width = 0.4\textwidth]{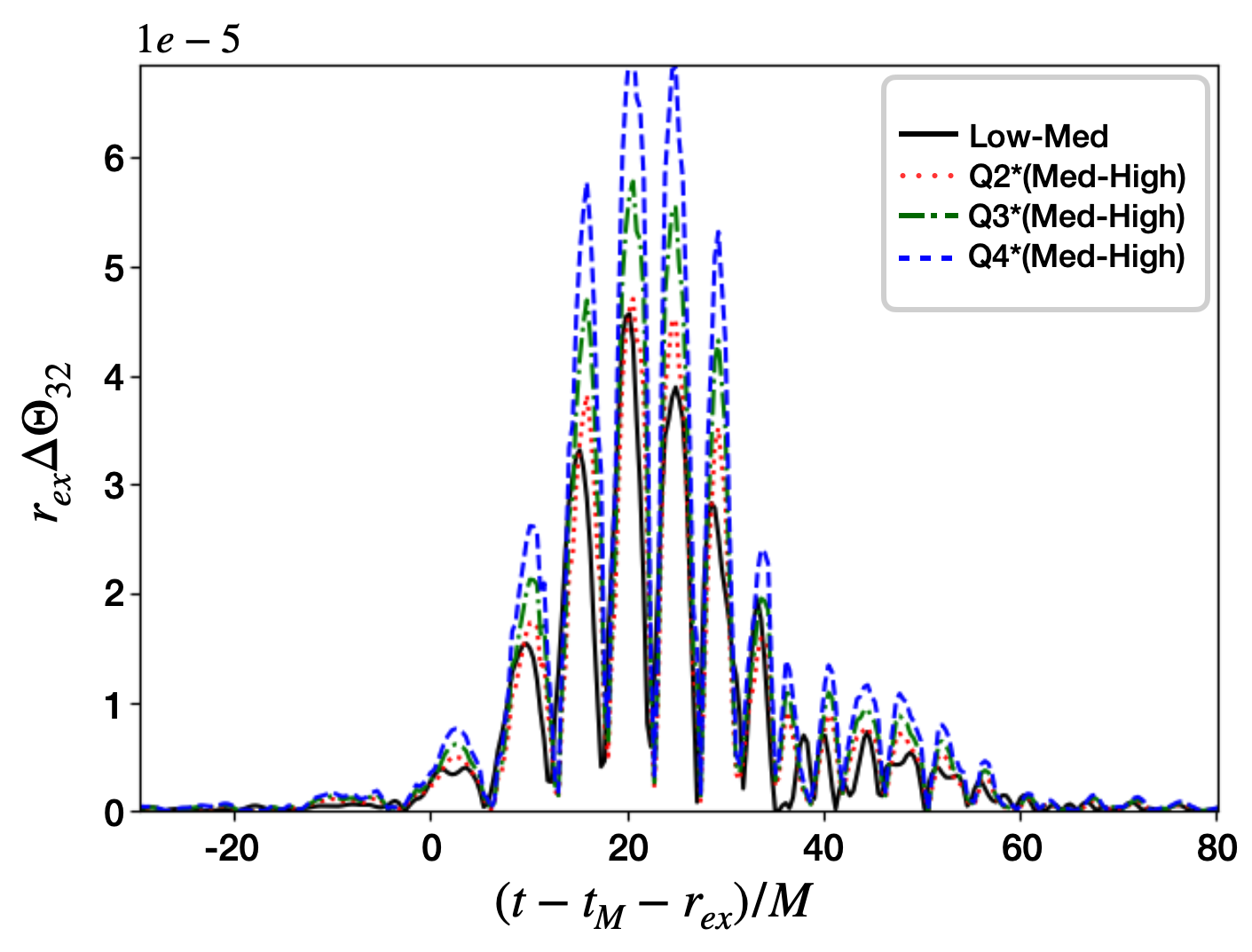}
    \caption{Same as Fig.~\ref{fig:Psi422_conv_test} but for $\Theta_{32}$.
    }
    \label{fig:Theta32_conv_test}
\end{figure}

\begin{figure}[t!]
    \centering
    \includegraphics[width = 0.4\textwidth]{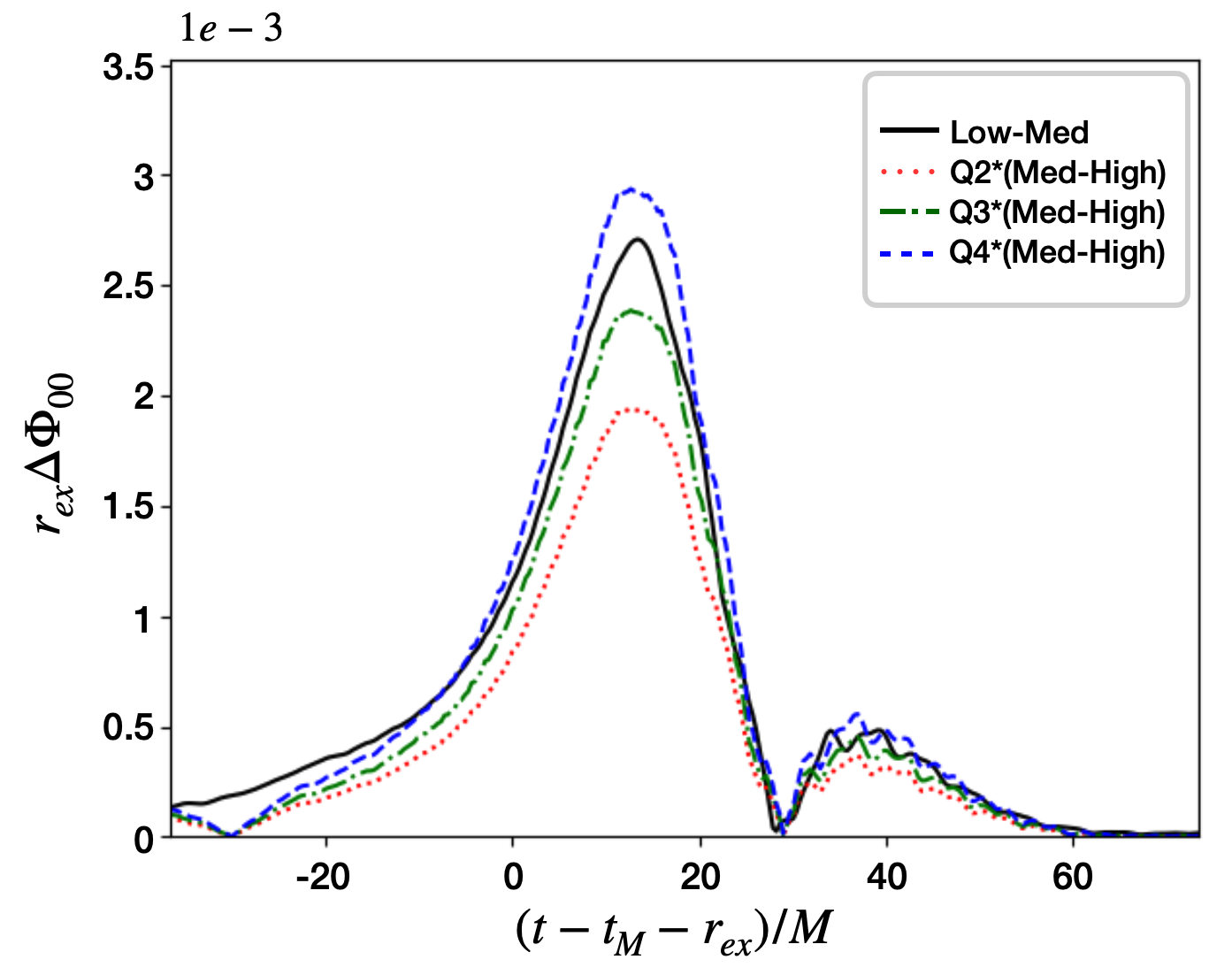}
    \caption{Same as Fig.~\ref{fig:Psi422_conv_test} but for $\Phi_{00}$.
    }
    \label{fig:Phi00_conv_test}
\end{figure}

\begin{figure}[t!]
    \centering
    \includegraphics[width = 0.4\textwidth]{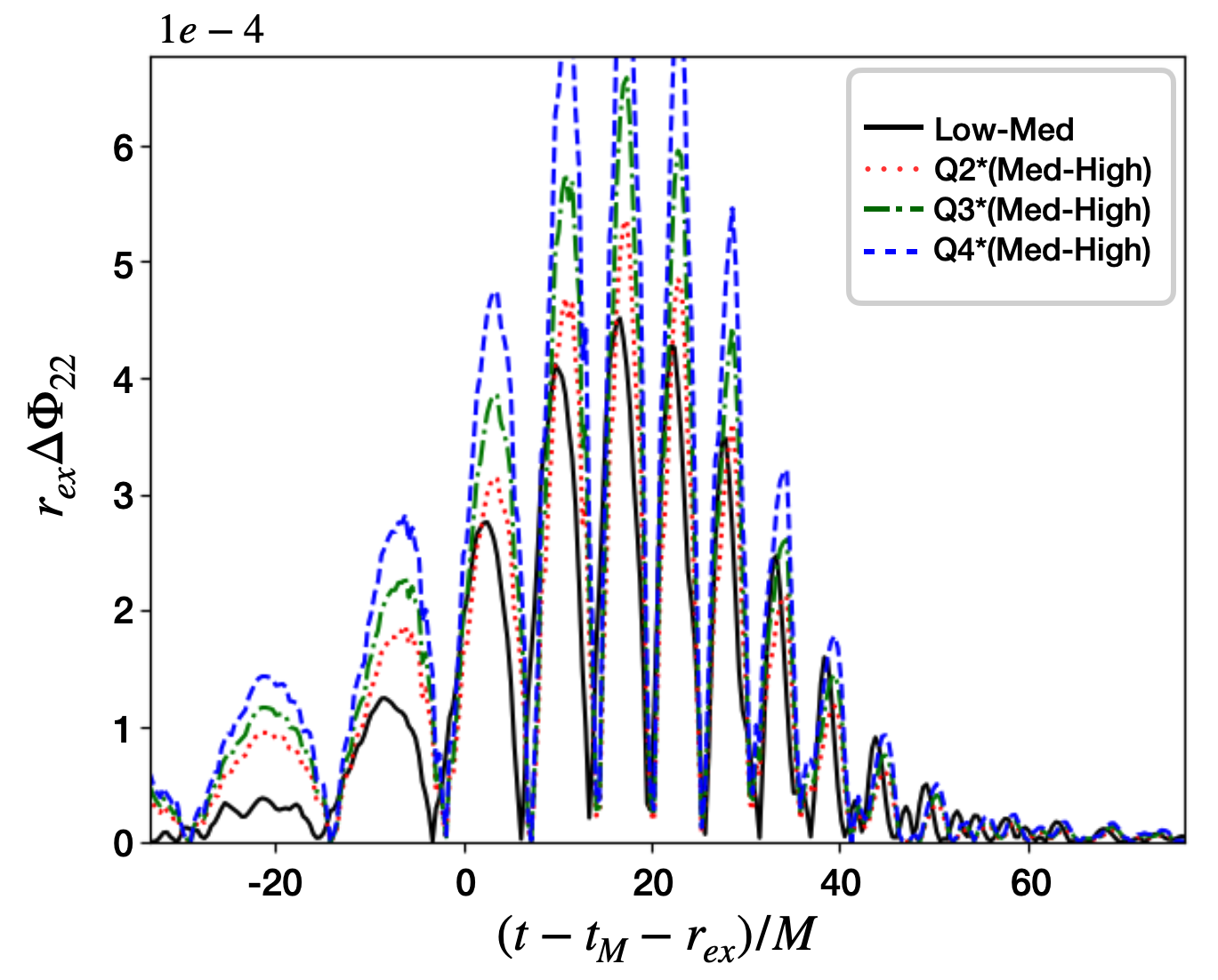}
    \caption{Same as Fig.~\ref{fig:Psi422_conv_test} but for $\Phi_{22}$.
    }
    \label{fig:Phi22_conv_test}
\end{figure}

To verify the accuracy of the \bbh{} results, we also perform a convergence study for a \bbh{} simulation with equal-mass, nonrotating BHs with coupling strength $\ha = 0.1$.
We use three resolutions with grid spacing on the coarsest refinement level of $dx_\textrm{l} = 1.6$, $dx_\textrm{m} = 1.28$, and $dx_\textrm{h} = 1.067$.
In Figs.~\ref{fig:Psi422_conv_test} - ~\ref{fig:Phi22_conv_test}, we test convergence for the Weyl scalar $\Psi_{4,22}$, the axion field $\Theta_{10}$ and $\Theta_{32}$, and the dilaton field $\Phi_{00}$ and $\Phi_{22}$.
For each quantity, we plot the difference between the low resolution and medium resolution simulations as well as the difference between the medium and high resolution simulations scaled by the factor $Q_n$.
By varying $n$, we find the convergence factor that best aligns the data.
The plots include $Q_2 = 1.84$, $Q_3 = 2.26$, and $Q_4 = 2.78$ indicating second, third, and fourth order convergence, respectively.
The convergence order differs slightly for each of the quantities plotted, but typically falls between three and four.

\section{$3+1$ formulation of quadratic gravity}\label{appsec:GeneralAxiDil3p1Eqs}

In this section, we present the general evolution equations for the axion and dilaton fields in quadratic or bi-scalar--tensor gravity.
We continue to work in the 
decoupling
approximation, i.e., considering the axion and dilaton fields in the background of a vacuum \gr{} spacetime.
To present the most general evolution equations (without the assumption that $\aGB=\aCS=\as$), we use dimensionless coupling parameters $\haCS = \aCS / M^2$ and $\haGB = \aGB / M^2$,
in analogy to $\ha = \as / M^2$ in the main text.
In the \CanudaAD code,
$\haCS$ and $\haGB$ correspond to the parameters \texttt{aCS}, \texttt{aGB}. 

We present vacuum Einstein's equations in their \bssn{} formulation in Sec.~\ref{appssec:GRBSSNform},
followed by the time evolution formulation of Eqs.~\eqref{eq:axionKG_decoup} and~\eqref{eq:dilatonKG_decoup}
in ADM-York form in Sec.~\ref{appssec:ADMYorkform}
and their \bssn{} form in Sec.~\ref{appssec:BSSNform}.

\subsection{Einstein's equations in \bssn{} form }\label{appssec:GRBSSNform}
For completeness, we present the evolution and constraint equations of the \gr{} background spacetime here.
We employ the \bssn{} formulation~\cite{Shibata:1995we,Baumgarte:1998te}
that consists of a clever constraint addition and
conformal rescaling of the evolution variables such that the resulting PDEs are strongly hyperbolic.
In \Canuda{} we use the $W$ version of the \bssn{} formulation, with variables
\begin{subequations}
\label{appeq:BSSNVars}
\begin{align}
W          = & \gamma^{-\frac{1}{6}}
\,,\quad
\tgam_{ij} =   W^{2} \gamma_{ij}
\,,\\
K          = & \gamma^{ij} K_{ij}
\,,\quad
\tA_{ij}   =   W^{2} A_{ij}
\,,\\
\tGam^{i}  = & \tgam^{kl} \tGam^{i}{}_{kl} = - \p_{j} \tgam^{ij}
\,,
\end{align}
\end{subequations}
where
$\gamma\equiv\det{\gamma_{ij}}$,
$\tgam_{ij}$ and $W$ are the conformal metric and factor, chosen such that $\tgam\equiv\det{\tgam_{ij}}=1$,
$\tA_{ij}$ is the conformal tracefree part of the extrinsic curvature,
and $\tGam^{i}$ is the conformal connection function.

\noindent{\bf{Constraints of the \gr background:}}
As is common in numerical relativity, we adopt a free evolution scheme. That is, we solve the constraints
only on the initial time slice to obtain initial data. 

During the evolution we monitor the constraints
to verify that they are satisfied.
Thus, they provide an excellent first benchmark test for the simulations.
In terms of the \bssn{} variables, Eqs.~\eqref{appeq:BSSNVars},
the Hamiltonian and momementum constraints in vacuum are given by
\begin{subequations}
\label{appeq:ConstraintsTenGabBSSN}
\begin{align}
\label{appeq:HamiltonianBSSN}
\H      = & R - \tA^{ij} \tA_{ij} + \frac{2}{3} K^2
        = 0
\,,\\
\label{appeq:MomentumBSSN}
\M_{i}  = & \tD^{j} \tA_{ij} - \frac{2}{3} \tD_{i} K - \frac{3}{W} \tA^{j}{}_{i} \tD_{j} W 
        =   0
\,,
\end{align}
\end{subequations}
where $\tD_{i}$ and $\tR_{ij}$ are the covariant derivative and Ricci tensor associated with the conformal metric $\tgam_{ij}$,
and we use the physical Ricci scalar as auxiliary variable,
\begin{align}
\label{appeq:RicciScalarBSSN}
R = & W^2 \tR + 4 W \tD^{i} \tD_{i} W - 6 \tD^{i} W \tD_{i} W
\,.
\end{align}

\noindent{\textbf{Evolution equations of the GR background:}}

After appropriate constraint addition to eliminate the Ricci scalar and divergence of the
(tracefree part of the)
extrinsic curvature, we obtain
\begin{subequations}
\label{appeq:EvolGabBSSN}
\begin{align}
\label{appeq:EvolBSSNW}
\dif_{t} W          = & \frac{1}{3}\alpha W K
\,,\\
\label{appeq:EvolBSSNtgam}
\dif_{t} \tgam_{ij} = & - 2 \alpha \tA_{ij}
\,,\\
\label{appeq:EvolBSSNtrk}
\dif_{t} K          = & - D^{i} D_{i} \alpha
        + \alpha \left( \tA^{ij} \tA_{ij} + \frac{1}{3} K^{2} \right)
\,,\\
\label{appeq:EvolBSSNAij}
\dif_{t} \tA_{ij}   = & - W^2 [D_{i} D_{j} \alpha ]^{\rm tf}
\\ &
        + \alpha\left( W^2 R^{\rm tf}_{ij} + K \tA_{ij} - 2 \tA_{ik} \tA^{k}{}_{j} \right)
\,,\nonumber\\
\label{appeq:EvolBSSNtGam}
\dif_{t} \tGam^{i}  = & - 2 \tA^{ik} \p_{k}\alpha
        + \tgam^{kl} \p_{k}\p_{l} \beta^{i}
        + \frac{1}{3} \tgam^{ik} \p_{k}\p_{l}\beta^{l}
\\ &
+ 2 \alpha \left( \tGam^{i}{}_{kl} \tA^{kl}
        - \frac{3}{W} \tA^{ik} \p_{k} W
        -\frac{2}{3} \tD^{i} K
\right)
\,,\nonumber
\end{align}
\end{subequations}
where
$X^{\rm tf}_{ij} = X_{ij} - \frac{1}{3} \gamma_{ij} \gamma^{kl}X_{kl}$ denotes the trace-free part of a tensor taken with respect to the {\textit{physical}} metric, 
$\dif_{t} = \p_{t} - \Lie_{\beta}$,
and $\Lie_{\beta}$ denotes the Lie derivative 
along the shift vector.
We introduced the abbreviations
\begin{subequations}
\begin{align}
\label{appeq:RijADMvsBSSN}
R_{ij}              = & \tR_{ij}
        + \frac{1}{W}\left( \tD_{i}\tD_{j}W + \tgam_{ij} \tD^{k}\tD_{k}W \right)
\\ &
        - \frac{2}{W^2} \tgam_{ij} \tD^{k}W \tD_{k}W
\,,\nonumber\\
D_{i} D_{j} \alpha  = & \tD_{i} \tD_{j}\alpha
        +\frac{2}{W} \tD_{(i}W \tD_{j)} \alpha
\\ &
        - \frac{1}{W} \tgam_{ij} \tD^{k} W \tD_{k} \alpha
\,,\nonumber
\end{align}
\end{subequations}
and we denote $D^{i}D_{i}\alpha=\gamma^{ij}D_{i}D_{j}\alpha$.
These equations are evolved in conjunction with the moving puncture gauge~\cite{Campanelli:2005dd,Baker:2005vv}.


\subsection{Scalar equations in ADM-York form}\label{appssec:ADMYorkform}

We now rewrite the
axion's and dilaton's equations of motion,
Eqs.~\eqref{eq:axionKG_decoup} and~\eqref{eq:dilatonKG_decoup},
as a set of time evolution equations in ADM-York form.
We follow the same procedure as discussed in Sec.~\ref{ssec:ScalarEvolutionEquations},
and find
\begin{subequations}
\label{appeq:AxiDil_gen_ADM}
\begin{align}
\label{appeq:dttheta_gen_ADM}
\dif_{\rm t} \Theta        = & -   \alpha \Ktheta
\,,\\
\label{appeq:dtphi_gen_ADM}
\dif_{\rm t} \Phi        = & -   \alpha \Kphi
\,,\\
\label{appeq:dtKtheta_gen_ADM}
\dif_{t} \Ktheta =  & - \alpha D^{i} D_{i} \Theta - D^{i}\alpha D_{i} \Theta 
\nonumber\\
+ & \alpha \left( K \Ktheta + \dVth - \frac{\haCS M^2}{4}\frac{\dfth}{\gph^2}\RCS \right)
\nonumber\\
+ & 2 \alpha\frac{\dgph}{\gph}\left( \frac{}{}\Ktheta \Kphi - D_i\Theta D^i\Phi \right)
\,,\\
\label{appeq:dtKphi_gen_ADM}
\dif_{t} \Kphi =  & - \alpha D^{i} D_{i} \Phi  - D^{i}\alpha D_{i} \Phi
\nonumber\\
 + & \alpha \left( K  \Kphi + \dVph - \frac{\haGB M^2}{4}\dfph \RGB \right)
\nonumber\\
- & \alpha \dgph\gph\left(\frac{}{} \Ktheta^2 - 2 \Vth - D_i\Theta D^i\Theta \right)
\,.
\end{align}
\end{subequations}
The topological invariants, $\RCS$ and $\RGB$, appearing in the right-hand-side of the above equations are defined in Eqs.~\eqref{eq:RCSRGBinEBWeyl} in terms of the gravito-electromagnetic tensors, $E_{ij}$ and $B_{ij}$.
The latter are defined in Eqs.~\eqref{eq:EijBijInGR}. 

To estimate the energy-momentum due to the axion and dilaton, it is useful to compute the projections of the effective energy momentum tensor given in Eq.~\eqref{eq:Tmneff}.
Before deriving those expressions, we re-write the auxiliary tensors
$\E_{a}$, $\F_{ab}$ and $\G_{ab}$ (see Eqs.~\eqref{eq:tensorAux}) in terms of the induced metric and extrinsic curvature.

The projections of the auxiliary tensor $\F_{ab}$ appear in the \dCS modifications, and are given by
\begin{subequations}
\label{appeq:Fab3p1}
\begin{align}
\F_{nn} = & \ddfth \Ktheta^2  
    + \frac{\haCS M^2}{4}\frac{\dfth{}^2}{\gph{}^2}\RCS
\\ &
    + \dfth \left( D_kD^k\Theta  - \Ktheta K - \dVth \right)
\nonumber\\ &
    - 2\dfth\frac{\dgph}{\gph} \left( \Ktheta\Kphi - D_k\Phi D^k\Theta \right)
\,,\nonumber\\
\F_{i}  = & \ddfth \Ktheta D_{i}\Theta
    + \dfth\left(D_{i}\Ktheta - K^{j}{}_{i} D_{i}\Theta \right)
\,,\\
\F_{ij} = & \ddfth D_{i}\Theta D_{j}\Theta
    + \dfth\left(D_{i}D_{j}\Theta - \Ktheta K_{ij} \right)
\,,
\end{align}
\end{subequations}
\begingroup
\allowdisplaybreaks
where we used 
Eq.~\eqref{appeq:dtKtheta_gen_ADM} to derive $\F_{nn}$.
In a vacuum \gr background, $R_{ab} = 0 = \nabla^e R_{ab}$, so terms multiplying auxiliary tensor $\E_{a}$ 
vanish and thus
do not contribute to the effective energy--momentum tensor.
Nonetheless, we state its projections here for completeness,
\begin{align}
\label{eq:Ea3p1}
\E_{nn} = & \dfth \Ktheta
\,,\quad
\E_{i}  =   \dfth \D_{i}\Theta
\,.
\end{align}

The projections of the auxiliary tensor $\G_{ab}$,
appearing in the contributions from the Gauss-Bonnet term,  are
\begin{subequations}
\label{appeq:Gab3p1}
\begin{align}
\G_{nn} = & \ddfph \Kphi^2 
    + \frac{\haGB M^2}{4}\dfph{}^2\RGB
\\ &
    + \dfph \left( D_kD^k\Phi - \dVph - \Kphi K \right)
\nonumber\\ &
    + \dfph\dgph\gph\left( \Ktheta^2 - 2 \Vth - D_k\Theta D^k\Theta \right)
\,\nonumber\\
\G_{i}  = & \ddfph \Kphi D_{i}\Phi
    + \dfph \left(D_{i}\Kphi - K^{j}{}_{i}D_{j}\Phi \right)
\,,\\
\G_{ij} = & \ddfph D_{i}\Phi D_{j} \Phi 
    + \dfph\left(D_{i}D_{j}\Phi - \Kphi K_{ij} \right)
\,,
\end{align}
\end{subequations}
where we used the scalar field evolution equation, Eq.~\eqref{appeq:dtKphi_gen_ADM}, to derive $\G_{nn}$.
In the following, we denote the traces $\F=\gamma^{ij}\F_{ij}$ and $\G=\gamma^{ij}\G_{ij}$.
\endgroup

With all auxiliary tensors in place, we can write down the projections of the effective energy-momentum tensor, Eq.~\eqref{eq:Tmneff}.
The (effective)
energy density $\rho^{\rm{eff}} = n^{a} n^{b} T^{\rm eff}_{ab}$,
energy-momentum flux $j^{\rm eff}_{i} = - \gamma^{a}{}_{i} n^{b}T^{\rm eff}_{ab}$,
and spatial stress-tensor
$S^{\rm eff}_{ij} = \gamma^{a}{}_{i} \gamma^{b}{}_{j} T_{ab}$
become,
\begin{widetext}
\begin{subequations}
\label{appeq:SplitEffTmunuVac}
\begin{align}
\label{appeq:TeffRho}
\rho^{\rm eff} & = 
2 M^2 \left(\haCS B^{ij}\F_{ij} - \haGB E^{ij}\G_{ij} \right) +  \frac{1}{2} \left( \Kphi^2 + 2\Vph + D_i\Phi D^i\Phi \right)
+ \frac{\gph^2}{2} \left( \Ktheta^2 + 2\Vth + D_i\Theta D^i\Theta \right)
\,,\\
\label{appeq:TeffJi}
j^{\rm eff}_{i} & = 
2 \haCS M^2 \left( B_{ij}\F^j - \epsilon_{ijk} E^{lj}\F_{l}{}^{k}\right) - 2 \haGB M^2\left(  E_{ij}\G^j  + \epsilon_{ijk} B^{lj}\G_{l}{}^{k})
\right) + \left( \Kphi D_i\Phi + \gph^2 \Ktheta D_i\Theta \right)
\,,\\
\label{appeq:TeffSij}
S^{\rm eff}_{ij} & =
2 \haCS M^2\left(
    B_{ij}\left(\F + \F_{nn} \right)
    + \gamma_{ij} B^{kl}\F_{kl}
    - 2 B^{k}{}_{(i} \F_{j)k}
    + 2 \epsilon_{(i|}{}^{kl} \F_{k} E_{|j)l}
\right)
\nonumber\\ & \quad
-2 \haGB M^2\left( 
    E_{ij}\left(\G + \G_{nn} \right)
    + \gamma_{ij} E^{kl}\G_{kl}
    - 2 E^{k}{}_{(i} \G_{j)k}
    - 2 \epsilon_{(i|}{}^{kl} \G_{k} B_{|j)l}
\right)
\nonumber\\ & \quad
+ \gph^{2}\left( D_i\Theta D_j\Theta + \frac{1}{2} \gamma_{ij} \left( \Ktheta^2 - 2\Vth - D_k\Theta D^k\Theta \right) \right)
+ \left( D_i\Phi D_j\Phi 
    +  \frac{1}{2} \gamma_{ij} \left( 
        \Kphi^2 - 2 \Vph 
        - D_k\Phi D^k\Phi \right) \right)
\,.
\end{align}
\end{subequations}
\end{widetext}

\subsection{Scalar equations in \bssn{} form}\label{appssec:BSSNform}

We implement the axion and dilaton equations in their \bssn{} form in \CanudaAD{,}
\begin{subequations}
\label{appeq:EvolScalarBSSNKinematic}
\begin{align}
\label{appeq:dtthetaBSSN}
\dif_{t} \Theta  = &  - \alpha \Ktheta
\,,\\
\label{appeq:dtphiBSSN}
\dif_{t} \Phi  = & - \alpha \Kphi
\,,\\
\label{appeq:dtKthetaBSSN}
\dif_{t} \Ktheta = & - W^2 \tD^{k}\alpha \tD_{k} \Theta
+ \alpha \left(  K \Ktheta - W^2 \tD^i \tD_{i} \Theta \right.
\nonumber\\
& \left. + W \tD^i \Theta \tD_{i} W + \dVth - \frac{\haCS M^2}{4} \frac{\dfth}{\gph^2} \RCS \right)
\nonumber\\
& + 2 \alpha \frac{\dgph}{\gph}\left(
\Ktheta \Kphi - W^2 \tD^k\Phi \tD_k\Theta
\right)
\,,\\
\label{appeq:dtKphiBSSN}
\dif_{t} \Kphi = & - W^2 \tD^{k}\alpha \tD_{k} \Phi
+ \alpha \left(  K \Kphi - W^2 \tD^i \tD_{i} \Phi \right.
\nonumber\\
& \left. + W \tD^i \Phi \tD_{i} W + \dVph - \frac{\haGB M^2}{4} \dfph \RGB \right)
\nonumber\\
& - \alpha \dgph\gph\left(
\Ktheta^2 - 2 \Vth - W^2 \tD^k\Theta\tD_k\Theta
\right)
\,,
\end{align}
\end{subequations}

The curvature invariants are expressed as,
\begin{subequations}
\begin{align}
\RCS = & - 16 \tE^{ij}\tB_{ij}
\,,\\
\RGB = & 8 \left( \tE^{ij}\tE_{ij} - \tB^{ij}\tB_{ij} \right)
\,,
\end{align}
\end{subequations}
where
\begin{subequations}
\begin{align}
\tE_{ij} = & W^2 E_{ij}  
\\
    = &  W^2 R^{\rm tf}_{ij} + \frac{1}{3} \left( \tgam_{ij}\tA^{kl}\tA_{kl} + K \tA_{ij} - 3\tA_{i}{}^{k}\tA_{kj} \right)
\,,\nonumber\\
\tB_{ij} = & W^2 B_{ij}
\\
    = & \tilde{\epsilon}_{(i|}{}^{kl} \tA_{|j)k}  \tD_{l}W
    + W \tilde{\epsilon}_{(i|}{}^{kl} \tD_{k}\tA_{|j)l}
\,.\nonumber
\end{align}
\end{subequations}

The auxiliary tensors become
\begin{subequations}
\label{appeq:FabWBSSN}
\begin{align}
\tscF_{nn} = & 
      \ddfth \Ktheta^2
    + \frac{\haCS M^2}{4} \frac{\dfth{}^{2}}{\gph^2}\RCS
\\ &
    + \dfth\left( W^2 \tD^{k}\tD_{k} \Theta - W \tD^{k}W \tD_{k}\Theta - K \Ktheta - \dVth
    \right)
\nonumber\\ &
    + 2 \frac{\dfth\dgph}{\gph} \left( W^2 \tD^{k}\Theta \tD_{k}\Phi -\Ktheta \Kphi 
    \right)
\,,\nonumber\\
\tscF_{i} = & 
    \ddfth \Ktheta \tD_{i}\Theta
    + \dfth \left(\tD_{i}\Ktheta - \tA^{j}{}_{i} \tD_{j}\Theta - \frac{1}{3} K \tD_{i} \Theta
    \right)
\,,\\
\tscF_{ij} = & W^2 \F_{ij}
\nonumber \\ = &
    W^2 \dfth \tD_{i}\tD_{j} \Theta
    + W^2 \ddfth \tD_{i}\Theta \tD_{j} \Theta
\\ &
    + W \dfth \left( 2 \tD_{(i} W \tD_{j)} \Theta - \tgam_{ij} \tD^{k} W \tD_{k} \Theta \right)
\nonumber \\ &
    - \dfth \Ktheta \left( \tA_{ij} + \frac{1}{3} \tgam_{ij} K \right)
\,,\nonumber 
\end{align}
\end{subequations}
and
\begin{subequations}
\label{appeq:GabWBSSN}
\begin{align}
\tscG_{nn} = & 
      \ddfph \Kphi^2
    + \frac{\haGB M^2}{4}\dfph{}^{2} \RGB 
\\ &
    + \gph \dgph \dfph \left( \Ktheta^2 - W^2 \tD^{k}\Theta \tD_{k}\Theta - 2 \Vth \right)
\nonumber \\ &
    + \dfph\left(W^2 \tD^{k}\tD_{k}\Phi - W \tD^{k} W \tD_{k} \Phi - K \Kphi - \dVph \right)
\,,\nonumber\\
\tscG_{i} = &
      \ddfph \Kphi \tD_{i} \Phi
    + \dfph\left( \tD_{i} \Kphi - \tA^{j}{}_{i} \tD_{j} \Phi - \frac{1}{3} K\tD_{i}\Phi \right)
\,,\\
\tscG_{ij} = & W^{2} \G_{ij} 
\nonumber \\ = &
      W^2 \dfph  \tD_{i}\tD_{j}\Phi
    + W^2 \ddfph \tD_{i}\Phi \tD_{j}\Phi
\\ &
    + W \dfph \left( 2 \tD_{(i}\Phi \tD_{j)} W - \tgam_{ij} \tD^{k} W \tD_{k} \Phi \right)
\nonumber \\ &
    - \dfph \Kphi \left( \tA_{ij} + \frac{1}{3} \tgam_{ij} K \right)
\,.\nonumber
\end{align}
\end{subequations}
We denote the traces
$\tscF{}=\tgam^{kl}\tscF_{kl}$
and $\tscG=\tgam^{kl}\tscG_{kl}$.

Finally, the projections of the effective energy-momentum tensor are
\begin{widetext}
\begin{subequations}
\label{appeq:SplitEffTmunuVacBSSN}
\begin{align}
\rho^{\rm eff} = & 
    \frac{1}{2}\left(W^2 \tD^{k}\Phi \tD_{k}\Phi + \Kphi^2 + 2 \Vph \right)
    + \frac{\gph{}^2}{2} \left(W^2 \tD^{k}\Theta \tD_{k}\Theta + \Ktheta^2 + 2 \Vth \right)
    + 2 M^2\left(\haCS \tB^{kl}\tscF_{kl} - \haGB \tE^{kl} \tscG_{kl} \right)
\,,\\
j^{\rm eff}_{i} = & 
    \Kphi \tD_{i} \Phi + \gph{}^2 \Ktheta \tD_{i} \Theta
    + 2\,\haCS M^2 \left( \tB_{i}{}^{j} \tscF_{j} - \frac{1}{W} \tilde{\epsilon}_{i}{}^{jk} \tE^{l}{}_{j}\tscF_{kl} \right)
    - 2\,\haGB M^2 \left(\tE_{i}{}^{j}\tscG_{j} + \frac{1}{W} \tilde{\epsilon}_{i}{}^{jk} \tB^{l}{}_{j}\tscG_{kl} \right)
\,,\\
S^{\rm eff}_{ij} = & 
       \tD_i\Phi \tD_j\Phi 
    +  \frac{\tgam_{ij}}{2 W^2} \left( \Kphi^2 - 2\Vph - W^2 \tD_k\Phi \tD^k\Phi \right)
    + \gph^{2} \left( \tD_i\Theta \tD_j\Theta 
        + \frac{\tgam_{ij}}{2W^2} \left( \Ktheta^2 - 2 \Vth - W^2 \tD_k\Theta \tD^k\Theta\right) \right)
\\ &
+ 2\, \frac{\haCS M^2}{W^{2}} \left( 
        \tB_{ij} (\tscF + \tscF_{nn} )
    - 2 \tB^{k}{}_{(i}\tscF_{j)k} 
    +   \tgam_{ij}  \tB^{kl}\tscF_{kl} 
    {\color{black}{+}} 2 W \tilde{\epsilon}_{(i|}{}^{kl} \tE_{|j)k} \tscF_{l}
\right)
\nonumber \\ &
- 2\, \frac{\haGB M^2}{W^{2}} \left( 
        \tE_{ij} (\tscG + \tscG_{nn}) 
    - 2 \tE^{k}{}_{(i}\tscG_{j)k} 
    +   \tgam_{ij} \tE^{kl}\tscG_{kl}
    {\color{black}{-}} 2 W \tilde{\epsilon}_{(i|}{}^{kl} \tB_{|j)k} \tscG_{l}
\right) 
\,.\nonumber
\end{align}
\end{subequations}
\end{widetext}

\bibliographystyle{apsrev4-2}
\bibliography{Refs_Axidilaton.bib,Refs_websites_software.bib}

\end{document}